\documentclass[journal]{IEEEtran}

\usepackage{orcidlink,amsmath,amssymb,amsfonts,algorithm,algpseudocode,booktabs,makecell,pifont,csquotes,comment,tabularx,diagbox,multirow,cite,fancyvrb,framed,listings,url,caption,subcaption,graphicx,pgfplots,adjustbox,relsize}
\usepackage[listings,skins]{tcolorbox}
\usepackage[binary-units]{siunitx}
\hypersetup{hidelinks}
\PassOptionsToPackage{hyphens}{url}
\usepackage{xcolor}
\definecolor{color1}{HTML}{00305d}
\definecolor{color2}{HTML}{7c166a}
\definecolor{color3}{HTML}{e20074}
\definecolor{color4}{HTML}{33305d}
\definecolor{TUDBlue100}{RGB}{0, 48, 93}
\definecolor{TUDBlue90}{RGB}{26, 69, 109}
\definecolor{TUDBlue80}{RGB}{51, 89, 125}
\definecolor{TUDBlue70}{RGB}{77, 110, 142}
\definecolor{TUDBlue60}{RGB}{102, 131, 158}
\definecolor{TUDBlue50}{RGB}{128, 152, 174}
\definecolor{TUDBlue40}{RGB}{153, 172, 190}
\definecolor{TUDBlue30}{RGB}{179, 193, 206}
\definecolor{TUDBlue20}{RGB}{204, 214, 223}
\definecolor{TUDBlue10}{RGB}{230, 234, 239}

\pgfplotsset{compat=1.17}
\usetikzlibrary{arrows,backgrounds,shapes,decorations.pathreplacing,calc,shadows.blur,positioning}

\tikzset{
	/tikz/render blur shadow/.code={
		\pgfbs@savebb
		\pgfsyssoftpath@getcurrentpath{\pgfbs@input@path}%
		\pgfbs@compute@shadow@bbox
		\pgfbs@process@rounding{\pgfbs@input@path}{\pgfbs@fadepath}%
		\pgfbs@apply@canvas@transform
		\colorlet{pstb@shadow@color}{white!\pgfbs@opacity!\my@shadow@color}%
		\pgfdeclarefading{shadowfading}{\pgfbs@paint@fading}%
		\pgfsetfillcolor{\my@shadow@color}%
		\pgfsetfading{shadowfading}%
		{\pgftransformshift{\pgfpoint{\pgfbs@midx}{\pgfbs@midy}}}%
		\pgfbs@usebbox{fill}%
		\pgfbs@restorebb
	},}
\tikzset{
	/tikz/shadow color/.store in=\my@shadow@color,
	/tikz/shadow color=gray,
}

\pgfkeys{/tikz/tikzit fill/.initial=0}
\pgfkeys{/tikz/tikzit draw/.initial=0}
\pgfkeys{/tikz/tikzit shape/.initial=0}
\pgfkeys{/tikz/tikzit category/.initial=0}

\tikzstyle{none}=[inner sep=0mm]
\tikzstyle{red dot}=[fill=red, draw=black, shape=circle]
\tikzstyle{box}=[fill=white, draw=black, shape=rectangle]
\tikzstyle{textbox}=[fill=none, draw=none, shape=rectangle]
\tikzstyle{TUDbox0}=[fill=color1!100, draw=color1!100, shape=rectangle,text=white]
\tikzstyle{TUDbox1}=[fill=color1!70, draw=color1!100, shape=rectangle,text=white]
\tikzstyle{TUDbox2}=[fill=color1!50, draw=color1!100, shape=rectangle]
\tikzstyle{TUDbox3}=[fill=color1!10, draw=color1!100, shape=rectangle]
\tikzstyle{TUDroundbox3}=[fill=white, draw=color1!100, shape=rectangle, rounded corners=3pt]
\tikzstyle{TUDroundbox0}=[fill=color1!100, draw=color1!100, shape=rectangle, rounded corners=3pt]
\tikzstyle{TUDroundbox1}=[fill=color1!70, draw=color1!100, shape=rectangle, rounded corners=3pt]
\tikzstyle{TUDroundbox2}=[fill=color1!50, draw=color1!100, shape=rectangle, rounded corners=3pt]
\tikzstyle{TUDroundbox2error}=[fill=color2!50, draw=color1!100, shape=rectangle, rounded corners=3pt]
\tikzstyle{TUDroundbox3}=[fill=color1!10, draw=color1!100, shape=rectangle, rounded corners=3pt]
\tikzstyle{decision}=[fill=color1!10, draw=color1!100, shape=diamond]
\tikzstyle{dot}=[fill=color1!50, draw=black, shape=circle]
\tikzstyle{arrow1}=[->]
\tikzstyle{arrow2}=[<->]
\tikzstyle{startstop} = [rectangle, rounded corners, minimum width=3cm, minimum height=1cm,text centered, draw=black, fill=red!30]
\tikzstyle{io} = [trapezium, trapezium left angle=70, trapezium right angle=110, minimum width=3cm, minimum height=1cm, text centered, draw=black, fill=blue!30]
\tikzstyle{process} = [rectangle, minimum width=3cm, minimum height=1cm, text centered, draw=black]
\tikzstyle{tikzfig}=[baseline=-0.25em,scale=0.5]

\usepackage[acronym,shortcuts]{glossaries}
\newacronym{tas}{TAS}{Time-aware Shaper}
\newacronym{dut}{DuT}{Device-under-Test}
\newacronym{tsn}{TSN}{Time-Sensitive Networking}
\newacronym{qos}{QoS}{Quality-of-Service}
\newacronym{cdf}{CDF}{Cumulative Distribution Function}
\newacronym{ccdf}{CCDF}{Complementary \gls{cdf}}
\newacronym{sfd}{SFD}{Start of Frame Delimiter}
\newacronym{tsa}{TSA}{Transmission Selection Algorithm}
\newacronym{vhdl}{VHDL}{Very High Speed Integrated Circuit Hardware Description Language}
\newacronym{cbs}{CBS}{Credit-based Shaper}
\newacronym{tcc}{TCC}{Time Coordinated Computing}
\newacronym{sep}{SEP}{Switched Endpoint}
\newacronym{vbr}{VBR}{Variable Bit Rate}
\newacronym{cbr}{CBR}{Constant Bit Rate}
\newacronym{pcp}{PCP}{Priority Code Point}
\newacronym{ipc}{IP-Core}{Intellectual Property Core}
\newacronym{soc}{SoC}{System on Chip}
\newacronym{vlan}{VLAN}{virtual LAN}
\newacronym{cots}{COTS}{Commercial-of-the-Shelf}
\newacronym{sdn}{SDN}{Software-defined Networking}
\newacronym{ti}{TI}{Tactile Internet}
\newacronym{urllc}{URLLC}{Ultra-Reliable Low-Latency Communication}
\newacronym{lsb}{LSB}{Least Significant Bit}
\newacronym{pdu}{PDU}{Packet Data Unit}
\newacronym{phy}{PHY}{Physical Layer Chip}
\newacronym{spq}{SPQ}{Strict Priority Queuing}
\newacronym{pdv}{PDV}{Packet Delay Variation}
\newacronym{lcfs}{LCFS}{Last-Come-First-Served}
\newacronym{cnc}{CNC}{Central Network Controller}
\newacronym{ptp}{PTP}{Precision Time Protocol}
\newacronym{kpi}{KPI}{Key Performance Indicator}
\newacronym{rms}{RMS}{root mean square}
\newacronym{pc}{PC}{Personal Computer}
\newacronym{etf}{ETF}{Earliest Time First}
\newacronym{5gs}{5GS}{5G System}
\newacronym{fifo}{FIFO}{First-In-First-Out}
\newacronym{tx}{TX}{transmitter}
\newacronym{vm}{VM}{Virtual Machine}
\newacronym{gm}{GM}{Grand Master}
\newacronym{bmc}{BMC}{Best Master Clock}
\newacronym{gptp}{gPTP}{generalized Precision Time Protocol}
\newacronym{mac}{MAC}{Medium Access Control Layer}
\newacronym{skb}{SKB}{Socket Buffer}
\newacronym{nic}{NIC}{Network Interface Card}
\newacronym{fpga}{FPGA}{Field-programmable Gate Array}
\newacronym{rt}{RT}{Real-time}
\newacronym{gcl}{GCL}{Gate Control List}
\newacronym{tdma}{TDMA}{Time-Division Multiple Access}
\newacronym{avb}{AVB}{Audio/Video Bridging}
\newacronym{tg}{TG}{Task Group}
\newacronym{rx}{RX}{receiver}
\newacronym{rfc}{RFC}{Request for Comments}
\newacronym{dpdk}{DPDK}{Data Plane Development Kit}
\newacronym{i40}{I4.0}{Industry~4.0}
\newacronym{gnss}{GNSS}{Global Navigation Satellite Systems}
\newacronym{ats}{ATS}{Asynchronous Traffic Shaping}
\newacronym{fp}{FP}{Frame Preemption}
\newacronym{psfp}{PSFP}{Per-Stream Filtering and Policing}
\newacronym{cts}{CUTS}{Cut-through Switching}
\newacronym{srp}{SRP}{Stream Reservation Protocol}
\newacronym{cqf}{CQF}{Cyclic Queuing and Forwarding}
\newacronym{ubs}{UBS}{Urgency-Based Scheduler}
\newacronym{frer}{FRER}{Frame Replication and Elimination for Reliability}
\newacronym{lan}{LAN}{Local Area Network}
\newacronym{wan}{WAN}{Wide Area Network}
\newacronym{ivn}{IVN}{Intra-Vehicle Networking}
\newacronym{be}{BE}{Best-Effort}
\newacronym{sms}{SMS}{Switch-Memory-Switch}
\newacronym{ppm}{ppm}{Parts Per Million}
\newacronym{sp}{SPQ}{Strict Priority Queueing}
\newacronym{capex}{CAPEX}{Capital Expenditure}
\newacronym{p2p}{P2P}{Peer-to-Peer}
\newacronym{ts}{TS}{Timestamp}
\newacronym{udp}{UDP}{User Datagram Protocol}
\newacronym{tcp}{TCP}{Transmission Control Protocol}
\newacronym{cpu}{CPU}{Central Processing Unit}
\newacronym{stream1}{$\Theta$}{stream set $\Theta$}
\newacronym{stream2}{$\Psi$}{stream set $\Psi$}
\newacronym{stream3}{$\Omega$}{stream set $\Omega$}
\newacronym{stream1h}{$\Theta_{high}$}{Generic~1}
\newacronym{stream1m}{$\Theta_{med}$}{Generic~2}
\newacronym{stream1l}{$\Theta_{low}$}{Generic~3}
\newacronym{stream2h}{$\Psi_{high}$}{Tactile}
\newacronym{stream2m}{$\Psi_{med}$}{Audio~\acrshort{vbr}}
\newacronym{stream2l}{$\Psi_{low}$}{Video~\acrshort{vbr}}
\newacronym{stream3h}{$\Omega_{high}$}{Spot CTRL}
\newacronym{stream3m}{$\Omega_{med}$}{Audio~\acrshort{cbr}}
\newacronym{stream3l}{$\Omega_{low}$}{Video~\acrshort{cbr}}

\newcommand{\coloronename}{\textcolor{color1}{blue}}
\newcommand{\colortwoname}{\textcolor{color2}{purple}}
\newcommand{\colorthreename}{\textcolor{color3}{magenta}}
\newcommand{\cmark}{\ding{51}}
\newcommand{\xmark}{\ding{55}}
\newcommand{\gclblk}{\colorbox{color1}{\textcolor{white}{X}}}
\newcommand{\gclblkGB}{\colorbox{gray!60}{\textcolor{white}{G}}}

\begin{document}
\title{{\normalsize This work has been submitted to the IEEE for possible publication in Transactions on Network and Service\vspace{-0.6cm} Management~(TNSM). Copyright may be transferred without notice, after which this version may no longer be accessible.\vspace{0.8cm}}
TSN-FlexTest: Flexible TSN Measurement Testbed (Extended Version)\thanks{An abridged preliminary version appeared in~\cite{9844050}.}
}

\author{Marian~Ulbricht\,\orcidlink{0000-0001-7536-3111},
        Stefan~Senk\,\orcidlink{0000-0003-0745-2264},
	    Hosein~K.~Nazari\,\orcidlink{0000-0003-1632-4782},
	    How-Hang~Liu\,\orcidlink{0000-0003-0158-360X},\\
	    Martin~Reisslein\,\orcidlink{0000-0003-1606-233X},
	    Giang~T.~Nguyen\,\orcidlink{0000-0001-7008-1537},
	    and
	    Frank~H.~P.~Fitzek\,\orcidlink{0000-0001-8469-9573}%
\thanks{Marian~Ulbricht and Stefan~Senk contributed equally to this work.}
\thanks{M.~Ulbricht, S.~Senk, H.~K.~Nazari, H.-H.~Liu, and F.~H.~P.~Fitzek are with Technische Universität Dresden (TUD), Deutsche Telekom Chair of Communication Networks}%
\thanks{M.~Reisslein is with the School of Electrical, Computer and Energy Eng., Arizona State Univ.,  Tempe, AZ 85287-5706, USA, Email: reisslein@asu.edu}%
\thanks{G.~T.~Nguyen is with Technische Universität Dresden, Junior Professorship of Haptic Communication Systems}%
\thanks{G.~T.~Nguyen and F.~H.~P.~Fitzek are with the Cluster of Excellence “Centre for Tactile Internet with Human-in-the-Loop” (CeTI) of TUD}%
}



\maketitle

\begin{abstract}
Robust, reliable, and deterministic networks are essential for a variety of applications.
In order to provide guaranteed communication network services, \acrfull{tsn} unites a set of standards for time-synchronization, flow control, enhanced reliability, and management.
We design the TSN-FlexTest testbed with generic commodity hardware and open-source software components to enable flexible \acrshort{tsn} measurements.
We have conducted extensive measurements to validate the TSN-FlexTest testbed and to examine \acrshort{tsn} characteristics.
The measurements provide insights into the effects of \acrshort{tsn} configurations, such as increasing the number of synchronization messages for the \acrlong{ptp}, indicating that a measurement precision of 15\,ns can be achieved.
The \acrshort{tsn} measurements included extensive evaluations of the \acrlong{tas} (\acrshort{tas}) for sets of \acrfull{ti} packet traffic streams.
The measurements elucidate the effects of different scheduling and shaping approaches, while revealing the need for pervasive network control that synchronizes the sending nodes with the network switches.
We present the first measurements of distributed \acrshort{tas} with synchronized senders on a commodity hardware testbed, demonstrating the same \acrlong{qos} as with dedicated wires for high-priority \acrshort{ti} streams despite a 200\% over-saturation cross traffic load.
The testbed is provided as an open-source project to facilitate future \acrshort{tsn} research.
\end{abstract}
\thispagestyle{empty} 
\begin{IEEEkeywords}
Ethernet, Industrial Communication, \acrlong{qos}, Testbed, \acrlong{tsn}.
\end{IEEEkeywords}

\section{Introduction}
\label{sec:introduction}
There are various reasons of unexpected behaviors in communication networks, such as changing numbers of connected devices, sudden spikes in network utilization, or even loss of entire transmission paths due to unpredictable events.
For a wide range of applications and domains it is necessary to avoid these unexpected behaviors.
Among others, the industrial, automotive, medical, and avionic domains have strict requirements on the underlying communication infrastructure~\cite{beh2021tim,fal2022dyn,gav2022con,hoe2021imp,sac2022res,s22041404}.

To mitigate problems in the communication infrastructure, deterministic data transmission is preferred.
Determinism in networks can be described as maintaining full-knowledge and control of packet-based transmissions.
However, achieving determinism in data communication is challenging.
Depending on the use case, extensive efforts must be made to ensure the network quality for critical services.
\gls{tsn} unites different mechanisms to achieve deterministic communication over Ethernet networks.
\gls{tsn} is managed by the IEEE~\gls{tsn}~\gls{tg} which defines the underlying standards.
\gls{tsn} emerged from the former \gls{avb}~\acrlong{tg} which had a focus on streaming services.
With the new \gls{tsn} \gls{tg}, the attention moved towards more generic applications, such as industrial communication.

\gls{tsn} mainly encompasses four categories: time-synchronization, latency and packet delay variation reduction through flow control, ultra-reliability, and resource management.
Each category improves specific aspects of data communication.
The standards can be flexibly combined to adapt to particular use cases.
Although the standards can be used individually, the greatest benefit is obtained by a judicious combination of selected features.
For instance, using time-aware traffic shaping algorithms usually requires tight time-synchronization.
More specifically, there is a \gls{tsn} standard that provides mechanisms to shape traffic based on a cyclic behavior, similarly to \gls{tdma}.
Packet transmissions can then be optimized by synchronizing end-stations with time-aware bridges.
Hence, employing a second \gls{tsn} standard for precise time-synchronization can enable new opportunities or enhance existing mechanisms for deterministic data transmissions.

Recent studies have mainly focused on integrating \gls{tsn} with wireless technologies, as well as frameworks for managing and optimizing \gls{tsn} core functionalities, such as the \gls{tas} \cite{timelySurvey}.
Also, enhancing end-to-end deterministic transmission \cite{C_based}, and the integration of \gls{tsn} with the software-defined networking (SDN) paradigm has been pursued~\cite{bal2021sdn,  bal2023fed, nay2017inc}.

\subsection{Limitations of Existing TSN Evaluation Frameworks}
\gls{tsn} focuses not only on the reliability of packet transmissions, but also on the timing aspects of the packet  transmissions. It is crucial to precisely measure timing-related metrics, such as one-way delay and \gls{pdv}. 
Three primary methods are used for evaluating \gls{tsn} protocols and systems: 
theoretical mathematical analysis, simulation and emulation, and hardware testbeds.
Mathematical analysis frameworks, such as~\cite{rwf155,rwf157}, have been developed to evaluate \gls{tsn}.
For instance, He et al.~\cite{rwf158ULL} analyzed the worst-case travel time of Ethernet frames with the \gls{cbs} in the \gls{avb} context, and the \gls{tas} from the \gls{tsn} domain. 
Guo et al.~\cite{rwfUPPAAL} studied strict minimum criteria in use scenarios. 
However, the mathematical analysis frameworks abstract the behaviors of \gls{tsn} systems compared to real-world systems.
Simulation frameworks, such as OMNET++~\cite{OMNET} and NS-3~\cite{ns3}, have been widely used in networking research, including \gls{tsn} research. 
The advantages of simulations include flexibility, reduced cost, and scalability. 
A main drawback of simulations in \gls{tsn} research is that they do not involve real experimental hardware networking components, making it impossible to showcase the applicability and demonstrate the developed \gls{tsn} systems and protocols with real operating networks.
Emulation frameworks and tools were developed to address the applicability issue of simulation software. 
Ulbricht et al.~\cite{Ulbr2105:Emulation} implemented \gls{tsn} in the Mininet emulation software~\cite{Mininet}, operating
the entire emulation and \gls{tsn} framework on Linux. Nodes communicate with one another via virtual networking interfaces. The scale of the emulated system is limited to the computing resources of the PC running the emulation. 
Therefore, timing-related emulation measurements of large and complex systems are unreliable.

Using dedicated hardware is, in most cases, more precise while having the same complexity as the targeted actual network deployment. 
Even though scalability is a limiting factor for designing hardware-based \gls{tsn} testbeds caused by limited resources, dedicated \gls{tsn} devices offer the full range of settings and parameters to be considered. 
However, testbeds have complexities: the measuring techniques, setup, and other elements can strongly influence the outcomes. 
Therefore, building a testbed with high precision and flexibility is challenging, yet desirable and facilitates reproducible experiments. 

As \gls{tsn} is a family of standards, \gls{tsn}-related testbeds have been built to study various isolated \gls{tsn} aspects, namely i) packet processing, ii) evaluating the Precision Time Protocol, iii) communication over-the-air; iv) \gls{tsn} performance; and v) \gls{tsn} management.
One of the primary goals of packet processing frameworks, such as MoonGen~\cite{rw1} and P4STA~\cite{rw2}, is to generate packets precisely and to precisely timestamp both transmitted and received packets. 
Even though MoonGen requires \gls{dpdk}-supported hardware, it offers a high degree of flexibility and high performance. 

Nevertheless, a testbed for studying \gls{tsn} needs more than packet generation and processing.
Next, several existing testbeds focus on time synchronization, primarily with \gls{ptp}, such as~\cite{nrw7, nrw8, nrw9}. However, they solely evaluate the synchronization precision without assessing \gls{tsn} mechanisms.
More comprehensive testbeds have been built to study the performance of \gls{tsn} in combination with wireless communication, such as WiFi and 5G~\cite{nrw1, nrw5, nrw17}.
A few testbeds target specific application contexts, such as for benchmarking \gls{tsn} performance of the controller on multi-domain networks~\cite{nrw3} and for intra-vehicle networking~\cite{nrw18}. 
However, they either neglect the per-packet delay or have limited precision due to inconsistent hardware and software timestamping.
Overall, the existing testbed studies focus on specific narrow sets of \glspl{kpi}. Also, the existing studies commonly consider only limited (coarse) precision levels and often neglect detailed investigations of the effects experienced by a \gls{tas} scheduled data stream.

\subsection{Contributions: Open-Source TSN-FlexTest Testbed}
We address a root impediment in \gls{tsn} research: the need for high-precision reproducible measurement studies in comprehensive \gls{tsn} testbeds. 
We design and develop TSN-FlexTest, an open-source, flexible, high-precision, yet affordable measurement testbed framework to comprehensively study \gls{tsn}. 
Our key innovation is a well-engineered combination of affordable \gls{cots} hardware with an automated measurement workflow from packet generation to data collection and visualizing measurement results.
Our open-source measurement framework facilitates the sharing of experiment scenarios and setups. 
Furthermore, the testbed workflow is organized into a centralized management and control entity, making the testbed workflow very easy to change. 
Last, introducing key abstraction entities, such as the \gls{dut} and the traffic generator, make the entire testbed flexible to change.
Striving towards high precision and affordability at the same time, we advocate the use of commodity hardware, which also facilitates reproducibility.

This paper makes the following two main contributions:
\begin{enumerate}
    \item \autoref{sec:testbed} first reviews the general design principles for a \gls{tsn} testbed, including the available options for hardware and software components and their trade-offs. 
    Subsequently, we detail the TSN-FlexTest testbed architecture, design, and implementation, including the workflow of our measurement framework.
    We also share the lessons we learned while developing our measurement framework. 
    In \autoref{sec:testbed-validation}, we validate our TSN-FlexTest testbed, which produced results with nanosecond precision. 
    A preliminary version of this contribution is presented in~\cite{9844050}, focusing on the testbed design. 
    We make TSN-FlexTest openly available at \href{https://github.com/5GCampus/tsn-testbed}{https://github.com/5GCampus/tsn-testbed}.
    \item The second main contribution of this paper is the investigation of critical aspects of \gls{tsn} leveraging the developed and validated TSN-FlexTest testbed in \autoref{sec:testbed-evaluation}. These are the significant differences of this extended article compared to the preliminary conference version~\cite{9844050}.
    There are four main findings:
    1) We evaluate a typical configuration in a net-neutral Internet, without specialized packet treatment and with limited resources in the backbone networks. We demonstrate that it is nearly impossible to ensure prescribed service quality in such a net-neutral Internet.
    2) We investigate the impacts of soft \gls{qos} with Strict Priority Queuing (\gls{spq}) and conclude that \gls{spq} cannot enforce strict guarantees.
    3) We measure the one-way packet latency for \gls{tas}. The results reveal inconsistent advantages of traffic shaping and hard scheduling. The \gls{tas} sometimes performs better, especially for the average latency. 
    4) We demonstrate that it is possible to synchronize a generic application running on \gls{cots} hardware with the \gls{tas} by scheduling on the transmitter side itself.
\end{enumerate}

\section{Background}  \label{sec:background}
\acrlong{tsn}, is a collection of standards that enable guaranteed latency and deterministic communication over wired Ethernet.
These standards can be categorized according to their functions: time-synchronization, flow control, ultra-reliability, and network management.
This background section highlights the main approaches of \gls{tsn} evaluation.
Next, background on high-precision time-synchronization, \gls{tsn} flow control, and \gls{tsn} \glspl{kpi} is provided.

\subsection{TSN Evaluation Methodologies}
There is a set of tools in order to evaluate these recent developments in the area of \gls{tsn}.
Each tool provides different advantages and disadvantages, such as low costs, high reproducibility, or significance for real-world deployments.

\subsubsection{Analysis, Simulation, and Emulation} 
Studies often employ a formal evaluation based on mathematical analysis that is typically coupled with synthetic test cases~\cite{zha2022qua}.
The coupled evaluation approach ensures reproducibility: Equations do not rely on hardware nor software, and evaluation results should be the same irrespective of the underlying computing hardware.
Depending on the degree of modeling details in the implementation of algorithms and abstracting \enquote{real-world} behaviors, simulation can offer a solid foundation for research.
Furthermore, simulation provides the ground to investigate various architectures, influences, and interdependencies, that would be difficult to examine in real testbeds.
This can be, e.g., due to missing hardware: setting up large networks is expensive and incurs high costs for setup and maintenance; also, interoperability issues may arise.
Furthermore, for recent developments as well as new designs and algorithms, hardware implementations may not be available yet.

An alternative to simulation and formal evaluation is emulation.
Emulation uses real implementations, but without high investments in hardware infrastructure.
The network emulator MiniNet, for instance, can be configured to include \gls{tsn} features.

\subsubsection{Hardware Testbed Measurement}
\gls{tsn} network behaviors can be studied through measurements on actual hardware testbeds~\cite{car2021exp,rw11,liu2022ind,taj2021opt,qua2020ope}.
Generally, testbed measurements offer authentic results in terms of closeness to real systems~\cite{hei2022net,kun2020con,mag2022exp,mar2019exp,mak2021ser,pen2022kqi,pol2022ass,tro2022res}.
The main drawback of testbed measurements is in most cases, besides the actual hardware costs, the complexity of the testbed system~\cite{der2021ena,fid2022mea,imt2022emp,mic2021sof}.
Not only software can have bugs, but the actual hardware implementation may have design flaws as well.
Investigating real hardware also introduces dependencies on vendors.
If there is no open hardware platform available, which is---seen with eyes of independent researchers---unfortunately common, then the hardware must be treated as is. Consequently, it is usually impossible to tweak, modify, or gain insight into the implementation specifics of the hardware from commercial vendors.
Only self-developments, e.g., with \gls{fpga}-based boards, offer the freedom to modify the hardware, but at a much higher cost of initial implementation.

Studying real-world systems with testbed measurements poses additional challenges.
First, researchers must rely on the capabilities of the \acrfull{dut}.
Hence, a \gls{dut} is often treated as a \textit{black box}.
Therefore, if the \gls{dut} itself does not offer any evaluation metrics, such as a packet counter, then measurements must be taken from the outside, necessitating additional measurement equipment.
Further complexity is introduced by the fact that measurement probes in networks are often distributed.
For instance, measuring time-dependent parameters, e.g., one-way delay, requires precise synchronization.
Also, measurement tools can be expensive.
If a certain use case should be investigated, not all tools may offer the degree of freedom of adaptation, resulting in similar dependencies on vendor implementation as for the \gls{dut}.

\subsection{High-precision Time-synchronization}
\label{sec:background-timesync}

\subsubsection{Overview}
In a \gls{tsn} network, network components need to be time-aware and have a shared understanding of time with high precision.
This knowledge of time might be either absolute or relative.
While absolute timing is typically used to associate data points with specific real-life occurrences, relative timing is primarily concerned with synchronizing various components and maintaining them synchronized to coordinate several processes that are running simultaneously.
The needed level of synchronization varies depending on the use case and ranges from tens of milliseconds to tens of nanoseconds.
The \acrfull{ptp} is recommended in the \gls{tsn} standards for achieving this level of synchronization.

\subsubsection{PTP Protocol}
The \gls{ptp} protocol can distribute precise timing, phase, and frequency over packet networks with sub-microseconds precision.
The time source in the \gls{ptp} protocol is known as the \gls{gm} clock.
The \gls{gm} clock exchanges time information with other nodes, which can themselves be a boundary, ordinary, or transparent clock.
The \gls{gm} in the \gls{ptp} protocol can be chosen either automatically or manually within a  \gls{ptp} time domain. 
The automated selection process leverages the \gls{bmc} algorithm.
Generally, the clock with the highest precision in the clock domain is the \gls{bmc}, which is detected by using ID, class, priority, or MAC address.
\gls{gm} clocks are frequently synchronized using \gls{gnss}.
Satellite constellations offering global or regional positioning, navigation, and timing services are referred to as \gls{gnss} (such as GPS or GLONASS).
Another option is employing highly precise oscillators (e.g., cesium or rubidium clock).
Ordinary clocks only have one port for receiving the time information, boundary clocks have two or more \gls{ptp} ports and can operate as a time source to other connected nodes.
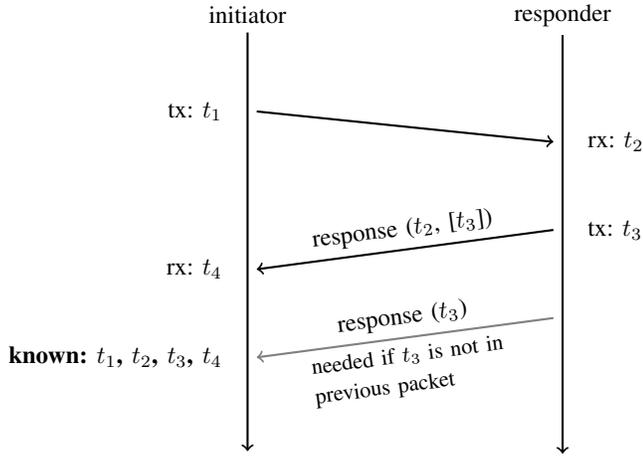
\begin{figure}[t]\centering
    \adjustbox{max width=0.49\textwidth}{
        \begin{tikzpicture}[thick]
            \node[style=textbox] at (0,0)(init) {initiator};
            \node[style=textbox, right=of init, xshift=2cm] (resp) {responder};
            \node[style=textbox, below=of init,label={[anchor=east,label distance=0.1cm]left:{tx: $t_1$}}](t1) {};
            \node[style=textbox, below=of resp,yshift=-0.4cm,label={[anchor=west,label distance=0.1cm]right:{rx: $t_2$}}](t2) {};
            \node[style=textbox, below=of t2,label={[anchor=west,label distance=0.1cm]right:{tx: $t_3$}}](t3) {};
            \node[style=textbox, below=of t3](t3a) {};
            \node[style=textbox, below=of t1,yshift=-1cm,label={[anchor=east,label distance=0.1cm]left:{rx: $t_4$}}](t4) {};
            \node[style=textbox, below=of t4,label={[anchor=east,label distance=0.1cm]left:{\textbf{known: $t_1$, $t_2$, $t_3$, $t_4$}}}](t5) {};
            \node[style=textbox, below=of init,yshift=-5cm](init-end) {};
            \node[style=textbox, below=of resp,yshift=-5cm](resp-end) {};
            \draw[->,color=black!100,line width=0.3mm] (init.south) -- (init-end.north) ;
            \draw[->,color=black!100,line width=0.3mm] (resp.south) -- (resp-end.north) ;
            \draw[->,color=black!100,line width=0.3mm] (t1.east) -- (t2.west) ;
            \draw[->,color=black!100,line width=0.3mm] (t3.west) -- (t4.east) node[sloped,midway,above,color=black,align=left] {response ($t_2$, [$t_3$])} ;
            \draw[->,color=black!50,line width=0.3mm] (t3a.west) -- (t5.east) node[sloped,midway,above,color=black,align=left] {response ($t_3$)} node[sloped,midway,below,color=black,align=left] {\small needed if $t_3$ is not in\\\small previous packet} ;
        \end{tikzpicture}
    }
	\caption{Illustration of \acrlong{ptp} synchronization according to IEEE~802.1AS~\cite{ieee802.1AS}: Via a message exchange in a master-slave system, two clocks can be synchronized with sub-microseconds precision using path-delay measurements.}
	\label{fig:ptp-message-exchange}
\end{figure}

To distribute time correctly across all clocks in the network, the path-delay between two clocks has to be measured precisely.
The synchronization thereby leverages timestamping, i.e., the exact time needs to be taken when a packet leaves at the \gls{tx} and arrives at the \gls{rx}.

Distributing time from the \gls{gm} is insufficient to synchronize clocks, and the influence of path delay on the synchronization should be minimized.
Path delay techniques leverage the time-stamps supplied in frames.
The path delay can be evaluated in a one-way mode, where the master node sends sync and follow-up messages, or in a two-way mode, where downstream clocks also send delay request packets to master clocks.
The path delay can be evaluated in the unicast mode, where the master clock generates sync and follow-up unicast messages, or in a mixed mode, where the master clock sends multicast messages but responses are unicast.
The path delay must be evaluated periodically since it is dynamic and fluctuates over time. The length of this period depends on the required level of precision.
This period length and other \gls{ptp} parameters can have an influence on the performance of the \gls{ptp} protocol.

\subsubsection{Timestamps---Step by Step}
To ensure sub-microsecond precision, some degree of hardware support is needed.
The \gls{rx} hardware timestamp is latched by the \gls{nic} when the packet arrives and can be used by the software layer later on.
The \gls{tx} timestamp can be implemented in three ways: one-step mode, two-step mode, and a combination of both one- and two-step modes.
In two-step mode, the \gls{nic} reports the \gls{tx} timestamp of the currently to-be-sent packet with the next frame, in a so-called \textit{follow-up} message.
This can be done solely in software.
In one-step mode, the \gls{nic} has the capability of correcting the time information of outgoing frames right before their transmission.
An offset register specifies the location within the Ethernet frame.
Finally, there is a 1.5-step mode, where the \gls{cpu} generate a dummy \textit{follow-up} message, and the \gls{nic} only updates the few fields to form a real \textit{follow-up} message.
Each method has its advantages and disadvantages:
While the one-step mode is less complex for the receiving clock implementation, the complexity of the sender \gls{nic} increases because of calculations to provide the timestamp at the point of time the frame will leave the \gls{phy}.
On the other hand, the two-step mode does not require the computation on the \gls{tx} clock.
Furthermore, the one-step mode can only be employed if the link speed does not exceed a certain threshold.
Because of the timestamp calculations that need to be done in real-time, at some point the link speed is higher than the hardware capabilities.

\subsubsection{Path Delay Evaluation}
\autoref{fig:ptp-message-exchange} illustrates the transmission steps for the \gls{ptp} time-synchronization and the corresponding path-delay calculation.
For calculating the path delay, four time values are needed.
The initiator and the responder of the path delay measurement exchange two to three messages, to obtain the packet runtime.
The initiator generates a message and stores its transmission time $t_1$. The responder replies to the message and includes the time $t_2$ when the first message was received.
Depending on the one- or two-step procedure, an additional message transfers the transmission time of the reply $t_3$ back to the initiator.
Using $d=[(t_2-t_1)+(t_4-t_3)]/2$, the four timestamps give the path delay between initiator and responder.

\subsubsection{The Art of Timestamp Provisioning}
\label{sec:timestampexplanation}
\gls{ptp} can be deployed on different hardware, such as
commercial devices, e.g., Ethernet switches, as well as \gls{cots} computer hardware and general-purpose operating systems, e.g., GNU Linux.
LinuxPTP~\cite{ptp4l} provides a commonly used \gls{ptp} implementation for the Linux platform.
LinuxPTP offers software and hardware support for timestamping.
Depending on the implementation of the hardware timestamping mechanism, different packet throughputs can be achieved.
The hardware-based timestamping in the \gls{rx} and \gls{tx} directions can be enabled by packet filters applied in the network driver layer (corresponding to the \gls{tx} direction) or in the \gls{nic}
(corresponding to the \gls{rx} direction).
Typical filters timestamp \textit{only \gls{ptp}} packets or timestamp \textit{all} packets.
However, filters are necessary because timestamping \textit{all} packets can at some point exceed the hardware capabilities.

To describe the bottlenecks, the \gls{tx} and \gls{rx} directions have to be considered separately.
\gls{tx} timestamping can be enabled operating system-wide by setting a flag in the \gls{nic}'s configuration space.
This enables the timestamping for all packets, or more precisely: packet by packet.
A bit in the packets descriptor data structure, which is used to pass the memory location of the data to the network hardware, can be set for this purpose.
If two-step \gls{tx} timestamping is enabled for all or just one packet, then the \gls{nic} will latch the clock time instant when the first bit is on wire (practically it is done before, with a time offset), and stores the value in internal memory.
This memory structure needs to be read by high-level software.
This timestamp storage can be a bottleneck for the \gls{tx} direction:
The software needs to read the timestamps before the limited memory in the \gls{nic} is overwritten~\cite{i210ds,rw1}.
Moreover, the high-level software is responsible to match the read-out timestamps with the \gls{skb} structure of the transmitted packet, which contains the pointers to the packet data, and reporting back the \gls{tx} timestamp to the higher layers of the network protocol stack.
In some implementations, the matching is supported by an ID that corresponds to the transmitted packet~\cite{tngit}.
An improvement to this procedure is one-step timestamping:
the \gls{nic} writes the latched \gls{tx} timestamp directly into the transmitted packet. After this operation, the Ethernet checksum is repaired. These operations need to be conducted on the fly, during the transmission process of the packet through the \gls{nic}.

In the receiving direction, there are also different implementations:
The most scalable solution is to deliver the \gls{rx} timestamp within the packets descriptor structure, so that the memory pointer and the timestamp information are stored in the same structure and do not need to be aggregated later on~\cite{tngit,i210ds}.
Other solutions may be limited in performance:
Some implementations provide the \gls{rx} timestamp in an extra register, which has to be read out by software fast enough to align the timestamp with the corresponding packet~\cite{i210ds}.

\subsubsection{Profiling}
\gls{ptp} is being developed since 2002~\cite{ieee1588-2002} and with an updated version in 2008~\cite{ieee1588-2008}, the IEEE Std.~1588 supports \textit{profiles.}
IEEE 1588-2019, i.e., PTPv2.1~\cite{ieee1588-2019}, was proposed \mbox{11 years} later. In contrast to the IEEE 1588-2002 and IEEE 1588-2008 which are not compatible, IEEE 1588-2019 is backward compatible with the previous versions, and new features are completely optional. The primary goal of IEEE 1588-2019 is to improve the \gls{ptp} protocol's precision, flexibility, and resilience. For instance, \gls{ptp}v2.1 introduces modular transparent clocks, hybrid multicast/unicast operation to improve flexibility, and a new range of security guidelines for improving robustness.
The IEEE 1588-2019 standard provides a variety of options,  not all of them are necessary for each use case.
As a result, each profile chooses a subset of parameters for achieving different degrees of precision, for various use cases and network topologies.
Each profile needs to set a default value and range for \gls{ptp} attributes and mention prohibited, permitted, and required options as well as allowed node types.
The profile settings can directly impact the performance of \gls{ptp}.

IEEE~802.1AS~\cite{ieee802.1AS} is a standard of the \gls{tsn} standards set that enables the network to meet the rigorous requirements of time-sensitive applications, such as tactile feedback, or audio and video data streaming.
The IEEE~802.1AS standard includes a \gls{ptp} profile, called \gls{gptp}.
A \gls{gptp} domain consists of end station and bridge nodes that can communicate \gls{ptp} messages directly with each other.

\subsection{Time-sensitive Flow Control}
\label{sec:background:flow-control}
According to \cite{8458130}, the \gls{tsn} flow control mechanisms can be categorized into three different categories: (i) traffic shaping, (ii) traffic scheduling, and (iii) \gls{fp}~\cite{ieeeframepreemption}.
A \gls{tsn} network uses a combination of them to fulfill the requirements of different traffic profiles. These methods are described in detail below, with an emphasis on comparing different traffic shapers.
\begin{figure}[t]
    \centering
    \adjustbox{max width=0.49\textwidth}{
        \begin{tikzpicture}[thick]
        	\node [style=TUDbox2,minimum width=1.6cm,label={Queue 0}] (q0-top) at (0,0) {};
            \node [style=textbox,minimum width=2cm,yshift=-1cm,xshift=1.3cm] (dots)at (0,0) {$\cdot\cdot\cdot$};
            \node [style=textbox,minimum width=2cm,yshift=-2.7cm,xshift=1.3cm] (dots)at (0,0) {$\cdot\cdot\cdot$};
            \node [style=textbox,minimum width=2cm,yshift=-3.7cm,xshift=1.3cm] (dots)at (0,0) {$\cdot\cdot\cdot$};
        	\node [style=TUDbox2,minimum width=1.6cm,below=0.01cm of q0-top.south, node distance=0cm] (q0-1)  {};
        	\node [style=TUDbox2,minimum width=1.6cm,below=0.01cm of q0-1.south, node distance=0cm] (q0-2)  {};
        	\node [style=TUDbox2,minimum width=1.6cm,below=0.01cm of q0-2.south, node distance=0cm] (q0-3)  {};
        	\node [style=TUDbox2,minimum width=1.6cm,below=0.01cm of q0-3.south, node distance=0cm] (q0-4) {};
        	\node [style=TUDbox2,minimum width=1.6cm,below=0.01cm of q0-4.south, node distance=0cm] (q0-5) {};
        	\node [style=TUDbox2,minimum width=1.6cm,below=0.01cm of q0-5.south, node distance=0cm] (q0-6)  {};
        	\node [style=TUDbox2,minimum width=1.6cm,below=0.01cm of q0-6.south, node distance=0cm] (q0-7) {};
          	\node [style=TUDbox3,minimum width=1.6cm,below=0.2cm of q0-7.south, node distance=0cm,align=center] (q0-sched) {\acs{tsa}};
        	\node [style=TUDbox3,minimum width=1.6cm,below=0.2cm of q0-sched, node distance=0cm] (q0-gate) {Gate = 0};
        	\node [style=none,below=0.5cm of q0-gate.south, node distance=0cm] (q0-end) {};
        	\node [style=TUDbox2,minimum width=1.6cm,label={Queue n-1},right=1cm of q0-top] (q2-top) {};
        	\node [style=TUDbox2,minimum width=1.6cm,below=0.01cm of q2-top.south, node distance=0cm] (q2-1)  {};
        	\node [style=TUDbox2,minimum width=1.6cm,below=0.01cm of q2-1.south, node distance=0cm] (q2-2)  {};
        	\node [style=TUDbox2,minimum width=1.6cm,below=0.01cm of q2-2.south, node distance=0cm] (q2-3)  {};
        	\node [style=TUDbox2,minimum width=1.6cm,below=0.01cm of q2-3.south, node distance=0cm] (q2-4) {};
        	\node [style=TUDbox2,minimum width=1.6cm,below=0.01cm of q2-4.south, node distance=0cm] (q2-5) {};
        	\node [style=TUDbox2,minimum width=1.6cm,below=0.01cm of q2-5.south, node distance=0cm] (q2-6)  {};
        	\node [style=TUDbox2,minimum width=1.6cm,below=0.01cm of q2-6.south, node distance=0cm] (q2-7) {};
        	\node [style=TUDbox3,minimum width=1.6cm,below=0.2cm of q2-7.south, node distance=0cm,align=center] (q2-sched) {\acs{tsa}};
        	\node [style=TUDbox3,minimum width=1.6cm,below=0.2cm of q2-sched.south, node distance=0cm] (q2-gate) {Gate = 1};
        	\node [style=none,below=0.5cm of q2-gate.south, node distance=0cm] (q2-end) {};
        	\node [style=TUDbox2,minimum width=1.6cm,label={Queue n},right=0.25cm of q2-top] (q3-top) {};
        	\node [style=TUDbox2,minimum width=1.6cm,below=0.01cm of q3-top.south, node distance=0cm] (q3-1)  {};
        	\node [style=TUDbox2,minimum width=1.6cm,below=0.01cm of q3-1.south, node distance=0cm] (q3-2)  {};
        	\node [style=TUDbox2,minimum width=1.6cm,below=0.01cm of q3-2.south, node distance=0cm] (q3-3)  {};
        	\node [style=TUDbox2,minimum width=1.6cm,below=0.01cm of q3-3.south, node distance=0cm] (q3-4) {};
        	\node [style=TUDbox2,minimum width=1.6cm,below=0.01cm of q3-4.south, node distance=0cm] (q3-5) {};
        	\node [style=TUDbox2,minimum width=1.6cm,below=0.01cm of q3-5.south, node distance=0cm] (q3-6)  {};
        	\node [style=TUDbox2,minimum width=1.6cm,below=0.01cm of q3-6.south, node distance=0cm] (q3-7) {};
            \node [style=TUDbox3,minimum width=1.6cm,below=0.2cm of q3-7.south, node distance=0cm,align=center] (q3-sched) {\acs{tsa}};
        	\node [style=TUDbox3,minimum width=1.6cm,below=0.2cm of q3-sched.south, node distance=0cm] (q3-gate) {Gate = 0};
        	\node [style=none,below=0.5cm of q3-gate.south, node distance=0cm] (q3-end) {};
        	\node [style=textbox,minimum width=1.6cm,label={Gate Control List},right=0.7cm of q3-top,yshift=-0.12cm] (gcl-top) {T00: 01...101};
        	\node [style=textbox,minimum width=1.6cm,below=0.01cm of gcl-top.south, node distance=0cm] (gcl-1)  {T01: 10...101};
        	\node [style=textbox,minimum width=1.6cm,below=0.01cm of gcl-1.south, node distance=0cm,dashed,draw=TUDBlue100,line width=0.3mm] (gcl-2)  {T02: 01...111};
        	\node [style=textbox,minimum width=1.6cm,below=0.01cm of gcl-2.south, node distance=0cm] (gcl-3)  {T03: 00...101};
        	\node [style=textbox,minimum width=1.6cm,below=0.01cm of gcl-3.south, node distance=0cm] (gcl-4) {T04: 01...111};
        	\node [style=textbox,minimum width=1.6cm,below=0.01cm of gcl-4.south, node distance=0cm] (gcl-5) {T05: 10...111};
        	\node [style=textbox,minimum width=1.6cm,below=0.01cm of gcl-5.south, node distance=0cm] (gcl-6)  {T06: 01...111};
        	\node [style=none,left=0.2cm of gcl-2.west, node distance=0cm] (gcl-path1) {};
        	\node [style=none,below left=0.5cm and -0.5cm of q3-gate.east, node distance=0cm] (gcl-path2) {};
        	\node [style=none,below left=0.5cm and -0.5cm of q2-gate.east, node distance=0cm] (gcl-path3) {};
        	\node [style=none,below left=0.5cm and -0.5cm of q0-gate.east, node distance=0cm] (gcl-path5) {};
        	\node [style=TUDbox1,minimum width=6.58cm,below=0cm of q0-end, node distance=0cm, anchor=north west,xshift=-1cm] (ts) {Transmission Selection};
        	\node [style=none,below=0.5cm of ts, node distance=0cm] (end) {};
            \draw[->,color=TUDBlue100,line width=0.3mm] (q0-7.south) -- (q0-sched.north) ;
            \draw[->,color=TUDBlue100,line width=0.3mm] (q0-sched.south) -- (q0-gate.north) ;
            \draw[->,color=TUDBlue100,line width=0.3mm] (q0-gate.south) -- (q0-end.north) ;
        	\draw[->,color=TUDBlue100,line width=0.3mm] (q2-7.south) -- (q2-sched.north) ;
            \draw[->,color=TUDBlue100,line width=0.3mm] (q2-sched.south) -- (q2-gate.north) ;
        	\draw[->,color=TUDBlue100,line width=0.3mm] (q2-gate.south) -- (q2-end.north) ;
        	\draw[->,color=TUDBlue100,line width=0.3mm] (q3-7.south) -- (q3-sched.north) ;
            \draw[->,color=TUDBlue100,line width=0.3mm] (q3-sched.south) -- (q3-gate.north) ;
        	\draw[->,color=TUDBlue100,line width=0.3mm] (q3-gate.south) -- (q3-end.north) ;
        	\draw[-,color=TUDBlue100,line width=0.2mm,dashed] (gcl-2.west) -- (gcl-path1.center) ;
        	\draw[-,color=TUDBlue100,line width=0.2mm,dashed] (gcl-path1.center) -- (gcl-path2.center) ;
        	\draw[-,color=TUDBlue100,line width=0.2mm,dashed] (gcl-path2.center) -- (gcl-path3.center) ;
        	\draw[-,color=TUDBlue100,line width=0.2mm,dashed] (gcl-path3.center) -- (gcl-path5.center) ;
        	\draw[->,color=TUDBlue100,line width=0.2mm,dashed] (gcl-path5.center) -- (q0-gate.south east) ;
        	\draw[->,color=TUDBlue100,line width=0.2mm,dashed] (gcl-path3.center) -- (q2-gate.south east) ;
        	\draw[->,color=TUDBlue100,line width=0.2mm,dashed] (gcl-path2.center) -- (q3-gate.south east) ;
        	\draw[->,color=TUDBlue100,line width=0.3mm] (ts.south) -- (end.center) ;
        \end{tikzpicture}
    }
    \caption{General illustration of queuing operation. In modern Ethernet switches, \glspl{nic}, or similar devices, there are usually multiple queues in which incoming packets are stored. This can be done based on different metrics, e.g., the \gls{pcp}, or with a matching approach, e.g., by using ingress/egress ports to queue mappings. Within each queue, a \gls{tsa} can be applied to change the order of packets. In most cases, \gls{fifo} queuing is employed. Between queues, a second \gls{tsa} can be used, e.g., \acrfull{spq}. Further it is possible to shape the traffic with a traffic shaper, e.g., with a \acrfull{tas}, which operates by deploying \textit{gates} in front of queues.}
    \label{fig:qbv}
\end{figure}
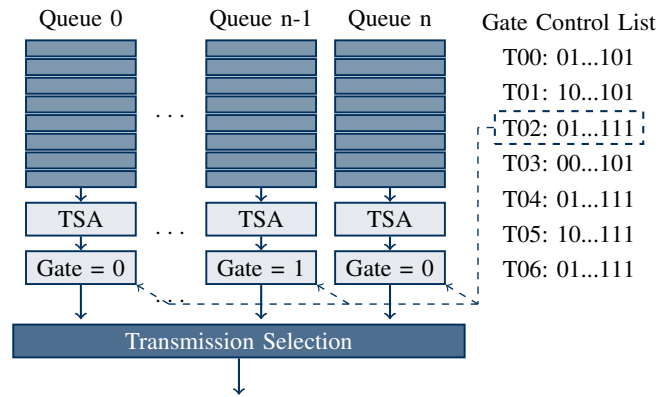

Traffic shapers strive to give appropriate packet transmission opportunities to streams of the different priority classes~\cite{hua2022fle,li2022ban,nas2019per,zha2022qua}.
The primary distinction between traffic shaping and traffic policy is that shapers do not discard packets but rather delay them.
\gls{tsn} traffic shapers aid in assigning resources to \gls{tsn} traffic streams, based on priorities, to as to isolate the streams from interference caused by cross traffic.
Shapers also attempt to limit the residence time of enqueued traffic.
Their benefit is often intended for traffic with higher priority, such as \acrfull{ti} data streams.
If these traffic classes, particularly those with hard deadlines, are mixed with non-scheduled traffic, there are no assurances for best-effort streams.
Different types of shapers have been introduced for \gls{tsn}.
They differ in terms of implementation complexity for software and hardware (e.g., buffer size or required chip logic), introduced \gls{pdv} (often also referred to as \textit{jitter}), and delay, as well as \textit{burstiness} reduction.

Traffic scheduling determines which frame to transmit based on a scheduling algorithm or policy.
The appropriate scheduling policy is usually determined by the use case requirements.
A scheduling algorithm may merely examine arrival time, as in \gls{fifo} and \gls{lcfs}. Alternatively, to spread bandwidth between traffic classes, a variable quantity of transmission opportunities, such as weighted round robin, might be considered. In some applications, strict priority is used to protect the highest-priority traffic, and the packet with the highest priority can be sent first.

\subsubsection{Credit Based Shaper}
\gls{tsn} inherits the \gls{cbs} (IEEE~802.1Qav~\cite{ieee802.1Qav}) from the \gls{avb}~\gls{tg}.
This concept was initially intended to give a guarantee of limited delay and \gls{pdv}.
\gls{cbs} distributes packets in a queue over time so that the limit of the allowed rate is not exceeded.
The standard defines two classes, A and B, each of which gets a credit. 
When a frame of a class waits for transmission, its credit grows at a rate known as \texttt{idle slope}, and it falls at a rate known as \texttt{send slope}.
\gls{cbs} was shown to give unsatisfactory delay assurances at high link utilization~\cite{rwf157, li2022ban}.
Furthermore, the maximum latency of \gls{cbs} varies according to topology and hop count.
Overall, up to seven network hops, IEEE~802.1Qav ensures worst-case latencies of less than \SI{2}{\milli\second} for class A and less than \SI{50}{\milli\second} for class B~\cite{ieee802.1Qav}.

\subsubsection{Time-Aware Shaper}
IEEE~802.1Qbv~\cite{ieee802.1Qbv} proposes a \acrfull{tas} based on the notion of \gls{tdma}, to solve the limitations of the \gls{cbs}.
\gls{tas} is based on a cyclic transmission with time-triggered window slots.
Thus, the nodes must have a precise knowledge of time.
IEEE~802.1Qbv establishes a set of queues for each \gls{tsn} bridge output port.
Queues are managed via \textit{gates} that may be opened or closed. 
Throughout a cycle, the \gls{gcl} specifies which queue at what precise moment can transmit its data. Each of these queues usually represents a level of priority.
In addition, the time length of each slot can be identified.
The cycle time is the accumulated time of all defined slots in a \gls{gcl}.
After the final entry, a new cycle of the \gls{gcl} begins.
The \gls{tas} enables network devices to precisely plan the transmission of enqueued frames.
\label{sec:preemption-description}
In general, for \gls{tas}, the output port is a shared resource that can only be used from one frame at once.
The scheduler decides if this resource is full and which frame should be selected next for transmission.
If a frame is still in transmission, then it blocks the line and no other packet can be scheduled.
This issue results in the case that long low prior frames can block high prior frames because they are already in transmission. The effect could be mitigated by inserting blocking slots into the \gls{tas} \gls{gcl} to block every traffic to be transmitted before the time slot of a high-priority stream. 
Unfortunately, these guard bands are wasted time, because in most cases no frame is transmitted within. 
\acrfull{fp} is a technique that solves the issue without inserting a guard band.
With preemption, the transmission on the Ethernet layer could be interrupted, another frame could be inserted, and the old transmission is continued afterward.
Therefore, special features in the \glspl{phy} are necessary. 

The main limitation of \gls{tas} is that it requires network-wide nearly perfect time-synchronization.
An unsynchronized stream endpoint may cause a delay, where the streams will have to wait for the next time-triggered window.
Moreover, \gls{tas} cannot control individual streams if they are with the same priority.
Therefore, to achieve the same level of per-flow \gls{qos}, additional mechanisms are required, such as \gls{psfp} and \gls{frer}~\cite{nas2019per}. 
A \gls{psfp} enabled switch can discard non-essential streams but schedule high-priority streams on a per-stream, per-priority, and per-frame basis, and \gls{frer} duplicates frames on multiple different paths and eliminates them at the receiver to achieve reliability.
In addition, it is challenging to design coordinated \gls{tas} schedules when multiple \gls{tsn} switches are in the same \gls{tsn} domain~\cite{9598852}.

\subsubsection{Cyclic Queuing and Forwarding}
The standard IEEE~802.1Qch~\cite{ieeeqch} proposes \gls{cqf}).
\gls{cqf} makes \gls{tsn} switch design easier by inserting static configurations into the \gls{tas} \gls{gcl}~\cite{cqfpractical}. In this method, multiple queues are used for each traffic class. All arriving frames are held in the closed queue of a traffic class, whereas previously received frames might be forwarded from the open queue of that class. The technique employs synchronized operations, allowing \gls{lan} bridges to organize cyclic frame transmissions to achieve zero congestion loss and constrained latency regardless of network structure.

The disadvantage of \gls{cqf} is the extra delay of at least one cycle for all incoming frames.
Furthermore, delay in transmission and processing can result in a frame being received at an incorrect cycle.
Although \gls{cqf} and \gls{tas} offer ultra-low latency for \gls{tsn} traffic, they rely on network-wide time-synchronization since packet transmission is enforced at periodic intervals.
Both are best suited for deterministic communication and isolation between traffic streams, but they make inefficient use of network resources.

\subsubsection{Asynchronous Traffic Shaper}
The \gls{ats}~\cite{IEEE802.1Qcr} does not require gates or a global clock.
Instead, it makes use of the idea of \textit{leaky} and token buckets, which is an algorithm that regulates discrete event rates.
\gls{ubs} is a feature that \gls{ats} adds to the bridge architecture in an effort to improve traffic flow.
It controls the \textit{burstiness} of the streams, redesigns the traffic at each hop with a per-stream token bucket scheduler, and gives urgent traffic precedence over other traffic classes.
A set of periods known as eligibility periods are assigned to frames of certain streams by \gls{ats} shapers.
This information is used in the scheduling process by the \acrfull{tsa}.
Even when running with high link usage and a combination of periodic and sporadic traffic, \gls{ats} can efficiently use the bandwidth with comparatively lower complexity than \gls{tas}.
Using \gls{ubs} in conjunction with \gls{ats} cannot replace the need for \gls{tas}, but \gls{ats} can be seen as an improvement over the \gls{cbs}~\cite{UBSTASATS}.
In ~\cite{9482597} it is shown that \gls{ubs} can enhance link usage by reducing contention, but it does not provide the same level of \gls{tas} protection for various priority streams from one another~\cite{UBSTASATS}.

\begin{table*}[t]
    \centering
    \caption{Comparison of \gls{tsn} shapers: Precise time-synchronization between entities relates to the necessity for precisely synchronized devices, otherwise performing poorly. Awareness for low priority indicates if the method can protect lower priority traffic against burstiness of highest priority traffic. Deterministic data transmission implies that the shaper can precisely guarantee the requirements of the highest-priority traffic.
    }
    \label{tab:compareShaper}
    \begin{tabularx}{\textwidth}{lcccccc}
        \toprule
        \textbf{Standard Name} & \makecell{Precise Time-Sync.\\Between Entities} & \makecell{Awareness for\\Low Priority} & \makecell{Deterministic\\Data Transmission}  & \makecell{Stream-level\\\acrshort{qos} Control} & \acrshort{pdv} Control & \makecell{Higher \acrshort{capex}}\\
        \midrule
        \acrfull{cbs} & \xmark & \cmark & \xmark & \cmark & \xmark & \xmark \\
        \acrfull{tas} & \cmark & \xmark & \cmark & \xmark & \xmark & \cmark \\
        \acrfull{cqf} & \cmark & \cmark & \xmark & \cmark & \cmark & \cmark \\
        \acrfull{ats} & \xmark & \cmark & \xmark & \cmark & \xmark & \xmark \\
        \bottomrule
    \end{tabularx}
\end{table*}
Shapers should be selected based on the application of the system and the required performance.
Generally, real-time \gls{tsn} systems can be categorized into three groups: (i)~event-triggered (e.g., \gls{ats}), (ii)~time-triggered (e.g., \gls{tas}), and (iii)~mixed systems.
Furthermore, traffic shaping can also be used in a combination with other standards to attain specific outcomes.
Table~\ref{tab:compareShaper} compares several shaper technologies based on various metrics.

\subsection{Key Performance Indicators for Time-Sensitive Networking}
In general, traffic in \gls{tsn} can be categorized either as time-triggered/cyclic or event-driven/acyclic.
80 percent of data communication in industrial applications can be grouped into the latter group~\cite{whiteicc}.
The requirements and expectations from a \gls{tsn} system vary depending on the use case and traffic profile of the network, however, they can be broadly characterized.

The IEC~61784-1:2019 standard~\cite{IECStandard} defines \glspl{kpi} for industrial communication networks:
\begin{itemize}
  \item Cycle time of transmitted frame
  \item End-to-end frame latency between devices
  \item Time-synchronization precision
  \item \gls{pdv} within data streams
  \item Packet-loss ratio
\end{itemize}
Specific performance has to be attained with regard to these metrics when developing a \gls{tsn} testbed.
Throughout, we use the terms "frame" and "packet" interchangeably.
We proceed to explain the importance and relevance of each indicator.

\subsubsection{Traffic Profile and Cycle Time}
In a converged network, the cycle time and frame size of high-priority traffic impact channel utilization and can delay low-priority packets~\cite{TTTECH}.
Real-time applications in industrial field bus communication require strict control of timing between controllers and controlled devices.
Control packets are typically short frames that are sent periodically at a fixed rate.
Motion control and robotics can have control loops in the sub-hundred microseconds range, and in factory automation, this range can be as short as a few microseconds~\cite{TrafficProfile}.

\subsubsection{End-to-end Frame Latency Between Devices}
The main target of \gls{tsn} is not about minimizing latency for all traffic classes, but rather focusing on providing a deterministic communication network for certain priority streams.
In some cases, the \gls{tsn} system therefore needs to delay lower priority traffic for the benefit of higher priority traffic classes.

\subsubsection{Time-synchronization}
The performance of a \gls{tsn} network is correlated with the degree of precision of its time-synchronization.
In the presence of clock inconsistencies, the entire process of packet time-stamping and resource allocation utilizing techniques that depend on precise time will be significantly impacted.
Therefore, \gls{ptp} is a continuous process in a \gls{tsn} network that aims to regulate clock drifts within a time domain.
The configuration of the \gls{ptp} profile based on available resources, requirements, and network topology has a considerable impact on the \gls{tsn} network as well.

\subsubsection{Packet Delay Variation within Data Streams}
\gls{pdv} should be maintained under control since it may have a negative effect on user experience and applications that depend on low delay variation. High \gls{pdv} in industrial applications correspond to unstable tool movements, and most machines will stop working in such circumstances \cite{pdvindustry}.
\gls{pdv} or late delivery of a packet may result in losing a packet or a packet being discarded by the application leading to an increase in the packet-loss ratio~\cite{whiteicc}.

\subsubsection{Packet-loss Ratio}
The packet-loss ratio is another crucial element for \acrlong{tsn} and almost zero packet loss reliability is required for \gls{tsn} Networks \cite{timelySurvey}.
For instance, two consecutive packet losses in industrial use cases may harm equipment.
Packet loss may be caused by high \gls{pdv} or equipment failure, but generally, the primary cause of packet loss is network congestion \cite{8412458}.

\section{Related work} \label{sec:related-work}
This section reviews the related work on \gls{tsn} network performance evaluation. 

\subsection{Formal Mathematical Analysis} 
Some of the \gls{tsn} core functionalities, such as traffic shaping, are tractable in formal mathematical analysis.
Mathematical analysis can explore a wide range of configurations for an equation-based model of the \gls{tsn} system.
As a result, performance can be analyzed more thoroughly than with simulations or empirical measurements for specific configurations.

Several studies have used formal mathematical analysis to evaluate \gls{tsn}, such as~\cite{rwf155,rwf157}.
He et al.~\cite{rwf158ULL} concentrate on the \gls{cbs} in \gls{avb} context, and the \gls{tas} from the \gls{tsn} domain.
They evaluated the worst-case travel time of Ethernet frames to ensure that deadline requirements are satisfied.
Guo et al.~\cite{rwfUPPAAL} describe formal methods for traffic shaping to validate strict minimum criteria in use scenarios.
They employed the \textit{UPPAAL verifier} which is a real-time system modeling, verification, and validation tool.
They also compared shapers in terms of utilization and delay.

Network calculus is a specific formal mathematical framework comprising a collection of mathematical principles for analyzing network performance.
Network calculus can be used for evaluating \gls{tsn} standards due to its capability to analyze the maximum buffer size and worst-case delay~\cite{9123308}.
Zhao et al.~\cite{zha2022qua} analyze the performance of \gls{tsn} shapers, such as \gls{ats}, \gls{cbs}, and \gls{sp}, as well as their combination, with network calculus.
Thomas et al.~\cite{9801077} use network calculus to evaluate the worst-case latency for \acrfull{frer}.
Li et al.~\cite{li2022ban} used network calculus to evaluate the performance of the \gls{srp} with \gls{cbs} and demonstrate that increasing reserved bandwidth does not always improve the tightness of the delay bound.
Generally, formal mathematical analysis avoids the issue of configuring specific simulation or measurement scenarios, but the formal modeling may tend to reflect pessimistic timing behaviors \cite{formalPessimestic}.

\subsection{Simulation and Emulation}
Simulation tools have played a pivotal role in \gls{tsn} evaluations since new protocols can be designed and validated quickly and with relatively low costs in simulation models.
The OMNeT++~\cite{OMNET} and INET~\cite{inet} frameworks have been used in several \gls{tsn} studies, e.g.,~\cite{tsimnet,CoRE4INET,7557870,nesting,8869206,hua2022fle}. 
OMNeT++ is a modular framework, implemented in C++, and is extensible by combining multiple libraries to create \gls{tsn} simulators.
A freely available open-source library known as INET extends OMNeT++ by adding protocols and mechanisms common to current Ethernet networks.
Moreover, INET encompasses wired and wireless connection layer models, e.g., Ethernet and internet stack protocols (such as IPv4, IPv6, TCP, and UDP), facilitating \gls{tsn} simulations.

Heise et al.~\cite{tsimnet} present an open-source framework that focuses on a \gls{tsn} industry profile, including \gls{fp} and \gls{psfp}. 
Importantly,~\cite{tsimnet} does not aim to model time-based algorithms or mechanisms.
In contrast, local clock models are implemented with the \gls{tsn} network components in the simulator introduced by Steinbach et al.~\cite{CoRE4INET}.
The clocks are synchronized using a fail-safe two-step synchronization protocol.

Specht et al.~\cite{7557870} introduced the \gls{ubs} to deal with asynchronous traffic, whereby \gls{ubs} fulfills the delay requirement even if time-synchronization fails.
Falk et al.~\cite{nesting} implemented a standard-compatible simulation model based on OMNeT++ and INET, which includes VLAN tagging, forwarding and queuing enhancements for time-sensitive streams (\acrlong{cbs}), \acrlong{fp}, and scheduled traffic (\gls{tas}).
Moreover, Huang et al.~\cite{hua2022fle} introduced an alternative to \gls{cbs} called time-aware cyclic-queueing (TACQ), and employed OMNeT++ and INET to demonstrate the benefits of TACQ for scheduling mix-flows and reducing the \gls{pdv} for isochronous traffic compared to \gls{cbs}.

Apart from OMNeT++, NS-3~\cite{ns3} is a well-known network simulator into which \gls{tsn} has recently been integrated~\cite{Guillermo,ns3tsn}.
The powerful modeling features for  wireless channels make NS-3 a promising candidate for combining 5G and \gls{tsn} in future simulation models~\cite{Guillermo,ns3tsn}.
However,~\cite{Guillermo,ns3tsn} mainly focus on \gls{tas} and neglect other \gls{tsn} features, such as time synchronization.

In summary, none of the existing \gls{tsn} simulators includes all \gls{tsn} features.
Although the latest INET library supports most of these features, the correctness of combining them has not yet been validated.
Moreover, the existing \gls{tsn} simulators lack a comprehensive framework that accommodates large-scale heterogeneous network architectures~\cite{8458130}.

Emulators provide a cost-efficient alternative to dedicated hardware testbeds by mimicking the underlying hardware and software environments to test code in a variety of settings. Thus, emulations  are commonly an important intermediate evaluation and developmental phase towards real-world experiments with actual hardware. 

One example of software for emulation is \textit{Mininet}~\cite{Mininet}.
Mininet uses Linux \textit{namespaces} to emulate a set of nodes on a host system. 
The connections between the nodes are made via virtual interfaces, so that the behavior is similar to a \gls{vm}-based virtual environment but with less overhead.
Mininet provides an \texttt{ssh} connection into each node, thus each node can be handled as a standalone Linux system.
Ulbricht et al.~\cite{Ulbr2105:Emulation} implemented \gls{tsn} in Mininet.
The scalability of network emulation is limited by the resources of the host system; therefore, the precision of emulated \gls{tsn} networks typically decreases as the number of used nodes increases.

\subsection{Measurements in Hardware Testbeds}
Simulation, mathematical analysis, and emulation are important stepping stones on the way towards designing real-world testbeds.
However, using dedicated hardware is in most cases more precise while having the same complexity as the targeted real network deployment.
Whereas simulation models reduce the system complexity, dedicated \gls{tsn} devices offer the full range of settings and parameters to be considered.
Caused by limited resources, scalability is a limiting factor for designing hardware-based \gls{tsn} testbeds.
Employing hardware testbeds comes with its own set of complexities: the measuring techniques, setup, and other elements can strongly influence the outcome.
For instance, simulation models usually have built-in tools for collecting evaluation data. However, aside from some exceptions, such as development kits or explicit measurement tools, commercial hardware is typically not designed with built-in evaluation tools.

Measurement evaluations of proposed \gls{tsn} methods have been conducted on dedicated hardware testbeds for several purposes, as reviewed next. 

\subsubsection{Packet Processing}
A set of software and hardware should be employed for conducting the intended evaluation in testbeds.
For example, when utilizing time-aware gates, the precision of producing scheduled traffic can be crucial. 
We briefly review options for generating and time-stamping packets.
MoonGen~\cite{rw1} is a high-speed packet generator that takes advantage of hardware capabilities to reliably regulate the pace of arbitrary traffic patterns on commodity hardware and to timestamp data packets with sub-microsecond precision.
For efficient operation, MoonGen requires the \acrfull{dpdk}, which comprises libraries to speed up workloads operating on a broad range of CPU architectures. 

P4STA~\cite{rw2} operates on programmable hardware, such as smartNICs and \gls{fpga}~\cite{lin2019sur,sha2020har,thy2022ope}, as well as P4-based technology~\cite{lia2022adv} to achieve measurement precisions of a few nanoseconds. 
Programmable network hardware is important for flexible and precise measurements~\cite{9722801}.
Runge et al.~\cite{rw3} and Beifuß et al.~\cite{rw4} focused on \gls{qos} aspects of packet processing in \gls{cots} hardware and investigated the effects of the Linux network stacks and related queuing issues.

\subsubsection{Evaluating the Precision Time Protocol (PTP)}
Some hardware testbed studies concentrate on improving the hardware and software aspects of time-synchronization, mostly with \gls{ptp}.
Clock synchronization precision in smart-grid systems is investigated in~\cite{nrw7}.
For these systems to operate reliably, timestamping is necessary, and it is crucial to maintain performance standards even when communication bandwidth is limited.
To achieve an optimal \gls{ptp} system for a \gls{wan}, Kassouf et al. investigate the effects of various network factors and clock settings on synchronization precision.
They report a \SI{6.38}{\nano\second} back-to-back master-slave deviation, compared to our 3.4\,ns standard deviation and 11\, ns maximum deviation, see \autoref{fig:ptp-measurements}.
Ferencz et al.~\cite{nrw8} built a testbed using inexpensive \gls{cots} components, such as generally available CPUs with x86~architecture and typical \glspl{nic}.
Using IEEE~1588v2 and hardware-assisted \gls{cots} devices, they aim to attain sub-\SI{1}{\micro\second} precision.
In a \gls{ptp}-limited performance evaluation, they observed a synchronization precision of \SI{482.71}{\nano\second}.
Similarly, Kyriakakis et al.~\cite{nrw9} employ hardware-assisted clocks for synchronization.
To increase the timestamp precision, they integrate a \gls{ptp} hardware-assist unit into a multicore \gls{fpga} platform.
Through studies on a testbed consisting of two \glspl{fpga} that implement the suggested design and are interconnected through a \gls{cots} switch, they examined the worst-case execution time, achieving a worst-case offset of \SI{138}{\nano\second}.

A completely centralized IEEE~802.1Qcc~\cite{ieee802.1qcc} architecture for configuring the \gls{gptp} profile is used by Thi et al.~\cite{nrw4} in an \gls{sdn}-based automotive and industrial application setting.
Thi et al. examined specialized hardware from the manufacturer NXP and limited their evaluations to temporal synchronization, ignoring packet scheduling.

The existing HW testbed studies on evaluating \gls{ptp} typically evaluate only the synchronization precision. 
In contrast, this TSN-FlexTest study evaluates the \gls{ptp} synchronization precision, and broadly evaluates \gls{tsn} mechanisms.

\subsubsection{Communication Over-the-Air}
The performance and design of \gls{tsn} in complex structures as well as their integration with other communication technologies, such as WiFi and 5G, have attracted the interest of both academia and industry.
Sudhakaran~et~al.~\cite{nrw1} designed an industrial testbed that
combines traditional \gls{tsn} in the wired domain with IEEE~802.11.
Sudhakaran~et~al.~\cite{nrw1} are primarily concerned with the feasibility and testing of an industrial collaborative robotics use case.
Kehl et al.~\cite{nrw17} built an over-the-air testbed on an industrial shop floor with a prototype of 5G-\gls{tsn} integration to investigate a typical industrial use case with cloud-controlled mobile robots.

Agarwal~et~al.~\cite{nrw5} investigated the benefit of employing \gls{tsn} in microgrid monitoring.
They used a commercial Cisco switch to conduct measurements on a laboratory-scale microgrid with four nodes.
Agarwal~et~al.~\cite{nrw5} found that \gls{tsn} can enhance \gls{qos} through increased data rates and traffic shaping while retaining connection reliability.
\subsubsection{TSN Measurement}
Bohm~et~al.~\cite{nrw3} investigated the controller performance of \gls{tsn} networks on a three-domain \gls{tsn} testbed.
They found that a negotiation mechanism for provisioning real-time end-to-end connections across various \gls{tsn} domains can be beneficial.
In contrast to the TSN-FlexTest measurement technique, Bohm~et~al.~\cite{nrw3} did not measure the latencies of individual packets and did not consider oversaturated network conditions.

Bosk et al.~\cite{nrw18} devised a methodology to assess \gls{tsn} networks, specifically in the context of \gls{ivn}.
Through systematic analysis of \gls{ivn} traffic patterns for \gls{tsn} and \gls{be} traffic, Bosk et al.~\cite{nrw18} evaluate different \gls{tsn} standards in a configuration using \gls{cots} hardware and open-source solutions.
They try to optimize the system to meet the \gls{ivn} application requirements.
To achieve this, Bosk et al. employ the EnGine~\cite{methodology} framework, which provides a reproducible and scalable \gls{tsn} experimentation environment.
However, the measurement environment in \cite{nrw18,methodology} mixes hardware- and software-based time stamping in combination with low-precision \texttt{iperf} packet generation which results in relatively long cycle times for the \gls{tas} configurations and underscore the need of a well-evaluated \gls{tsn} measurement methodology.

\subsubsection{Time-sensitive Networking Management}
Several testbed measurement studies investigated the impact of dynamic reconfiguration of the \gls{tsn} network and optimizing \gls{tsn} main functionalities, e.g., shapers, frame preemption, and stream reliability, as reviewed in the following.
Jiang et al.~\cite{8869206} provided not only the \gls{tsn} simulation but also built a hardware testbed.
The simulation results verify that the real-world testbed matches the \gls{tsn} specification, even when they have just $\mu s$ resolution.
A self-configuring testbed for \gls{qos} management has been developed by Garbugli et al.~\cite{nrw2} and employed for testing numerous scenarios with varying packet sizes.
However, the details of the testbed, such as the specifications of the switches in the testbed, are not disclosed.
Groß et al.~\cite{nrw10} proposed a specialized hardware extension for standard Ethernet controllers.
The flexible and scalable architecture proposed in ~\cite{nrw10} can run solely in hardware, solely in software, or in both hardware and software.
The proposed \gls{fpga}-based component can reduce CPU load and software complexity and is particularly beneficial for applications with mixed cyclic and acyclic traffic types.  

Coleman et al.~\cite{nrw11} addressed the performance impact of the degree of precision of network and CPU clock synchronization on a real-time \gls{tsn} network. 
Coleman et al. presented an enhancement utilizing the PCIe bus interface to improve the network-to-CPU synchronization and assessed the improvement using \gls{cots} hardware.
Vlk et al.~\cite{nrw14} developed a method for increasing the schedulability and throughput of time-triggered traffic in IEEE~802.1Qbv. 
This technique relaxes the scheduling constraints and throughput of time-triggered traffic in a \gls{tsn} network while maintaining the deterministic nature and timing guarantees.
Related routing techniques for IEEE~802.1Qbv have been examined by Nayak et al.~\cite{nay2018rou}.

Miranda et al.~\cite{nrw16} demonstrate the setup and operation of a cloud-based Linux testbed for \gls{tsn} experimentation.  
Using a \gls{cnc} controller prototype, \gls{tsn} bridges and nodes can be initialized and managed in the Linux testbed. 

\subsubsection{Improved Hardware Design for Deterministic Communication}
A few research studies have focused on hardware designs to improve the \gls{tsn} performance. 
A very early discussion on \gls{tsn} testbeds in~\cite{nrw6} provides some guidelines for building a measurement testbed and notes some pitfalls while building a testbed. 
Pruski et al.~\cite{nrw12} elaborate hardware constraints in switch input/output queues and recommend a design architecture for high-capacity switches.
The proposed switch simulation can decrease residency time and resource costs by supporting \gls{fp}, \gls{cbs}, and \gls{tsn}, as well as other scheduling methods.
A time-triggered \gls{sms} design for memory-efficient \gls{tsn} switches is presented by Li~et~al.~\cite{nrw13}.
\gls{sms} can achieve efficiency by facilitating complicated \gls{tsn} tasks with scalable schedulability and fault-tolerance support. In addition, some measurements using NetFPGA hardware have been reported in~\cite{nrw6}. 

The studies on designing and evaluating \gls{tsn} switches based on \gls{fpga}, e.g., \cite{nrw13,nrw6}, are interesting because \gls{fpga} can achieve high performance,  programmability, and customization with a lower entry cost. However, \gls{fpga} is not cost-efficient for mass production. Also, the \gls{fpga} design procedure is time-consuming and requires a relatively high effort for achieving high precision levels. In contrast, the proposed TSN-FlexTest testbed is based on \gls{cots} hardware, which can be purchased and set up with relatively low development effort, while achieving a high degree of precision.

Overall, the existing testbed studies focus on specific narrow sets of \glspl{kpi}. Also, the existing studies commonly consider low precision levels and often lack detailed investigations of the effects that are experienced by a \gls{tas} scheduled data stream.
With our TSN-FlexTest testbed, we evaluate the measurement methods against the \gls{ptp} synchronization precision and we extend the measurement method with exact time stamping to a resolution that is in the same range as the \gls{ptp} synchronization to enable comprehensive  detailed measurements of the \gls{tsn} mechanisms, such as \gls{tas}.

\section{TSN-FlexTest Testbed Design} \label{sec:testbed}
This section provides insights into the hardware and software selected for the testbed. Also, the procedure flow and the used test data are described.

\subsection{Testbed Overview}
\label{sec:tb-overview}
The purpose of the described \gls{tsn} testbed is the examination of the behavior of different network stream patterns in one single \gls{tsn} switch.
\begin{figure}[tb]
    \centering
    \adjustbox{max width=\columnwidth}{
        \begin{tikzpicture}[thick]
        	\node [style=TUDbox1,minimum height=1.5cm,minimum width=2cm] (dut) at (0, 0) {\gls{dut}};
            \node [style=textbox,minimum height=0.1cm,minimum width=0.1cm,above left=0cm of dut,anchor=center,xshift=0.2cm] (dutport1) {};        
            \node [style=textbox,minimum height=0.1cm,minimum width=0.1cm,above right=0cm of dut,anchor=center,xshift=-0.2cm] (dutport2) {};
            \node [style=textbox,minimum height=0.1cm,minimum width=0.1cm,below left=0cm of dut,anchor=center,xshift=0.2cm] (dutport3) {};        
            \node [style=textbox,minimum height=0.1cm,minimum width=0.1cm,below right=0cm of dut,anchor=center,xshift=-0.2cm] (dutport4) {};
            \node [style=textbox,minimum height=0.1cm,minimum width=0.1cm,left=0cm of dut,anchor=center] (dutport0) {};
            \node [style=TUDbox2,minimum height=0.1cm,minimum width=0.1cm,above=0.7cm of dutport1] (node1) {node1};
            \node [style=TUDbox2,minimum height=0.1cm,minimum width=0.1cm,above=0.7cm of dutport2] (node2) {node2};
            \node [style=TUDbox2,minimum height=0.1cm,minimum width=0.1cm,below=0.7cm of dutport3] (node3) {node3};
            \node [style=TUDbox2,minimum height=0.1cm,minimum width=0.1cm,below=0.7cm of dutport4] (node4) {node4};
            \node [style=TUDbox2,minimum height=0.1cm,minimum width=0.1cm,left=0.75cm of dutport0] (node0) {node0};
            \node [style=TUDbox3,right=0cm of node0,anchor=west] (node0port0) {\tiny T};
            \node [style=TUDbox3,minimum height=0.1cm,minimum width=0.1cm,below=0cm of node1,anchor=north] (node1port0) {\tiny T};
            \node [style=TUDbox3,minimum height=0.1cm,minimum width=0.1cm,below=0cm of node2,anchor=north] (node2port0) {\tiny T};
            \node [style=TUDbox3,minimum height=0.1cm,minimum width=0.1cm,above=0cm of node3,anchor=south] (node3port0) {\tiny T};
            \node [style=textbox,minimum height=0.1cm,minimum width=0.1cm,above=0cm of node4,anchor=center] (node4port0) {};
            \node [style=textbox,minimum height=0.1cm,minimum width=0.1cm,left=0cm of node0,anchor=center] (node0port1) {};
            \node [style=textbox,minimum height=0.1cm,minimum width=0.1cm,above=0cm of node1,anchor=center] (node1port1) {};
            \node [style=textbox,minimum height=0.1cm,minimum width=0.1cm,above=0cm of node2,anchor=center] (node2port1) {};
            \node [style=textbox,minimum height=0.1cm,minimum width=0.1cm,below=0cm of node3,anchor=center] (node3port1) {};
            \node [style=textbox,minimum height=0.1cm,minimum width=0.1cm,below=0cm of node4,anchor=center] (node4port1) {};
            \node [style=TUDbox3,minimum height=0.1cm,minimum width=0.1cm,left=0.6cm of node0] (mntswitch) {MNGT Switch};
            \node [style=TUDbox2,minimum height=0.1cm,minimum width=0.1cm,left=0.6cm of mntswitch] (ctlpc) {CTRL PC};
            \node [style=textbox,minimum height=0.1cm,minimum width=0.1cm,right=0cm of ctlpc,anchor=center] (ctlpcport0) {};
            \draw[->,color=TUDBlue100,line width=0.3mm] (dutport0.center) -- (node0port0.east) {};
            \draw[<-,color=TUDBlue100,line width=0.3mm] (dutport1.center) -- (node1port0.south) {};
            \draw[<-,color=TUDBlue100,line width=0.3mm] (dutport2.center) -- (node2port0.south) {};
            \draw[<-,color=TUDBlue100,line width=0.3mm] (dutport3.center) -- (node3port0.north) {};
            \draw[<-,color=TUDBlue100,line width=0.3mm] (dutport4.east) -- (node4port0.east) {};
            \draw[<-,color=TUDBlue100,line width=0.3mm] (dutport4.west) -- (node4port0.west) {};
            \node [style=textbox,minimum height=0.1cm,minimum width=0.1cm,above=2cm of mntswitch] (help1) {};
            \node [style=textbox,minimum height=0.1cm,minimum width=0.1cm,above left=0.1cm of help1] (help2) {};
            \node [style=textbox,minimum height=0.1cm,minimum width=0.1cm,below=2cm of mntswitch] (help3) {};
            \node [style=textbox,minimum height=0.1cm,minimum width=0.1cm,below left=0.1cm of help3] (help4) {};
            \node [style=textbox,minimum height=0.1cm,minimum width=0.1cm,above=0.0cm of mntswitch,anchor=center,xshift=-0.35cm] (help5) {};
            \node [style=textbox,minimum height=0.1cm,minimum width=0.1cm,below=0.0cm of mntswitch,anchor=center,xshift=-0.35cm] (help6) {};
            \draw[<->,dashed,color=TUDBlue100,line width=0.3mm] (mntswitch) -- (node0port1.center) {};
            \draw[->,dashed,color=TUDBlue100,line width=0.3mm] (help1.center) -| (node1port1.center) {};
            \draw[->,dashed,color=TUDBlue100,line width=0.3mm] (help2.center) -| (node2port1.center) {};
            \draw[->,dashed,color=TUDBlue100,line width=0.3mm] (help3.center) -| (node3port1.center) {};
            \draw[->,dashed,color=TUDBlue100,line width=0.3mm] (help4.center) -| (node4port1.center) {};
            \draw[->,dashed,color=TUDBlue100,line width=0.3mm] (help1.center) -- (mntswitch) {};
            \draw[->,dashed,color=TUDBlue100,line width=0.3mm] (help2.center) -- (help5.center) {};
            \draw[->,dashed,color=TUDBlue100,line width=0.3mm] (help3.center) -- (mntswitch) {};
            \draw[->,dashed,color=TUDBlue100,line width=0.3mm] (help4.center) -- (help6.center) {};
            \draw[<->,dashed,color=TUDBlue100,line width=0.3mm] (ctlpcport0.center) -- (mntswitch) {};
        \end{tikzpicture}
    }
    \caption{The \gls{tsn} testbed architecture is comprised of five nodes in a star topology: one sink (node0) and four source nodes are used to evaluate the \acrfull{dut}. 
    The solid line arrows indicate the direction of the (payload) data communication.
    A control PC manages the measurements and is connected (see dashed line arrows) via a management switch with the nodes. 
    "T" represent the logical timestamping units.
    Node4 generates cross traffic with \texttt{MoonGen} on two links at once.
    All link bitrates are \SI{1}{\giga\bit\per\second}.}
    \label{fig:tsn-testbed-architecture}
\end{figure}
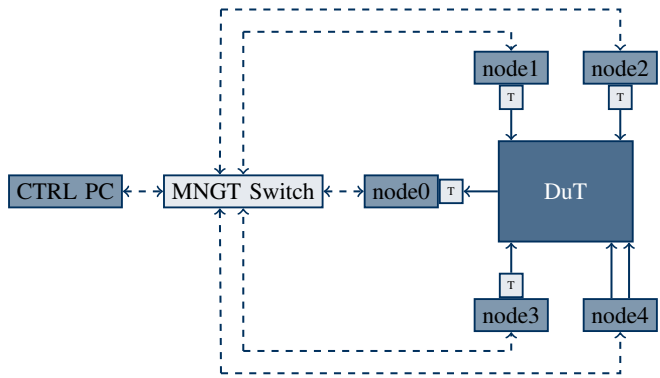
\autoref{fig:tsn-testbed-architecture} shows the testbed topology.
Five dedicated nodes are connected to one \gls{tsn} switch, which is the \gls{dut}, with \SI{1}{\giga\bit\per\second} links of negligible length.
A node is an end-station in a \gls{tsn} domain, e.g., talker or listener, and is comprised of \gls{cots} hardware, including an x86~CPU and multiple \glspl{nic}.
Node1, node2, and node3 replay and transmit different, high-priority packet streams through the \gls{dut} to one sink node (node0) which captures all received traffic.
Node4 generates two streams of interfering cross traffic.
All five generated packet streams from the transmitting nodes share the same bottleneck by traversing the output port of the \gls{dut} to the sink node (node0).
On this bottleneck, different scheduling and shaping strategies can be applied.

The topology is designed for the unidirectional measurement of the packet latency from the time instant when a transmitting node (node1, node2, or node3) transmits the first bit of a packet onto the physical link to the time instant when the first byte of the packet is received by the receive-NIC on the receiving node (node0). 
The testbed thus enables the measurement of the sojourn (residence) time of a packet inside the switch \gls{dut}, which always operates in cut-through mode in our evaluations, with a precision of a few tens of nanoseconds (see Section~\ref{sec:evaluation-ptp}). 
The packet sojourn time in the switch \gls{dut} accounts for the switch packet processing delay and packet queueing delay, and the packet transmission delay for the cut-through portion of the packet (see \autoref{sec:evaluation-cut-through-test}); in our setup, the link propagation delays are negligible due to the short physical cables. 
For the measurement, all devices are synchronized using \gls{ptp}, whereby one-step timestamping is used for one-way delay measurements.
The synchronization traffic is transferred in-band to be closely aligned with real \gls{tsn} scenarios where the main benefit of \gls{tsn} is the convergence of link technologies.

\subsection{Analysis of Available and Utilized Hardware}
\label{sec:testbed-hardware}
In preparation for the testbed design, we conducted an extensive market analysis of the available \gls{tsn} hardware components, which is summarized in~\autoref{tab:AV_HW}. 
We consider \gls{tsn} \glspl{ipc}, \glspl{soc}, \glspl{sep}, switches, and \glspl{nic}. 
Some vendors provide combined setups as \gls{tsn} kits.
We categorize the devices by available \gls{tsn} features. 
All devices support 802.1Qbv and \gls{ptp} time synchronization. 
The \texttt{Intel} \glspl{nic} does not have a dedicated hardware \gls{tas}, but supports the hardware-accelerated \gls{tx}-transmission time, which enhances all software schedulers with additional hardware support and improves their precision.

In comparison to industry-grade products, basic COTS (generic commodity) hardware components in combination with Linux facilitate the flexible extraction of measurement data. 
On the other hand, industry-grade hardware products provide typically state-of-the-art reliable performance.
For example, the \texttt{TrustNode} has an industry-grade version and a research version. The research version, which we purchased, has a highly precise and reliable packet time stamping method which makes additional time synchronization for source and sink unnecessary \cite{Ulbr2105:Emulation}. Unfortunately, with enabled timestamping of every packet, the device was not stable enough to handle two \SI{1}{\giga\bit\per\second} streams of \texttt{MoonGen} which makes the \texttt{TrustNode} unsuitable for our test setup.

The integration and debugging of \glspl{ipc} is a very complex and resource-consuming task and requires \gls{vhdl} programming skills and a \gls{vhdl} based device.
Therefore, the \gls{ipc} solution should be considered as a last resort if simpler methods are not applicable.
Also, the availability is a blocking point for hardware selection, \texttt{Intel} has proposed a special \gls{tsn} \gls{cpu} with \gls{tcc} \cite{inteltcc} which will probably reduce the effects of \gls{cpu} load on timing and measurements, as described in \autoref{sec:related-work}. Unfortunately, the device was not yet available for order.
Thus, the testbed hardware selection is a trade-off of feature completeness, measurement capability, stability, and availability. 
\begin{table*}[!t]
    \centering
    \caption{Summary of available \gls{tsn} \gls{ipc}, \gls{soc}, \glspl{sep}, \gls{tsn} kits, switches, and NICs.
    Features marked as (Y) are only available on devices with hardware-assisted \gls{tx}-injection on the \glspl{nic}. This table extends \cite{timelySurvey} and is up-to-date as of 10/2022.}
    \label{tab:AV_HW}
    \begin{tabularx}{\textwidth}{c l l l c c c c c c c c c }
    \toprule
    \textbf{ID} & \textbf{Type}  & \textbf{Manufacturer} & \textbf{Product Name} &\multicolumn{9}{c}{\textbf{Supported features}}\\ 
                &                &                       &                       &\gls{tas}&\gls{ats}&\gls{fp}&\gls{psfp}&\gls{ptp}&\gls{cbs}&\acrshort{srp}&\gls{cqf}&\\ 
    \midrule
    1& \gls{ipc}        &Fraunhofer IPMS                &IP-Core for TSN                    &Y &N 									&Y &Y   &Y  &Y	&Y  &Y\\
    2& \gls{soc}        &Broadcom                       &BCM53570                           &Y &N 									&Y &Y   &Y  &N	&Y  &Y\\
    3& \gls{sep}        &Hilscher                       &netX 90 Starter-kit                &Y &N 									&Y &N   &Y  &N	&N  &N\\
    4& \gls{sep}        &TTTech  Industrial             &Edge IP Solution EVM               &Y &N 									&Y &Y   &Y  &Y	&N  &N\\
    4& \gls{tsn} Kit    &Analog Device                  &RAPID-TSNEK-v0001                  &Y &N 							        &Y &Y   &Y  &N	&Y  &N\\
    6& \gls{tsn} Kit    &Kontron                        &KBox C-102-2  TSN starter kit      &Y &N 									&Y &N   &Y  &N	&Y  &N\\
    7& \gls{tsn} Kit    &NXP                            &LS1021A-TSN-R                      &Y &N 									&N &Y   &Y  &Y	&N  &N\\
    8& Switch          &Advantech                      &EKI-8510G-2FI                      &Y &N 									&Y &Y   &Y  &N	&N  &N\\
    9& Switch          &Cisco                          &IE-4000                            &Y &N 									&N &N   &Y  &N	&N  &N\\
    10& Switch          &Comcores                       &Manticore                          &Y &N 									&Y &Y   &Y  &Y	&N  &N\\
    11& Switch          &FibroLAN                       &Falcon-RX/G                        &Y &N 									&Y &Y   &Y  &N	&N  &N\\
    12& Switch           &Hirschmann                     &RSPE                               &Y &N 								    &N &N   &Y  &N	&N  &N\\
    13& Switch           &Hirschmann                     &BOBCAT                             &Y &N 									&N &N   &Y  &N	&N  &N\\
    14& Switch          &InnoRoute                      &TrustNode industrial               &Y &N 									&N &Y   &Y  &N	&N  &N\\
    15& Switch           &Kontron                        &KSwitch D10 MMT Series             &Y &N                                   &Y &N   &Y  &Y &N  &N\\
    16& Switch          &Marvell                        &88Q6113                            &Y &N 									&N &N   &Y  &Y	&N  &N\\
    17& Switch          &\acrshort{soc} Engineering     &MTSN Kit 1G Multiport              &Y &N 									&Y &Y   &Y  &Y	&Y  &N\\
    18& \gls{nic}       &ASIX Electronics               &AXM57104                           &Y &N 									&N &Y   &Y  &Y	&Y  &N\\
    19& \gls{nic}       &Intel                          &I210                               &(Y) &(Y)~\cite{pfefferle2021ieee}      &N &N   &Y  &(Y)&Y  &N\\
    20& \gls{nic}       &Intel                          &I225                               &(Y) &(Y)~\cite{pfefferle2021ieee}      &N &N   &Y  &(Y)&Y  &N\\
    21& \gls{nic}       &Kontron                        &PCIe-0400                          &Y   &(Y)~\cite{pfefferle2021ieee}      &Y &N   &Y  &(Y)&Y  &N\\
    22& \gls{nic}/\gls{sep}       &InnoRoute            & Raspberry RealTime HAT            &Y &(Y)~\cite{pfefferle2021ieee}        &N &N   &Y  &(Y)&Y  &N\\
    \bottomrule
    \end{tabularx}
\end{table*}
\begin{table}[bt]
    \centering
    \caption{Utilized Hardware in the Testbed.}
    \label{tab:utilized-hw}
    \begin{tabularx}{\columnwidth}{l X}
    \toprule
    \textbf{Component} & \textbf{Description}\\ 
    \midrule
    CPU                                     & Intel Core i7 6700\\
    Mainboard                               & Gigabyte H270N-WIFI\\
    RAM                                     & $2 \times$\SI{16}{\giga\byte}\\
    \acrshort{nic} Source (Prio)            & Intel I210\\
    \acrshort{nic} Source (cross traffic)   & Intel X520\\
    \acrshort{nic} Sink (onboard)           & Intel I211\\
    Operating System                        & Ubuntu 20.04.3 (GNU/Linux 5.11.4-rt11)\\
    \acrshort{dut}                          & FibroLAN Falcon-RX/G (SW:8.0.2.4)\\
    \bottomrule
    \end{tabularx}
\end{table}

As summarized in \autoref{tab:utilized-hw},
we selected the \texttt{Intel} \gls{nic} as flexible COTS device, and the stable and available switch from \texttt{FibroLan}.
Building on the insights from \autoref{sec:related-work}, we designed the testing nodes with enough resources to minimize \gls{cpu} scheduling effects on the measurements.

\subsection{Analysis of Available and Utilized Software} 

\subsubsection{Precise Time-Synchronization}
\label{sec:design-ptp}

A stable timebase is an important requirement for a viable \gls{tsn} testbed.
There are two main approaches for measuring the timing of network packets: The measurement system can be centralized in one location in the network; thus, the transmitted and received packets can be easily correlated in the same device without any requirements to synchronize clocks \cite{ris20215g,rischke2022empirical}.
The centralized setup is limited in that it does not allow measurements on different devices which are typical for \gls{i40} or other \gls{tsn} environments. Conducting all evaluations on one device also creates new issues regarding the shared resources for \gls{rx} and \gls{tx}.

The distributed method allows realistic measurement at different network locations. The probe locations require a common timebase which can be provided by clock synchronization. In comparison to the centralized method, the synchronization precision is the most important issue in the distributed configuration.
As described in \autoref{sec:background-timesync}, the \gls{ptp} protocol can provide a precise and stable synchronization through two clocks via Ethernet.

For validating the reliability of our testbed we conducted several investigations regarding the clock stability.
Emmerich et al.~\cite{rw1} investigated the synchronization capabilities of the \textit{Intel 82599} \gls{nic}.
The internal clock of the \textit{Intel 82599} is \SI{64}{\bit} wide, whereby the \gls{lsb} represents \SI{1}{\nano\second}. A timer event periodically increments the clock value. Unfortunately, this period depends on the link speed and for \SI{1}{\giga\bit\per\second} is fixed to \SI{64}{\nano\second}~\cite{i82599ds}.
Emmerich et al.~\cite{rw1} measured that the reported clock values from the \gls{nic} are consistently multiples of \SI{64}{\nano\second}.

Newer \glspl{nic} feature improved clock resolution. 
The \textit{I210} \gls{nic}, which evolved from the \textit{Intel 82599}, has an internal \SI{96}{\bit} wide timer which is incremented every \SI{8}{\nano\second}, independent of the link speed. \cite{i210ds}
Due to the larger cock register, the \textit{I210} has \SI{32}{\bit} in the sub nanosecond range, allowing the clock to be adjusted in fine-granular steps. Every \SI{8}{\nano\second} the \gls{nic} adds an increment value to the clock register. This increment value can be slightly greater or smaller than \SI{8}{\nano\second}.
The value itself is provided during the clock synchronization process from \texttt{ptp4l} via the \textit{ptp\_adjfreq()} kernel interface callback function.
By manipulating the increment value, the clock can be forced to run slightly faster or slower.
The clock control is thus controlled by the \texttt{ptp4l} clock servo.
Due to this implementation, the investigated clock values reported from the \textit{I210} \gls{nic} are not multiples of the clock granularity, but can differ in multiples of \SI{1}{\nano\second}.

There are several effects that influence the \gls{ptp} precision.
As described in \autoref{sec:background-timesync}, \texttt{ptp4l} periodically measures the path delay and adjusts the local hardware clock based on the received master clock values corrected by the path delay.
Thus, the jitter of the path delay and the corresponding servo-loop settings are important factors for the \gls{ptp} precision.

\subsubsection{Packet Generation}
\label{sec:packgen}
There are basically two types of traffic generators: a) (raw) socket based, and b) \gls{dpdk} based. Both approaches have pros and cons: using sockets and generating packets in user space is highly flexible and allows packet generation and other network usage (e.g., \gls{ptp}) at once. The disadvantage of sockets are the limited performance for generating enough traffic to fully load an Ethernet link. Emmerich et al.~\cite{emmerich2017mind} and Wong et al.~\cite{wong2018evaluation} found that not all packet generators can saturate a link, especially if small packets are used. 
If precision is not important, the usage of \gls{tcp} or \gls{udp} sockets within \texttt{iperf} is a cheap solution for generating traffic \cite{nrw18}. Even if \texttt{iperf} has not an optimal performance as packet generator \cite{wong2018evaluation}.

A more powerful approach is provided by \gls{dpdk}, by making the configuration memory space of the \gls{nic} accessible from user space. This enables the user to write Ethernet data into memory and to pass the memory pointer directly to the \gls{nic}'s \gls{tx} queue without utilizing the Linux network stack.
The problem with \gls{dpdk} is that it replaces the \gls{nic}'s driver with a user space pass-trough driver and disconnects the \gls{nic} from the Linux network stack. Thus, common software, such as \texttt{ptp4l}, cannot be used on this interface anymore.

Next to the achievable data rate, the generator's precision is an important \gls{kpi}. A fixed cyclic stream should transmit the packets periodically exactly according to the predefined cycle time. However, due to \gls{cpu} load or other software effects, the packets transmission time can have a random or constant offset. \gls{dpdk} based traffic generators achieve higher periodicity precision than socket-based generators \cite{emmerich2017mind}. 
MoonGen \cite{rw1} (which was compared with \texttt{tcpreplay} in \cite{9844050}) provides a \texttt{lua} framework for generating high data rate packet load with the help of \gls{dpdk}. We use MoonGen for generating high-load cross traffic on a dedicated node (node4) that does not require \gls{ptp} synchronization. 
For the replay of our test traffic we employ \texttt{tcpreplay} \cite{9844050}, which uses raw sockets to replay arbitrary traffic from prerecorded files.

\subsubsection{Scheduler -- GCL Configuration}
\label{sec:taprio-description}
Since version 5.4, the Linux kernel provides several features for hardware timestamping and hardware-assisted packet transmission. 
One feature is an (additional) transmission timestamp.
In contrast to the \gls{tx} timestamp, this is the time when the frame should be transmitted, i.e., a value in the future.
In theory, the packet with an additional transmission timestamp is handed over from the network driver to the \gls{nic} hardware, which stores the packet in the transmission queue and will release the packet when the transmission time is reached.

Dealing with limited resources in the hardware layer of the \gls{nic} this procedure needs to be supported by the higher layers of the network protocol stack. The \gls{nic} hardware cannot store a large amount of data for transmission, the internal memory is limited to a few packets. The \gls{nic} queue is a \gls{fifo} buffer; therefore, the packets cannot be reordered.
To respect the limited hardware resources, the Linux network stack has to be aware of these characteristics.
The qdisc environment provides an interface in the Linux network protocol stack to insert and manage queuing and packet scheduling. One loadable module is the \gls{etf} scheduler which reorders incoming packets according to their transmission timestamp.
To respect the limited hardware queue size, the packets need to be temporarily stored in a software queue and handed to the hardware just shortly before the transmission time. This time offset needs to be determined experimentally for each \gls{nic} and system configuration.
To set the transmission time of each packet, one possible module is \textit{ta-prio} which implements a software \gls{tas} for scheduling packets.

\subsection{Testbed General Procedure Flow}
The \gls{tsn} testbed supports the automatic test of the packet transmission capabilities of a dedicated \gls{dut}.
Several nodes are controlled to replay measurement traffic and send it through the \gls{dut}.
Each transmission node inserts a precise timestamp into the payload section of \gls{tx} packets, following Section~\ref{sec:background-timesync}. 
Additionally, nodes can be configured to replay a cross traffic to simulate a busy link.
One node (node0 in Fig.~\ref{fig:tsn-testbed-architecture}) captures the measurement packets after they have passed the \gls{dut}.
By comparing the \gls{tx} and \gls{rx} timestamps, the packets transit time through the DuT can be evaluated.
A detailed description of the testbed setup is provided in~\cite{9844050}. 

\subsubsection{Packet Header/Our Streams}
\label{sec:streamdescription}
\begin{table}[t]
    \centering
    \caption{Application layer (payload) traffic stream cycle duration and audio or video frame size mean (and standard deviation), as well as average bitrate (at link layer, incl. headers), of high-, medium-, and low-priority TSN streams of stream sets $\Theta$, $\Psi$, and $\Omega$ [with corresponding Priority Code Points (PCPs) 6, 5, and 4]; generic streams and best-effort cross traffic send maximum-sized Ethernet frame payload. 
    }
    \label{tab:streams}
    \scriptsize{
    \begin{tabularx}{\columnwidth}{lcccrrr}
    \toprule
    \textbf{Stream} & \textbf{Traffic}  & \textbf{PCP}  & \textbf{Perio-}   & \textbf{t\textsubscript{Cycle}}   & \textbf{Appl. Data}  & \textbf{Bitr.}\\   
    \textbf{Set}    & \textbf{Type}     &               & \textbf{dicity}   &  \textbf{[ms]}                    & \textbf{Size} [B] & \textbf{[b/s]}\\    
    \midrule
    \acrshort{stream1h} & \acrlong{stream1h} & 6 & cyclic   & 0.2   & 1472\,(0.0)       & 61\,M\\
    \acrshort{stream1m} & \acrlong{stream1m} & 5 & cyclic   & 0.3   & 1472\,(0.0)       & 41\,M\\
    \acrshort{stream1l} & \acrlong{stream1l} & 4 & cyclic   & 0.5   & 1472\,(0.0)       & 24\,M\\
    \hline
    \acrshort{stream2h} & \acrlong{stream2h} & 6 & cyclic   & 1     & 82\,(0.0)        & 1.1\,M\\
    \acrshort{stream2m} & \acrlong{stream2m} & 5 & cyclic   & 24    & 480\,(0.0)        & 177\,k\\
    \acrshort{stream2l} & \acrlong{stream2l} & 4 & cyclic   & 16.67 & 8336.4\,(24283.8) & 4.1\,M\\  
    \hline
    \acrshort{stream3h} & \acrlong{stream3h} & 6 & acyclic  & --    & 501.6\,(984.4)    & 175\,k\\
    \acrshort{stream3m} & \acrlong{stream3m} & 5 & cyclic   & 20    & 153.9\,(46.6)     & 82\,k\\
    \acrshort{stream3l} & \acrlong{stream3l} & 4 & cyclic   & 25    & 62412.5\,(60355.9)& 20\,M\\   
    \hline
                        & Cross Traffic      & 0 & acyclic  & --    & 1472\,(0.0)      & 2\,G\\
    \bottomrule
    \end{tabularx}
    }
\end{table}
To determine the packet handling behaviors of the \gls{dut}, the selection of characteristic test data is important. 
In the current scenario, we use three types of test data: a) generic generated cyclic streams, b) generated data streams which reflect typical \gls{i40} characteristics, and c) real captured data of a robot control environment, as summarized in \autoref{tab:streams}. 

The generic generated streams (\acrlong{stream1}) have maximum sized frames, but distinct packet generation cycle times to represent three cyclic \gls{tsn} streams.
The cycle times are prime numbers to enforce statistical collisions of all streams, even without cross traffic.

The \acrlong{stream2} corresponds to a typical \gls{i40} scenario where a machine delivers tactile data in combination with video and audio streams for human control and mechanical error monitoring~\cite{kothuru2018application}.
The audio and video data are lossy compressed therefore the data is cyclic and bursty with different burst sizes.

For \acrlong{stream3}, we captured the data streams of the well-known robot \textit{spot}~\cite{spotrobot}. 
The streams contain the control data stream of the spot robot in combination with the spot camera stream.
For audio, we included a lossy compressed voice audio file.

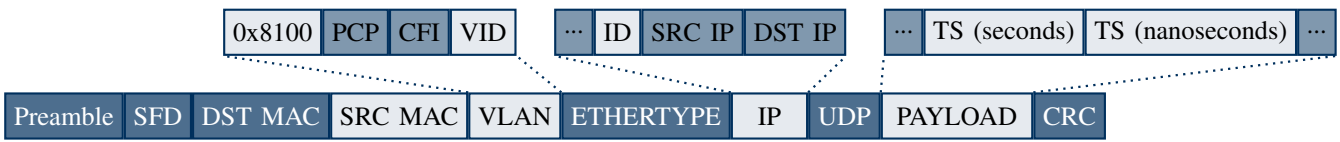
\begin{figure*}[tb]
    \centering
    \begin{tikzpicture}[thick]
    	\node [style=TUDbox1,minimum height=0.6cm] (dstmac) at (0, 0) {DST MAC};
        \node [style=TUDbox1,minimum height=0.6cm, left=0cm of dstmac.west] (sfd) {\acs{sfd}};
        \node [style=TUDbox1,minimum height=0.6cm, left=0cm of sfd.west] (sfd) {Preamble};
    	\node [style=TUDbox3,minimum height=0.6cm, right=0cm of dstmac.east] (srcmac) {SRC MAC};
        \node [style=TUDbox3,minimum height=0.6cm, right=0cm of srcmac.east] (vlan) {VLAN};
        \node [style=TUDbox3,minimum height=0.6cm, above=0.8cm of vlan.west,xshift=-2.6cm] (vlanethertype) {0x8100};
        \node [style=TUDbox2,minimum height=0.6cm, right=0cm of vlanethertype.east] (pcp) {PCP};
        \node [style=TUDbox2,minimum height=0.6cm, right=0cm of pcp.east] (cfi) {CFI};
        \node [style=TUDbox3,minimum height=0.6cm, right=0cm of cfi.east] (vid) {VID};
        \draw [dotted,color=TUDBlue100,line width=0.3mm] (vlan.north west) -- (vlanethertype.south west) {};
        \draw [dotted,color=TUDBlue100,line width=0.3mm] (vlan.north east) -- (vid.south east) {};
        \node [style=TUDbox1,minimum height=0.6cm, right=0cm of vlan.east] (etype) {ETHERTYPE};
        \node [style=TUDbox3,minimum height=0.6cm, right=0cm of etype.east,minimum width=1cm] (ip) {IP};
        \node [style=TUDbox2,minimum height=0.6cm, right=0.5cm of vid.east] (startip) {...};
        \node [style=TUDbox3,minimum height=0.6cm, right=0cm of startip.east] (idip) {ID};
        \node [style=TUDbox2,minimum height=0.6cm, right=0cm of idip.east] (srcip) {SRC IP};
        \node [style=TUDbox2,minimum height=0.6cm, right=0cm of srcip.east] (dstip) {DST IP};
        \draw [dotted,color=TUDBlue100,line width=0.3mm] (ip.north west) -- (startip.south west) {};
        \draw [dotted,color=TUDBlue100,line width=0.3mm] (ip.north east) -- (dstip.south east) {};
        \node [style=TUDbox1,minimum height=0.6cm, right=0cm of ip.east] (udp) {UDP};
        \node [style=TUDbox3,minimum height=0.6cm, right=0cm of udp.east, minimum width=2cm] (data) {PAYLOAD};
        \node [style=TUDbox2,minimum height=0.6cm, right=0.5cm of dstip.east] (startdata) {...};
        \node [style=TUDbox3,minimum height=0.6cm, right=0cm of startdata.east] (ts1) {TS (seconds)};
        \node [style=TUDbox3,minimum height=0.6cm, right=0cm of ts1.east] (ts2) {TS (nanoseconds)};
        \node [style=TUDbox2,minimum height=0.6cm, right=0cm of ts2.east] (enddata) {...};
        \node [style=TUDbox1,minimum height=0.6cm, right=0cm of data.east] (crc) {CRC};
        \draw [dotted,color=TUDBlue100,line width=0.3mm] (data.north west) -- (startdata.south west) {};
        \draw [dotted,color=TUDBlue100,line width=0.3mm] (data.north east) -- (enddata.south east) {};
    \end{tikzpicture}
    \caption{Frame structure of high-priority test data streams.
             The data is sent using VLAN-tagged Ethernet frames.
             The priority is set by modifying the \gls{pcp} according to \autoref{tab:streams}.
             Inside the IP header, the ID field is leveraged to provide a sequence number.
             The payload holds the \acrfull{ts}, which is split into integer and decimal part for seconds and nanoseconds, respectively.
             }
    \label{fig:packetstructure}
\end{figure*}

To be replayed in the \gls{tsn} testbed, the packet header needs to be modified.
\autoref{fig:packetstructure} shows the packet structure of stream type \acs{stream1}, \acs{stream2}, and \acs{stream3}.
The testbed uses several fields in the packet header for matching.
The source \gls{mac} address is used for matching the source node name of a received packet.
The \gls{dut} itself is configured to apply different queuing and scheduling strategies depending on the packet's \gls{vlan} \gls{pcp} ID.
According to \autoref{sec:timestampexplanation}, each transmitted packet is timestamped automatically by the \gls{tx} network hardware, which adds a \SI{64}{\bit} timestamp into the packet payload.

\subsubsection{Measurement Procedure}
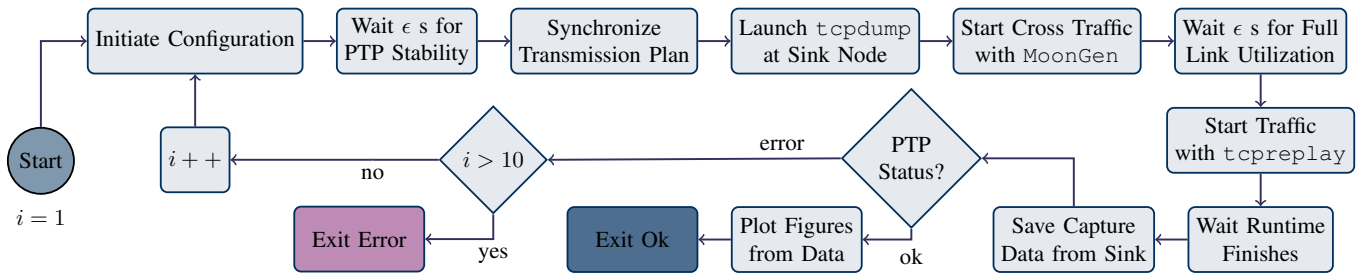
\begin{figure*}
    \adjustbox{max width=\textwidth}{
        \begin{tikzpicture}[thick]
            \node [style=dot,minimum width=1cm] (start) at (0,0) {Start};
            \node [style=textbox,minimum width=1cm,below=0.1cm of start] (startlabel) {$i=1$};
            \node [style=TUDroundbox3,minimum width=2cm,minimum height=1cm,above right=1cm of start.north,align=center,yshift=0.1cm] (initconf)  {Initiate Configuration};
            \node [style=TUDroundbox3,minimum width=2cm,minimum height=1cm,right=0.5cm of initconf.east,align=center] (waite)  {Wait $\epsilon$ s for\\PTP Stability};
            \node [style=TUDroundbox3,minimum width=2cm,minimum height=1cm,right=0.5cm of waite.east,align=center] (synctrans)  {Synchronize\\Transmission Plan};
            \node [style=TUDroundbox3,minimum width=2cm,minimum height=1cm,right=0.5cm of synctrans.east,align=center] (tcpdumpstart)  {Launch \texttt{tcpdump}\\at Sink Node};
            \node [style=TUDroundbox3,minimum width=2cm,minimum height=1cm,right=0.5cm of tcpdumpstart.east,align=center] (moonstart)  {Start Cross Traffic\\with \texttt{MoonGen}};
            \node [style=TUDroundbox3,minimum width=2cm,minimum height=1cm,right=0.5cm of moonstart.east,align=center] (waitlink)  {Wait $\epsilon$ s for Full\\Link Utilization};
            \node [style=TUDroundbox3,minimum width=2cm,minimum height=1cm,below=0.5cm of waitlink.south,align=center] (starttrans)  {Start Traffic\\with \texttt{tcpreplay}};
            \node [style=TUDroundbox3,minimum width=2cm,minimum height=1cm,below=0.5cm of starttrans.south,align=center] (waitcapture)  {Wait Runtime\\Finishes};
            \node [style=TUDroundbox3,minimum width=2cm,minimum height=1cm,left=0.5cm of waitcapture.west,align=center] (savedata)  {Save Capture\\Data from Sink};
            \node [style=TUDroundbox3,minimum width=2cm,minimum height=1cm,left=2cm of savedata.west,align=center] (plot)  {Plot Figures \\from Data};
            \node [style=decision,minimum width=1cm,minimum height=1cm,above left=1cm of savedata.west,align=center] (ptpdecision)  {PTP\\Status?};
            \node [style=TUDroundbox1,minimum width=2cm,minimum height=1cm,left=0.5cm of plot.west,align=center] (ok)  {Exit Ok};
            \node [style=decision,minimum width=1cm,minimum height=1cm,above left=1cm of ok.west,align=center,yshift=0.1cm] (errordecision)  {$i>10$};
            \node [style=TUDroundbox2error,minimum width=2cm,minimum height=1cm,below left=1cm of errordecision.west,align=center,xshift=0.5cm] (error)  {Exit Error};
            \node [style=TUDroundbox3,minimum width=1cm,minimum height=1cm,below=0.5cm of initconf.south,align=center,yshift=-0.3cm] (ipp)  {$i++$};
            \node [style=textbox,minimum width=1cm,minimum height=1cm,left=0.5cm of errordecision.west,align=center] (h1)  {};
            \draw[->,color=color4,line width=0.3mm] (start.north) |- (initconf.west) ;
            \draw[->,color=color4,line width=0.3mm] (initconf.east) -- (waite.west) ;
            \draw[->,color=color4,line width=0.3mm] (waite.east) -- (synctrans.west) ;
            \draw[->,color=color4,line width=0.3mm] (synctrans.east) -- (tcpdumpstart.west) ;
            \draw[->,color=color4,line width=0.3mm] (tcpdumpstart.east) -- (moonstart.west) ;
            \draw[->,color=color4,line width=0.3mm] (moonstart.east) -- (waitlink.west) ;
            \draw[->,color=color4,line width=0.3mm] (waitlink.south) -- (starttrans.north) ;
            \draw[->,color=color4,line width=0.3mm] (starttrans.south) -- (waitcapture.north) ;
            \draw[->,color=color4,line width=0.3mm] (waitcapture.west) -- (savedata.east) ;
            \draw[->,color=color4,line width=0.3mm] (savedata.north) |- (ptpdecision.east) ;
            \draw[->,color=color4,line width=0.3mm] (ptpdecision.south) |- (plot.east) node[sloped,midway,below,color=black,align=left] {ok};
            \draw[->,color=color4,line width=0.3mm] (plot.west) -- (ok.east) ;
            \draw[->,color=color4,line width=0.3mm] (ptpdecision.west) -- (errordecision.east) node[pos=0.2,above,color=black,align=left] {error};
            \draw[->,color=color4,line width=0.3mm] (errordecision.south) |- (error.east) node[sloped,midway,below,color=black,align=left] {yes};
            \draw[-,color=color4,line width=0.3mm] (errordecision.west) -| (h1.center) node[pos=0.2,midway,below,color=black,align=left] {no};
            \draw[->,color=color4,line width=0.3mm] (h1.center) |- (ipp.east) ;
            \draw[->,color=color4,line width=0.3mm] (ipp.north) -- (initconf.south) ;
        \end{tikzpicture}
    }
    \caption{Flowchart of the automated measurement process:
             The process flow should ensure that all measurements run successfully even if there are a few errors, e.g., due to timeouts during the \gls{ptp}-synchronization.
             Therefore, each test is performed at most ten times until it stops with an error.
             The process includes steps for initialization, synchronization, the measurement itself, as well as storing and analyzing the data.
             }
    \label{fig:flowchart}
\end{figure*}
The flowchart in \autoref{fig:flowchart} illustrates the test-bed measurement procedure. The testbed software, including data stream generators, are openly available from \url{https://github.com/5GCampus/tsn-testbed}.
The measurement procedure is structured into reconfiguration, traffic replay, and record results for plotting.
In the reconfiguration phase, the control PC resets all network connections and sets the nodes into a dedicated state. Depending on the test configuration, the software \gls{tas} is configured at the transmitting nodes.
To finish the preparation, the control \gls{pc} transfers the \gls{tx} pcaps to the nodes.

In the measurement phase, the control \gls{pc} starts the capturing process at the receiving node and initiates the pcap replay at the transmitting nodes.
If the replay or capture has finished, the received data is recorded and transferred to the control PC for further processing. We spotted that some \gls{dut} drop the \gls{ptp} connection sporadically. Because the \gls{ptp} synchronization is the backbone of the testbed we check the log for \gls{ptp} errors or to high clock derivations during the measurement and repeat the measurement if an error was detected. 

\section{TSN-FlexTest Testbed Validation}
\label{sec:testbed-validation}
Generally, a measurement is only as good as the metering setup. This section evaluates the measurement precision of our TSN-FlexTest testbed.
\subsection{Evaluating the Precision Time Protocol}
\label{sec:evaluation-ptp}
\begin{figure}[t]
    \centering
    \includegraphics[width=0.8\columnwidth]{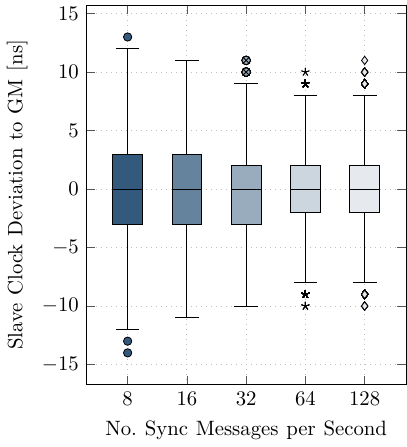}
    \caption{Boxplot of slave clock deviation to the \gls{gm} in nanoseconds as a function of number of synch messages per second over a measurement period of \SI{30}{\minute}.}
    \label{fig:ptp-measurements}
\end{figure}
\begin{lstlisting}[
  caption=Highlighted part of PTP configuration file with relation to the gPTP profile.,
  label=lst:PTPConf,language=sh,
  float=tb
]
[PTP config file]
#Default Data Set#
gmCapable		1
twoStepFlag		1
slaveOnly		1
domainNumber		0
[...]
#Port Data Set#
logAnnounceInterval	0
logSyncInterval		-6
operLogSyncInterval	0
announceReceiptTimeout	3
syncReceiptTimeout	3          #from gPTP
neighborPropDelayThresh	800        #from gPTP
min_neighbor_prop_delay -20000000  #from gPTP
BMCA  ptp
[...]
#Run Time Options#
logging_level	7
verbose	0
[...]
#Servo Options#
pi_proportional_const	      0.0
pi_integral_const	      0.0
pi_proportional_scale	      0.0
pi_proportional_exponent     -0.3
pi_proportional_norm_max      0.7
pi_integral_scale             0.0
[...]
max_frequency	900000000
[...]
#Transport Options#
[...]
#Default Interface Options#
clock_type		OC
network_transport	L2       #from gPTP
delay_mechanism		P2P      #from gPTP
time_stamping		hardware
[...]
\end{lstlisting}

As noted in \autoref{sec:design-ptp}, we designed the TSN-FlexTest testbed as a distributed architecture. 
In the distributed setting, clock synchronization via \gls{ptp} is a key factor for precise delay measurement.

For validating the testbed precision we conducted measurements for several servo settings and examined the influence of the 
cross traffic on the \texttt{ptp4l} precision.
To extract the clock synchronization metrics, we patched \texttt{ptp4l} to provide the actual clock deviation values and not only an \gls{rms} value.
Previous measurements demonstrated that the synchronization metrics of \texttt{ptp4l} are as precise as physical measurements with an oscilloscope~\cite{Ulbr2109:Precise}.
\autoref{fig:ptp-measurements} shows the results of the clock precision measurement.
For simplicity and to avoid clutter in this evaluation, we manipulate only one \texttt{ptp4l} servo parameter, namely the synchronization frequency with the number of sync messages per second in the range from $[8, \ldots, 128]$. 

We measured the clock deviation between one slave node and the \gls{dut} switch for several configurations with and without cross traffic in order to investigate the independence of the clock deviation from the cross traffic. 
Due to the activated hardware time stamping, the transmission of additional (cross traffic) packets increases the queuing delay of the \gls{ptp} packets.
However, this additional queueing delay is compensated for by the timestamps (which are applied after the \gls{tx} queue and before the \gls{rx} queue).
Thus, cross traffic does \textit{not} affect the testbed precision.
Generally, the transmission channel (link) between a node and switch has a jitter value, which needs to be compensated by the servo.
The results in \autoref{fig:ptp-measurements} indicate that the synchronization frequencies of 8 to 128 sync messages per second result in a reasonably small maximum clock deviation of less than \SI{\pm 14}{\nano\second}, indicating a precision of approximately \SI{30}{\nano\second}. 
As the synchronization frequency to 16 sync messages/s does not exhibit any clock deviation outliers, we use 16 synch messages/s for the remainder of this article.

\subsection{Sending Cyclic Data Precisely}
\label{sec:evaluation-cycle-time-measurement}
\begin{table*}[!t]
    \centering
    \caption{
        Measurement results of achievable cycle time precision: deviations from expected cycle time (of \SI{1}{\milli\second}) in nanoseconds; with 16 synch messages/second. 
        Type characterizes the software that generates (or subsequently shapes) the packets on sending nodes 1, 2, and 3.
    }
    \label{fig:cycle_time_comparison}
    \begin{tabularx}{0.9\textwidth}{llXrrrrrr}
    \toprule
    \textbf{Type} & \textbf{Tool} & \textbf{Version} & \textbf{Min} & \textbf{Mean} & \textbf{Median} & \textbf{Stdev} & \textbf{P99\%} & \textbf{Max}\\
    \midrule
    \textbf{Linux NAPI} & tcpreplay & 4.3.2     & $-$997176   & 149881    & 160763    & 29630     & 166339    & 447074\\
    \textbf{Linux NAPI} & tcpreplay & 4.3.4     & $-$998312   & 1         & 48        & 11001     & 15960     & 2254136\\
    \textbf{DPDK}       & MoonGen   & 25c61ee   & $-$6848     & 19        & 0         & 138.7     & 392       & 6895\\
    \midrule
    \textbf{Sender \acrshort{tas} (\SI{15}{\micro\second})}  & TA\_PRIO & Kernel 5.15 & $-$997904 & 18 & $-$126       & 19205   & 24822   & 2958199\\
    \textbf{Sender \acrshort{tas} (\SI{500}{\micro\second})}  & TA\_PRIO & Kernel 5.15 & $-$998784 & 336 & 2       & 1023658   & 2467535   & 6460670\\
    \bottomrule
    \end{tabularx}
\end{table*}
As described in \autoref{sec:packgen} different packet generation approaches have different advantages and precision levels.
In \autoref{fig:cycle_time_comparison} we extend the measurements from~\cite{9844050}.
The results in \autoref{fig:cycle_time_comparison} clearly indicate that all transmission techniques that use the Linux network stack (socket based and \texttt{TA\_PRIO}) are less precise than sending packets via \gls{dpdk} based generators e.g., \texttt{MoonGen}.
The substantial difference between the two \texttt{tcpreplay} versions is caused by a bug in the lower version, which results in a constantly too short cycle, due to the implementation this error increases with the runtime.

\subsection{Cut-through Switching: Effects of Ethernet Frame Size}
\label{sec:evaluation-cut-through-test}
\begin{figure}[ht]
    \centering
    \includegraphics{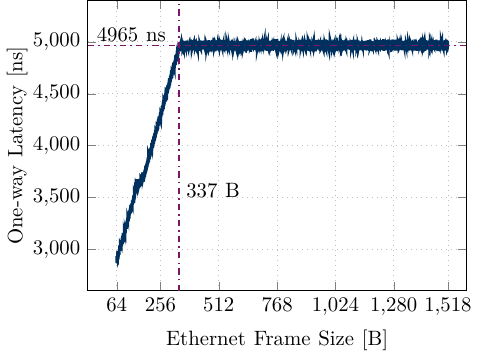}
    \caption{
    Packet latency of cut-through switch DuT as a function of packets size. 
    The switch forwards packets smaller than \SI{337}{\byte} when they have been completely received; for larger packets, only \SI{337}{\byte} are accumulated before forwarding.
    }
    \label{fig:cutthrough}
\end{figure}
\autoref{fig:cutthrough} shows the \gls{dut} one-way packet delay for cut-through switching.
We generated a random homogeneous data-set with packets of sizes from \SI{64}{\byte} to \SI{1522}{\byte} and sent them through the \gls{dut}. With the testbed, we measured the packet residence time in the \gls{dut} and plot the measured delay as a function of the packet size.
Packets of size \SI{64}{\byte} to \SI{337}{\byte} experience a linearly increasing delay. This is because the FibroLAN switch completely receives and stores these small packets  before forwarding, incurring the equivalent of the transmission delay (which the packet experienced as the sending node transmitted the packet bits into the physical wire) as the packet bits are accumulated at the switch, 
For larger packets, the considered FibroLAN switch accumulates \SI{337}{\byte} of data before starting to forward the packet; thus, incurring a constant equivalent transmission delay component for accumulating \SI{337}{\byte} of data.
The cut-through experiment should be conducted for every new \gls{dut} which is considered in the testbed, because the forwarding characteristics for different packet sizes depend on the \gls{dut}'s internal architecture and will affect all delay measurements~\cite{liss2017architecture}.

\subsection{Comparisons with Time-Sensitive Network Simulations}
\label{sec:evaluation-simulation}
The validation of a physical testbed with simulated data seems to be unconventionally, because hardware measurements are true by definition; however, this is only valid if the hardware is configured in the right way.
Simulations have often less complex settings and the setup is less complicated than configuring hardware measurements and retrieving hardware measurement results.
So, achieving similar results with hardware and software is a good indicator for data reliability.

\begin{figure*}
    \begin{subfigure}[t]{0.3\textwidth}
        \centering
        \includegraphics{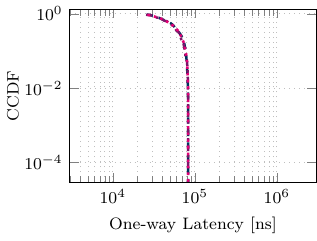}
        \caption{\gls{gcl} 1 w/o guard band.}
        \label{fig:simulation-tas1}
    \end{subfigure}
    \hfill
    \begin{subfigure}[t]{0.3\textwidth}
        \centering
        \includegraphics{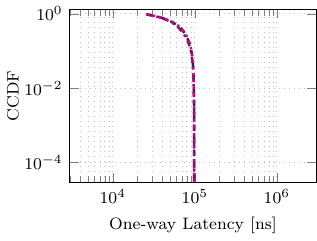}
        \caption{\gls{gcl} 1 w/ guard band.}
        \label{fig:simulation-tas1-gb}
    \end{subfigure}
    \hfill
    \begin{subfigure}[t]{0.3\textwidth}
        \centering
        \includegraphics{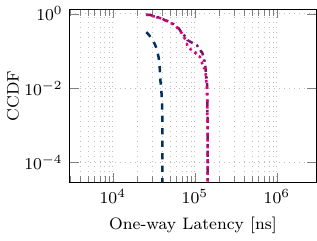}
        \caption{\gls{gcl} 2 w/o guard band.}
        \label{fig:simulation-tas2}
    \end{subfigure}
    
    \begin{subfigure}[t]{0.3\textwidth}
        \centering
        \includegraphics{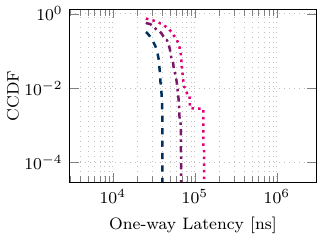}
        \caption{\gls{gcl} 3 w/o guard band.}
        \label{fig:simulation-tas3}
    \end{subfigure}
    \hfill
    \begin{subfigure}[t]{0.3\textwidth}
        \centering
        \includegraphics{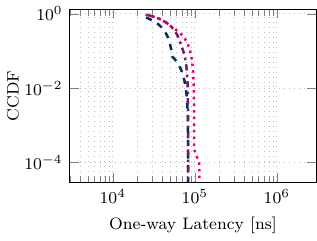}
        \caption{\gls{gcl} 4 w/o guard band.}
        \label{fig:simulation-tas4}
    \end{subfigure}
    \hfill
    \begin{subfigure}[t]{0.3\textwidth}
        \centering
        \includegraphics{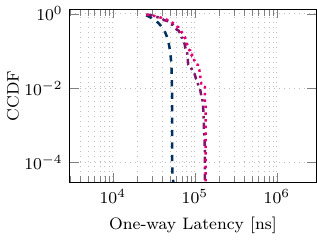}
        \caption{\gls{gcl} 4 w/ guard band.}
        \label{fig:simulation-tas4-gb}
    \end{subfigure}
    \caption{Simulation results for generic \acrlong{stream1}. The three colors correspond to three streams, where $\coloronename$ is stream \acrshort{stream1h}, $\colortwoname$ is \acrshort{stream1m}, and $\colorthreename$ is \acrshort{stream1l}, respectively. Moreover, the configuration of \gls{gcl} 1 to 4 corresponds to the \gls{gcl}s in \autoref{tab:gcl-configurations} from left to right. In addition, w/o means without, and w/ represents with. These conventions apply analogously to the subsequent figures.}
    \label{fig:simulation-results}
\end{figure*}

To validate our testbed, we implement the same network architecture as in \autoref{fig:tsn-testbed-architecture} based on \mbox{OMNet++ 6.0} with the INET framework \mbox{version 4.4}. 
Four different \gls{gcl} configurations are measured to compare with the behavior of our real testbed.

\subsubsection{Simulation Setup}
\begin{table*}[t]
\setlength{\tabcolsep}{1pt}
    \centering
    \caption{The four figures show the four considered \acrfull{gcl} configurations (1, 2, 3, and 4, from left to right) with optional guard bands, i.e., the measurements are conducted with and without guard bands. Each \gls{gcl} configuration consists of time slots (\textbf{Time}-axis) and different streams with associated priorities (\textbf{Priority}-axis). Each time slot is equi-sized since all frames in the \acrlong{stream1} have the same frame size of \SI{1522}{\byte}. However, the slot size itself can vary in length; i.e., in measurement \textit{A}, a slot is \SI{15}{\micro\second} long; in measurement \textit{B}, it is $3\times\SI{15}{\micro\second}=\SI{45}{\micro\second}$. This allows to detect impacts of too short or too long slot sizes and to draw conclusions for possible timing issues. The time where each stream is permitted to send is indicated by a dark colored box. Configurations~2 and~3 do not need a dedicated time slot for a guard band, since the to-be-protected \textit{high} priority stream is always allowed to send data. 
    }
    \label{tab:gcl-configurations}
    {\renewcommand{\arraystretch}{1.5}%
    \begin{tabular}{c|c|c|c|c|c|}
        \toprule
        \backslashbox[1.5cm]{\textbf{Priority}}{\textbf{Time}}& $t_1$     & $t_2$     & $t_3$     & $t_4$     & $t_5$ \\
        \midrule
        high        & \gclblk  &           &           &           & \gclblkGB \\
        \hline
        medium      &           & \gclblk  &           &           & \gclblkGB \\
        \hline
        low         &           &           & \gclblk  &           & \gclblkGB \\
        \hline
        cross traffic    &           &           &           & \gclblk  & \gclblkGB \\
        \bottomrule
    \end{tabular}%
    \hfill
    \begin{tabular}{c|c|c|c|c|}
        \toprule
        \backslashbox[1.5cm]{\textbf{Priority}}{\textbf{Time}}  & $t_1$     & $t_2$     & $t_3$     & $t_4$ \\
        \midrule
        high        & \gclblk   & \gclblk   & \gclblk   & \gclblk \\
        \hline
        medium      &           & \gclblk   &           & \\
        \hline
        low         &           &           & \gclblk   & \\
        \hline
        cross traffic    &           &           &           & \gclblk \\
        \bottomrule
    \end{tabular}%
    \hfill
    \begin{tabular}{c|c|c|c|c|}
        \toprule
        \backslashbox[1.5cm]{\textbf{Priority}}{\textbf{Time}}  & $t_1$     & $t_2$     & $t_3$     & $t_4$ \\
        \midrule
        high        & \gclblk   & \gclblk   & \gclblk   & \gclblk \\
        \hline
        medium      &           & \gclblk   & \gclblk   & \gclblk \\
        \hline
        low         &           &           & \gclblk   & \gclblk \\
        \hline
        cross traffic    &           &           &           & \gclblk \\
        \bottomrule
    \end{tabular}%
    \hfill
    \begin{tabular}{c|c|c|c|c|c|c|c|c|}
        \toprule
        \backslashbox[1.5cm]{\textbf{Priority}}{\textbf{Time}}  & $t_1$     & $t_2$     & $t_3$     & $t_4$     & $t_5$     & $t_6$     & $t_7$ & $t_8$\\
        \midrule
        high        & \gclblk   &           & \gclblk   &           & \gclblk   &           & \gclblk   & \gclblkGB \\
        \hline
        medium      &           & \gclblk   &           & \gclblk   &           & \gclblk   &           & \gclblkGB \\
        \hline
        low         &           &           & \gclblk   &           & \gclblk   &           &           & \gclblkGB \\
        \hline
        cross traffic    &           &           &           & \gclblk   &           &           &           & \gclblkGB \\
        \bottomrule
    \end{tabular}}%
\end{table*}
\label{sec:explain-gcl-configurations}
The \gls{gcl} configurations are shown in \autoref{tab:gcl-configurations}, where the \SI{15}{\micro\second} slot size is used, including the Guard Band.
The first configuration is a simple round-robin. 
The second configuration always opens the gate to the highest priority traffic, and other traffic holds the transmission time equally. 
In the third configuration, the \gls{gcl} gate duration time corresponds to the priority, whereby the higher the priority, the longer the gate remains open. 
The fourth configuration modifies the second configuration by compensating for high-priority traffic and avoiding the opening of more than two queues simultaneously. 

The simulation settings are aligned with the generic generated streams (\acrlong{stream1}) to verify our testbed. 
Since our main focus is on the real testbed measurements, only the generic stream set is simulated.
To simulate the cross traffic, packets with uniformly distributed sizes [\SI{64}{\byte} to \SI{1522}{\byte}] are generated according to a Poisson process. 
The simulated \gls{pcp} value of stream \acrshort{stream1l} is different from the value of the real testbed (4) since, in the INET framework \mbox{version 4.4}, \gls{pcp} value with five and four are mapped to the same queue.
Therefore, we set the simulated \gls{pcp} value three to fulfill the one-to-one mapping between streams and queues so that the \gls{tas} can control each stream.
The cut-through switching is not enabled in the simulation due to a bug in the current INET version.

We summarize the clock and synchronization settings as follows.
A \gls{gm} clock is configured to align with the simulation time, and other clocks have a constant drifting rate uniformly selected from $-100$\,\gls{ppm} to 100\,\gls{ppm} in the initial phase of the simulation.
A clock with one \gls{ppm} constant drift means it would be one second faster if one million seconds passed by. 
To synchronize the devices with the \gls{gm}, the INET framework \mbox{version 4.4} provides the \gls{gptp} tools according to IEEE 802.1AS.
The switch is configured to transmit the synchronization and peer-delay measurement messages every \SI{0.5}{\milli\second}.

The \gls{cdf} characterizes the distribution of discrete random variables, allowing one to derive the probability of a random variable smaller than a specific value.
However, we are mainly interested in how frequently the latency exceeds a prescribed level; therefore, the simulation results are presented as \gls{ccdf} curves.

Each generic stream measurement scenario was run for 30 minutes (covering $3.6 \cdot 10^{7}$ cycles of duration 50\,$\mu$s).
For selected simulated data and spot robot scenarios, we conducted pilot measurements over one minute (covering $1.2 \cdot 10^{6}$ cycles of duration 50\,$\mu$s) and over ten minutes (covering $1.2 \cdot 10^{7}$ cycles of duration 50\,$\mu$s). After confirming that the one minute and 10 minute measurement runs gave equivalent measurement results, we ran each simulated data and spot robot measurement scenario for one minute.
The corresponding validating simulations were run for one minute ($1.2 \cdot 10^{6}$ cycles of duration 50\,$\mu$s) and 5 seconds ($1 \cdot 10^{5}$ cycles of duration 50\,$\mu$s), which gave equivalent result; then the simulations were run for 5 seconds.

\subsubsection{Results}
In \autoref{fig:simulation-results}, the y-axis represents the complementary cumulative probability, and the x-axis represents the one-way packet latency of the streams in the \gls{dut}.
An ideal switch without latency variations corresponds to a vertical line, see \autoref{fig:baseline-generic}. 
If the switch applies an increasing constant delay to all packets, then the vertical line will move to the right.
However, in most cases, the delay of each packet changes, and small delays have a higher probability (i.e., are more likely to occur).
These packet latency variations bend the vertical line into a curved shape, see e.g.,~ \autoref{fig:simulation-tas1}.
The curve may not be smooth when packet delays are not continuously distributed.
For example, a step shape arises in \autoref{fig:simulation-tas3}.
because packet bursts are transmitted within consecutive time slots.
Generally, the bottom part of the \gls{ccdf} curve signifies the maximum packet latency, whereas the top denotes the minimum packet latency.

In \autoref{fig:simulation-tas1}, the packet delay curves completely overlap because each stream has the same transmission time slot.
However, in \autoref{fig:Generic-CT-GCL1-1}, a tail of \acrshort{stream1h} appears.
The cut-through switching may cause this tail phenomenon in our testbed since \acrshort{stream1h} streams should wait until the lower priority stream finishes transmission even when the gate is open.
This tail issue can be addressed by increasing the time slot size.
Therefore, the simulation result of \autoref{fig:simulation-tas1} is similar to the testbed measurement result presented in \autoref{fig:Generic-CT-GCL1-3}.
Moreover, the difference between \autoref{fig:simulation-tas1} and \autoref{fig:simulation-tas1-gb} is the guard band implementation, where \autoref{fig:simulation-tas1-gb} shifts slightly to the right compared to \autoref{fig:simulation-tas1}.
Once the guard band is introduced, all the streams must wait a certain time to obtain an open gate again, which increases the latency.
The same behavior occurred in the testbed measurements depicted in \autoref{fig:Generic-CT-GCL1-3} and \autoref{fig:Generic-CT-GCL1-3-GB}.

We turn now to \autoref{fig:simulation-tas2} and \autoref{fig:simulation-tas3}, which were obtained for the same configuration as the testbed measurement results in \autoref{fig:Generic-CT-GCL2-1} and \autoref{fig:Generic-CT-GCL3-1}.
The testbed results generally follow the simulation results, whereby \acrshort{stream1m} and \acrshort{stream1l} overlap for \gls{gcl} 2, but separate for the \gls{gcl} 3 configuration.
The behavior of the step shape in both \autoref{fig:Generic-CT-GCL2-1} and \autoref{fig:Generic-CT-GCL3-1} will be explained in the following section and can be smoothed by enlarging the slot size.
Therefore, the simulation results of \autoref{fig:simulation-tas2} and \autoref{fig:simulation-tas3} are more similar to \autoref{fig:Generic-CT-GCL2-3} and \autoref{fig:Generic-CT-GCL3-3}.

Next, we consider \autoref{fig:simulation-tas4} and \autoref{fig:simulation-tas4-gb}.
Compared to the testbed results, the ranges between streams are close to each other. 
This is because the cross traffic frame size varies in the simulation, whereas the full Ethernet frame size is utilized in the testbed all the time.
A large frame size leads to a blocking effect for other streams and further increases their latency.
However, the sequence or the priority of streams still holds, which means that the order of the latency curves from left to right follows the order of \acrshort{stream1h},  \acrshort{stream1m}, and \acrshort{stream1l}. 
The order is presented at the top of \autoref{fig:simulation-tas4} and is readily observed in \autoref{fig:Generic-CT-GCL4-1}, which also validates the correct operation of our testbed.

Applying a guard band has a significant influence to \gls{gcl} 4, as observed by comparing \autoref{fig:simulation-tas4} and \autoref{fig:simulation-tas4-gb}.
This is because the guard band protects the highest priority stream so that \acrlong{stream1} can transmit at the beginning of \gls{gcl} 4 without interference from other streams.
The same result can be observed from the testbed measurements in the comparison of \autoref{fig:Generic-CT-GCL4-1} and \autoref{fig:Generic-CT-GCL4-1-GB}, where the tail effect is mitigated at the bottom side of
 \autoref{fig:Generic-CT-GCL4-1-GB}.

We conclude this section by noting that our testbed works properly compared to the simulation results. 
However, we only emphasized the similarity and differences between the simulation and the testbed in this section.
The comparison of the four \gls{gcl} configurations and their shortage, how to improve the latency, and the impact of different types of input streams would be covered in the following sections.

\section{TSN Evaluation with TSN-FlexTest Testbed}
\label{sec:testbed-evaluation}
In this section, the TSN-FlexTest testbed is used to evaluate \acrfull{tsn}.
We examine how the \acrfull{pdv} can be reduced by leveraging multiple \gls{tsn} mechanisms.
We investigate the stream characteristics with respect to frame size, packet frequency, burstiness and the impact of different scheduler settings, e.g., \gls{tas} slot size and guard band.
Also, we compare \gls{tsn} performance for the generic data streams with recorded real-world data sets.

In order to provide \gls{qos} \textit{guarantees} for network streams and to reduce the \gls{pdv} with \acrlong{tsn}, there are multiple possible options:
First, the architecture can be modified, e.g., a dedicated connection can be established.
With dedicated connections, the delay increases caused by other network participants (streams) can almost completely be avoided, which provides a \textit{baseline} benchmark for the following improvements.
Second, if additional dedicated connections are either not applicable, e.g., due to physical circumstances, or economically not viable, then existing technologies for data prioritization can be used.
For instance, the Ethernet \gls{vlan} tag provides the \gls{pcp} field with \SI{3}{\bit} to distinguish 8~network packet priorities.
In conjunction with \gls{spq}, it is possible to prioritize packets (see \autoref{sec:background:flow-control}).
We subsequently refer to the \gls{pcp} approach as \textit{soft} \gls{qos}.
Finally, there is a third choice for enhanced network control:
With a \gls{tas}, it is possible to achieve deterministic transmission behavior, which we refer to as \textit{hard} \gls{qos} in the following.

We have conducted extensive measurements in the TSN-FlexTest testbed to elucidate the advantages and disadvantages of the different technologies.
Furthermore, we examine for the first time the \acrlong{tas} with synchronized senders using \gls{cots} hardware.
While \gls{tas} with synchronized senders has been
employed with industry-grade equipment, to the best of our knowledge, we are the first to conduct detailed measurements for \gls{tas} with synchronized senders in a testbed built from basic \gls{cots} hardware.
Our measurements demonstrate that \gls{tas} with synchronized senders improves the performance of conventional "switch-only" \gls{tas} to provide near-optimal performance on generic commodity (basic \gls{cots}) hardware. Our measurements indicate that \gls{tas} with synchronized senders on generic commodity hardware is comparable to the baseline while still allowing for the sharing of the network links by multiple streams.

The following measurements primarily use \acrlong{stream1}, allowing better insights into the transmission and forwarding characteristics, since effects within the stream set itself are negligible.
In the discussion, we only involve \acrlong{stream2} and \acrlong{stream3} for specific peculiarities and refer the interested reader for the corresponding figures to the Appendix.

\subsection{Baseline}
\label{sec:evaluation-baseline}
\begin{figure*}
    \begin{subfigure}[t]{0.3\textwidth}
        \centering
        \includegraphics{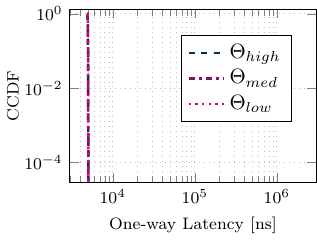}
        \caption{Generic,\acrlong{stream1}.}
        \label{fig:baseline-generic}
    \end{subfigure}
    \hfill
    \begin{subfigure}[t]{0.3\textwidth}
        \centering
        \includegraphics{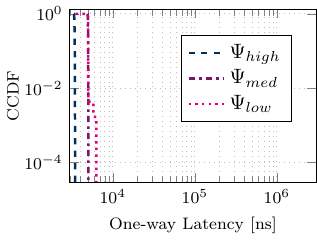}
        \caption{Real data, \acrlong{stream2}.}
        \label{fig:baseline-real}
    \end{subfigure}
    \hfill
    \begin{subfigure}[t]{0.3\textwidth}
        \centering
        \includegraphics{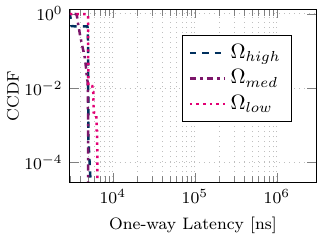}
        \caption{Spot data, \acrlong{stream3}.}
        \label{fig:baseline-spot}
    \end{subfigure}
    \caption{Baseline measurement of three stream sets for testbed validation.
    The figures show from left to right the independent replay of \acrlong{stream1}, \acrlong{stream2}, and \acrlong{stream3}, each in a combined plot. Specifically, for each of the three stream sets, each of the three streams in the set was replayed individually by a single transmission node (without any transmissions from the other nodes). The figures thus show the absolute minimum delay for traversing the \gls{dut}, whereby this minimum \gls{dut} traversal delay is mainly governed by the queuing delay, while switch processing delay and transmission delay are small and nearly constant for the constant frame size. Collisions (resource contention) can arise due to the in-band PTP synchronization, whereby PTP frames are sent with the highest priority (7, which is higher than the highest test stream PCP of 6), and thus the PTP frames can delay the data frame transmissions.
    }
    \label{fig:baseline}
\end{figure*}
The baseline measurement considers only \textit{one} sender (actually, three senders, but only one is sending in a given measurement run) and one receiver with the \gls{dut} connecting both.
The baseline scenario replays all streams, defined in \autoref{tab:streams}, one by one, without any cross traffic.
The packet latencies for the three streams within each stream set are combined together in one plot, although in fact three separate measurements have been performed.
According to the stream priorities, it is expected that the colored lines in the \gls{ccdf} always follow the order: $\coloronename < \colortwoname < \colorthreename$.
An additional objective is that the \coloronename curve should show a minimal deterministic latency, since the \gls{gcl} configurations are optimized towards the highest priority.
This objective is valid for all following \glspl{ccdf}.

\autoref{fig:baseline} shows the measured packet latency, i.e., packet residence time inside the \gls{dut}.
An identical residence time for all packets in \acrlong{stream1} can be observed in \autoref{fig:baseline-generic}, indicated by a vertical line.
With a standard deviation of \SI{43.5}{\nano\second}, \SI{41.5}{\nano\second}, and \SI{38.9}{\nano\second} for stream~\acrshort{stream1h}, \acrshort{stream1m}, and \acrshort{stream1l} respectively, the one-way delay distribution is very narrow (the values are in the range of the \gls{phy} tolerances), an almost deterministic behavior.
It is also expected that all three streams exhibit the same distribution since only the packet rate varies.
In contrast, \autoref{fig:baseline-real} and \autoref{fig:baseline-spot} exhibit varying gradients of the packet latencies.
This can be explained by the frame size:
Whereas the streams in \acrlong{stream1} have a fixed frame size of \SI{1522}{\byte}, and the streams \acrshort{stream2h} and \acrshort{stream2m} have a constant frame sizes of \SI{128}{\byte} and \SI{526}{\byte}, respectively (see \autoref{tab:streams}), all other streams have variable frame sizes.
Therefore, the \glspl{ccdf} of \acrlong{stream2} and \acrlong{stream3} exhibit the actual distribution of the frame sizes and the corresponding transmission delays.
For instance, a few frames of stream \acrshort{stream2l} are smaller than \SI{337}{\byte} (see \autoref{sec:evaluation-cut-through-test}) and therefore are forwarded faster.
Additionally, the video stream has packet bursts, i.e., a lot of frames are transmitted in a short period of time, which causes a filled queue.
Though the \gls{dut} could handle the stream, the in-band time-synchronization causes a slightly longer delay for a fraction of the frames.
A shift between stream~\acrshort{stream2m} and~\acrshort{stream2l} by \SI{1286}{\nano\second} can be seen above the 99 percentile -- this equals to an Ethernet frame of about \SI{140}{\byte}.
The same applies to the other streams with a variable frame size.

\subsection{No Prioritization (Internet Scenario)}
\label{sec:evaluation-no-prioritization}
\begin{figure*}
    \begin{minipage}[t]{0.49\textwidth}
        \begin{subfigure}[t]{0.49\linewidth}
            \centering
            \includegraphics{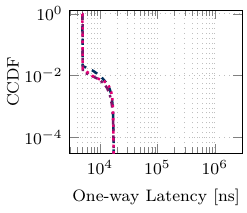}
            \caption{without cross traffic}
            \label{fig:Generic}
        \end{subfigure}%
        \begin{subfigure}[t]{0.49\linewidth}
            \centering
            \includegraphics{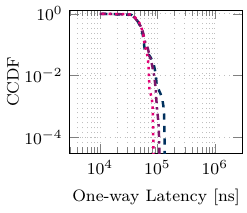}
            \caption{with cross traffic}
            \label{fig:Generic-CT}
        \end{subfigure}
        \caption{Measurement of \acrlong{stream1} with disabled prioritization settings at the \gls{dut} (all streams are treated equally).
        The streams are sent at the same time, so that packet collisions at the end of the \gls{dut}'s queue cause additional delay which is increased in case of added cross traffic. Because of the higher frequency of \acrlong{stream1h}, this stream is affected more often by collisions than others and has an higher delay than \acrlong{stream1l}.}
        \label{fig:evaluation-no-qos}
    \end{minipage}
    \hfill
    \begin{minipage}[t]{0.49\textwidth}
        \begin{subfigure}[t]{0.49\linewidth}
            \centering
            \includegraphics{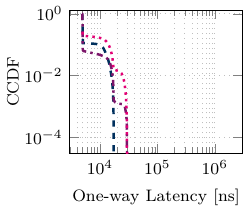}
            \caption{without cross traffic}
            \label{fig:Generic-SPQ}
        \end{subfigure}%
        \begin{subfigure}[t]{0.49\linewidth}
            \centering
            \includegraphics{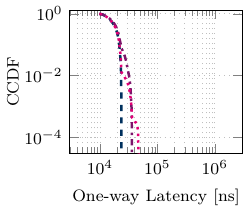}
            \caption{with cross traffic}
            \label{fig:Generic-CT-SPQ}
        \end{subfigure}
        \caption{Measurement of \acrlong{stream1} with \acrfull{spq} at \gls{dut}.
                 Similarly to \autoref{fig:evaluation-no-qos}, packet collisions cause additional queuing delay.
                 With \gls{spq}, the \gls{dut} scheduler preserves the stream priorities defined in \autoref{tab:streams}, i.e., the queue with the highest priority is emptied first, then the second highest priority queue, and so on.
                 Added cross traffic increases the queuing delay, but generally lowers the overall one-way latency for high-priority streams compared to \autoref{fig:Generic-CT}.
                 }
        \label{fig:evaluation-soft-qos}
    \end{minipage}%
\end{figure*}
After establishing a baseline, we investigated the one-way delay in a typical \textit{internet} scenario, i.e., the data streams are transmitted simultaneously with the \textit{same} priority.
In this case, the streams compete equally for resources in the \gls{dut}.
\autoref{fig:evaluation-no-qos} shows two measurements: \autoref{fig:Generic} illustrates the case where only the three streams of \acrlong{stream1} are transmitted simultaneously, and \autoref{fig:Generic-CT} shows the more typical case with added interference caused by cross traffic.
In all following measurements, the cross traffic consists of full Ethernet frames, with a throughput of \SI{2}{\giga\bit\per\second} (200\% loaded link), and a variable inter-packet delay based on a Poisson distribution.
In this internet scenario, the packets of the three simultaneously transmitted streams will have collisions and are affected by queuing latency at the egress port towards node0 at the \gls{dut}.
This scenario reveals the scheduling behavior of the \gls{dut} if packets from several streams need to be transferred through the same output port (our synthetically created bottleneck).
Thereby, we can observe the queuing mechanisms of the \gls{dut} in an oversaturation scenario that forces the switch to drop packets.
The \gls{dut} is configured to ignore the \gls{vlan} \gls{pcp}.
We observe from \autoref{fig:evaluation-no-qos} that the one-way delay is not aligned with the priority set in the packet headers.
In \autoref{fig:Generic}, we can identify that for 97.96\% of all sent packets, the one-way latency is identical to the baseline measurement (see \autoref{fig:baseline-generic}).
Above, we can notice that there is a latency increase, independent of the stream priority.
We observe a clear step caused by statistical collisions of frames, i.e., arriving frames compete for resources.
These \textit{collisions (resource contentions)} need to be considered for applications with hard real-time requirements.
Statistical collision of frames results in steps in the \gls{ccdf}, because sometimes a frames is forwarded immediately, whereas sometimes it has to wait.
This \textit{blocking} issue can be solved by frame preemption, see \autoref{sec:preemption-description}.
Moreover, we observe this staircase formation in all following figures where collisions cannot be prevented.
Additionally, cross traffic affects all streams equally and increases the packet delay in general (see \autoref{fig:Generic-CT}).
Now, only a very small part of the sent packets are immediately forwarded by the \gls{dut} and almost all other packets are delayed by up to \SI{135.4}{\micro\second}, \SI{109.6}{\micro\second}, and \SI{85.3}{\micro\second} for stream~\acrshort{stream1h}, \acrshort{stream1m}, and~\acrshort{stream1l} respectively.

The frame rate, i.e., the frequency of frame transmissions [frames/second], of a given stream influences the statistical incidence of collisions.
If no special scheduling is applied, then frames of streams with a high frame rate have to wait more often.
This is the reason why the stream latencies in \autoref{fig:Generic-CT} are ordered in the opposite direction as their priorities.

This is the typical configuration in a \textit{net-neutral} internet, without specialized packet treatment and with limited resources in the backbone networks.
This is especially true due to \textit{rush hours} in the internet, e.g., in the evening or at large events (black Friday, popular broadcasting events).
For \gls{ti} scenarios~\cite{fit2021tac,pro2020com,xia2019red}, it is nearly impossible to ensure a prescribed service quality in such a net-neutral internet.

\subsection{Soft Quality-of-Service with Strict Priority Queuing}
\label{sec:evaluation-soft-qos}
In a first step towards providing QoS, we examine the improvement of the service quality achieved by applying queuing disciplines that take the stream priorities into consideration.
One common approach is \acrfull{spq}, defined by IEEE~802.1Q, where always the queue that holds the highest priority packets is emptied first.
\autoref{fig:evaluation-soft-qos} shows this \gls{spq} approach with \acrlong{stream1} for measurements without and with added cross traffic.
We observe from \autoref{fig:Generic-SPQ}, that the distribution of the one-way latency has the lowest latency values for the highest stream priority, the second lowest latency for the second highest stream priority, and the highest latency values for the lowest stream priority.
Thus, we conclude that \gls{spq} achieves a certain priority isolation between the three streams.
However, \autoref{fig:Generic-CT-SPQ} shows that the overall one-way latency still increases from \SI{12.0}{\micro\second}, \SI{12.4}{\micro\second}, and \SI{14.2}{\micro\second} to \SI{16.7}{\micro\second}, \SI{20.8}{\micro\second}, and \SI{17.8}{\micro\second} for stream~\acrshort{stream1h}, \acrshort{stream1m}, and~\acrshort{stream1l} respectively, when cross traffic is added.
This is an increase by 44\% on average for all three streams.
Therefore, we characterize the \gls{spq} approach as \textit{soft} \gls{qos}, as it is not possible to enforce strict guarantees.
This can be explained by the fact that lower priority frames can be scheduled and selected for transmission when a higher priority frame arrives at the \gls{dut}.
In this case, the lower priority frame is not interrupted and the higher priority frame must wait until the lower priority frame is transmitted, which for full regular Ethernet frames with \SI{1522}{\byte} takes around \SI{12.3}{\micro\second} in the worst case.
This effect can be observed in \autoref{fig:Generic-SPQ}: without cross traffic, the highest priority stream~\acrshort{stream1h} is delayed from \SI{4894}{\nano\second} (minimum value) to at most \SI{18540}{\nano\second}, a growth by about one full Ethernet frame transmission.

\subsection{Hard Quality-of-Service with Time-aware Traffic Shaping}
\label{sec:evaluation-hard-qos}
\begin{figure*}
    \begin{subfigure}[b]{0.245\textwidth}
    \centering
    \includegraphics{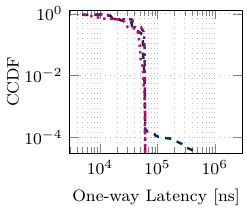}
    \caption{\gls{gcl} 1, Slot Size 1}
    \label{fig:Generic-CT-GCL1-1}
    \end{subfigure}
    \hfill
    \begin{subfigure}[b]{0.245\textwidth}
        \centering
        \includegraphics{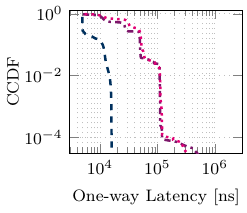}
        \caption{\gls{gcl} 2, Slot Size 1}
        \label{fig:Generic-CT-GCL2-1}
    \end{subfigure}
    \hfill
    \begin{subfigure}[b]{0.245\textwidth}
        \centering
        \includegraphics{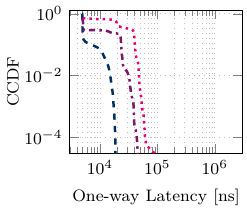}
        \caption{\gls{gcl} 3, Slot Size 1}
        \label{fig:Generic-CT-GCL3-1}
    \end{subfigure}
    \hfill
    \begin{subfigure}[b]{0.245\textwidth}
        \centering
        \includegraphics{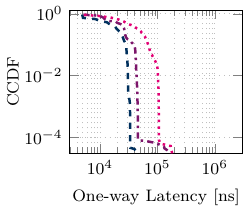}
        \caption{\gls{gcl} 4, Slot Size 1}
        \label{fig:Generic-CT-GCL4-1}
    \end{subfigure}
    
    \begin{subfigure}[b]{0.245\textwidth}
        \centering
        \includegraphics{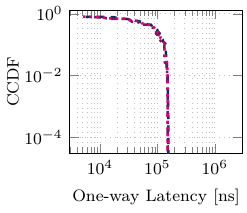}
        \caption{\gls{gcl} 1, Slot Size 3}
        \label{fig:Generic-CT-GCL1-3}
    \end{subfigure}
    \hfill
    \begin{subfigure}[b]{0.245\textwidth}
        \centering
        \includegraphics{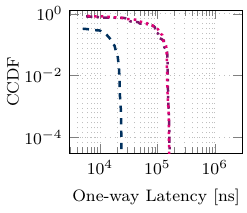}
        \caption{\gls{gcl} 2, Slot Size 3}
        \label{fig:Generic-CT-GCL2-3}
    \end{subfigure}
    \hfill
    \begin{subfigure}[b]{0.245\textwidth}
        \centering
        \includegraphics{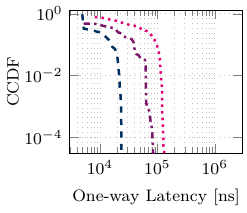}
        \caption{\gls{gcl} 3, Slot Size 3}
        \label{fig:Generic-CT-GCL3-3}
    \end{subfigure}
    \hfill
    \begin{subfigure}[b]{0.245\textwidth}
        \centering
        \includegraphics{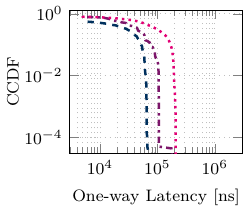}
        \caption{\gls{gcl} 4, Slot Size 3}
        \label{fig:Generic-CT-GCL4-3}
    \end{subfigure}
    
    \begin{subfigure}[b]{0.245\textwidth}
        \centering
        \includegraphics{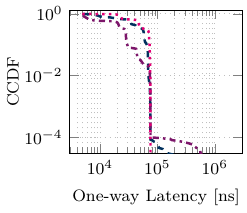}
        \caption{\gls{gcl} 1, w/ GB, Slot Size 1}
        \label{fig:Generic-CT-GCL1-1-GB}
    \end{subfigure}
    \hfill
    \begin{subfigure}[b]{0.245\textwidth}
        \centering
        \includegraphics{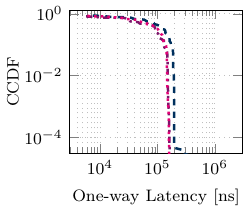}
        \caption{\gls{gcl} 1, w/ GB, Slot Size 3}
        \label{fig:Generic-CT-GCL1-3-GB}
    \end{subfigure}
    \hfill
    \begin{subfigure}[b]{0.245\textwidth}
        \centering
        \includegraphics{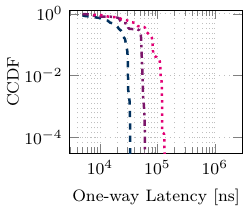}
        \caption{\gls{gcl} 4, w/ GB, Slot Size 1}
        \label{fig:Generic-CT-GCL4-1-GB}
    \end{subfigure}
    \hfill
    \begin{subfigure}[b]{0.245\textwidth}
        \centering
        \includegraphics{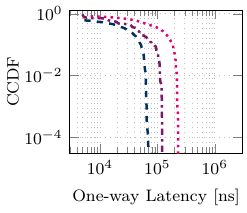}
        \caption{\gls{gcl} 4, w/ GB, Slot Size 3}
        \label{fig:Generic-CT-GCL4-3-GB}
    \end{subfigure}
    \caption{Measurement of \acrlong{stream1} with enabled \acrfull{tas} at the \gls{dut}.
             We tested with multiple generic \acrfull{gcl} configurations (see \autoref{tab:gcl-configurations}) and slot sizes in order to examine the \gls{tas} effects.
             The slot size refers to the time unit a gate of the corresponding queue is opened---a slot size of~1 equals \SI{15}{\micro\second} (see row~1), and~3 corresponds to \SI{45}{\micro\second} (see row~2).
             We conducted measurements without (w/o) and with (w/) enabled Guard Band (GB) for protecting the highest priority (see the last row).}
    \label{fig:generic-hard-qos}
\end{figure*}
\begin{figure}[bt]
    \centering
    \includegraphics{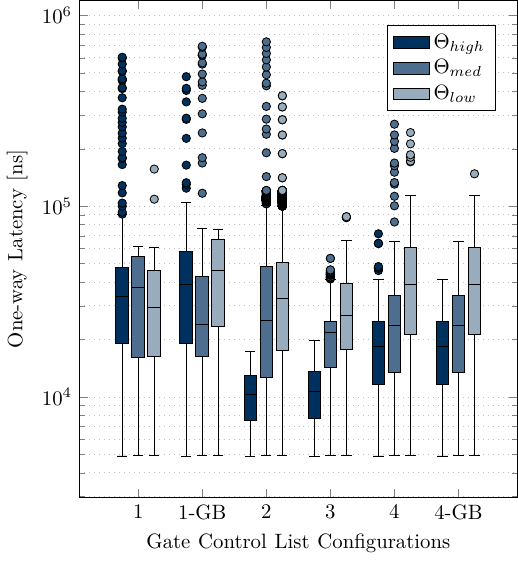}
    \caption{Boxplot of one-way delay measurements for different \gls{gcl} configurations with a window size of~1 for \acrlong{stream1}.
             The x-axis labeling refers to the six \gls{gcl} configurations in \autoref{tab:gcl-configurations}.
             This boxplot allows a direct comparison between all combinations with insights into particular statistical values, such as outliers, which are of high importance for \gls{tsn}.
             }
    \label{fig:gcl-comparison-generic-stream}
\end{figure}
\subsubsection{Overview}
In a second step towards providing QoS, we measured the one-way packet latency for \gls{tas}.
\autoref{fig:generic-hard-qos} shows measurements for the \gls{gcl} configurations in \autoref{sec:explain-gcl-configurations} and \autoref{tab:gcl-configurations}.
We strive to gain insights for the different \gls{tas} parameters settings.
Therefore, we systematically created four different \gls{gcl} configurations and, where appropriate, we added a guard band to protect the highest priority stream.
We considered \acrlong{stream1}, and we added cross traffic in \textit{all} twelve measurements.
Furthermore, we varied the slot size (the time a gate of the respective queue is open): a value of~1 corresponds to our unit of \SI{15}{\micro\second} which in theory should allow transmission of one full Ethernet frame.
Since we estimated that frames are not sent out at the sender with the highest precision, we wanted to investigate further the effect of increased slot size.
Therefore, we also considered \SI{45}{\micro\second} which corresponds to a slot size of~3.

\subsubsection{GCL Configuration 1}
We observe in Figures~\ref{fig:Generic-CT-GCL1-1}, \ref{fig:Generic-CT-GCL2-1}, \ref{fig:Generic-CT-GCL3-1}, and \ref{fig:Generic-CT-GCL4-1} the fundamental effects of the \gls{gcl} settings:
\autoref{fig:Generic-CT-GCL1-1} has a round-robin-like configuration in which every stream priority gets the same amount of time.
The goal of this configuration is to allocate equal bandwidth to all streams, without accounting for the different stream characteristics.  
However, we observe that the blue dashed curve has an increased tail-latency. This is due the higher number of sent frames than in the other two streams (see \autoref{tab:streams}).
If the \gls{gcl} configuration is composed of only short slots for each single stream, then streams with higher frame frequency are more affected by blocked time slots than other streams---a frame from the previous cycle which was buffered at the switch now uses the next cycle.
Therefore the probability of outliers increases.
The measurements with \acrlong{stream2} and \acrlong{stream3} in \autoref{fig:real-CT-GCL1-1} and \autoref{fig:real-CT-GCL1-3} (see Appendix) clearly demonstrate that this \gls{gcl} configuration imposes less delay to sporadic cyclic streams, and higher delays to bursty streams with higher data rates.
Also, the comparison of \autoref{fig:real-CT-GCL1-1} and \autoref{fig:real-CT-GCL1-3} demonstrates the negative effect of the short slots.

\subsubsection{GCL Configuration 2}
Configuration 2, which is considered \autoref{fig:Generic-CT-GCL2-1}, allows the highest priority to always transmit data (the gate is never closed), but other streams must wait for their window (similarly to configuration 1).
Stream~\acrshort{stream1h} is always preferred and should experience the lowest delay.
In \autoref{fig:Generic-CT-GCL2-1}, we observe the effects in a much lower one-way latency for the highest priority stream~\acrshort{stream1h} (compared to \autoref{fig:Generic-CT-GCL1-1}), however with a discontinuous gradient which indicates a relatively high \gls{pdv}.
Due to the small slots, there are collisions where the tails of forwarded streams block the next time slot.
Increasing the slot size, see \autoref{fig:Generic-CT-GCL2-3}, solves this issue for high-priority traffic; however, increases the overall latency of the lower priority streams.
The effects on \acrlong{stream2} and \acrlong{stream3} are similar: the prioritization for the high priority stream is guaranteed, and a larger slot size decrease the \gls{pdv} (see Figures~\ref{fig:real-dataset} and~\ref{fig:spot-dataset}).

\subsubsection{GCL Configuration 3}
\autoref{fig:Generic-CT-GCL3-1} considers configuration~3 which gives more transmission opportunities to higher priority streams, striving to isolate the stream priorities.
For \acrlong{stream1} and \acrlong{stream3} this works in the expected manner (see \autoref{fig:gcl-comparison-generic-stream} and \autoref{fig:spot-dataset}, respectively).
In \acrlong{stream2}, the large and bursty second priority stream has an effect on the high priority stream:
Due the short \SI{15}{\micro\second} slots, the stream~\acrshort{stream2m} can block the high priority stream from being transmitted. 
Increasing the slot size can be a solution, but reduces the bandwidth for the low priority, high-bursty stream~\acrshort{stream2l}, as shown in \autoref{fig:real-CT-GCL3-3} (vs. \autoref{fig:real-CT-GCL3-1}).
This can even increase the delay of stream~\acrshort{stream2l}, when the frame rate is higher than the \gls{gcl} cycle time, in which case the \gls{dut} needs to buffer more and more frames.

\subsubsection{GCL Configuration 4}
Configuration 4 is similar to configuration~3, but distributes the transmission opportunities over time (the shape resembles a tree).
The target was to provide a priority isolation between the streams, respecting their stream characteristics.
The measurements confirm this behavior except for \acrlong{stream3} with a short slot size, see \autoref{fig:spot-CT-GCL4-1}.
The overall one-way delay increases since there are more slots in which the corresponding stream is not allowed to send and forced to wait.
Nevertheless, we observe a clear distinction between the stream priorities.

\subsubsection{Comparison of GCL Configurations}
\autoref{fig:gcl-comparison-generic-stream} facilitates the comparison of the GCL configurations in one plot.
Figures~\ref{fig:Generic-CT-GCL1-1}, \ref{fig:Generic-CT-GCL2-1}, \ref{fig:Generic-CT-GCL3-1}, \ref{fig:Generic-CT-GCL4-1}, \ref{fig:Generic-CT-GCL1-1-GB}, and \ref{fig:Generic-CT-GCL4-1-GB} are combined in \autoref{fig:gcl-comparison-generic-stream}.
We observe that configuration~3 has the least outliers and clearest differentiation between the stream priorities.
This is to be expected, since configuration~3 increases the number of transmission opportunities linearly with the priority.
Configuration~4 is similar to configuration~3 and achieves similar one-way latencies, although the overall one-way latency increases compared to configuration 3.
It is important to note that the y-axis uses logarithmic scaling.
Therefore, a larger box in the box plots indicates a vastly increased latency variation.
Furthermore, the logarithmically scaled boxplot reveals outliers more clearly.
Especially in configurations~1, 1 with guard band, 2, and 4 without guard band, there are (numerous) outliers.

In summary, for a service with clear priority distinction, the measurement results indicate that a configuration that is similar to the third configuration is recommended.
If only the latency and \gls{pdv} of the highest priority stream is of importance, then configuration~2 can perform slightly better, while degrading the performance of the remaining stream more significantly.

\subsubsection{Slot Size}
With increased slot size (see second row in \autoref{fig:generic-hard-qos}), the overall delay increases since the gates are opened and closed for a longer period and frames from another queue have to wait longer for the next send window.
This can be seen, e.g., in \autoref{fig:Generic-CT-GCL4-1} and~\ref{fig:Generic-CT-GCL4-3}, where the average latency increases from \SI{18.4}{\micro\second} to \SI{34.0}{\micro\second} for stream~\acrshort{stream1h}.
On the other hand, the tail-latency can be slightly decreased due to a more relaxed time frame where the gate is open (see \autoref{fig:Generic-CT-GCL4-1}).
In general, we observe from all four measurements with an increased slot size that the shapes of the \gls{ccdf} curves are \textit{smoother}.
That means the \gls{pdv} is distributed over a longer period, since the slot size increases and therefore the slots are more contiguous than with a higher frequent \gls{gcl}.

The measurements with different slot sizes show that the \gls{gcl} slots need to be aligned with the maximum burst size of the streams.
Otherwise, the streams can block each other, which cannot be resolved with guard bands in all cases.
In conclusion, our setting with a slot size of one is not suitable for every stream set, but clearly demonstrates the importance of this slot size parameter.

\subsubsection{Guard Band}
Theoretically, a guard band should reduce the effect, that the tail of long frames blocks the next time slot.
In the bottom row in \autoref{fig:generic-hard-qos}, we investigate the effects of an added guard band.
The guard band should protect the highest priority stream~\acrshort{stream1h} from interference from the lower priority streams (\acrshort{stream1m} and~\acrshort{stream1l}) and is configured to always have a slot size of 1, regardless of the slot length for the test data, since the guard band should only cover the worst case: A full lower-priority Ethernet frame is enqueued for transmission right before the gate closes.
If this occurs at the end of a cycle (usually the highest priority gets the first transmission window in a cycle), this could interfere and delay the next high priority frame.
In the four \gls{gcl} configurations, it is only necessary to add guard bands for configurations~1 and~4, since the gate for the highest priority is always kept open in the other two configurations.

We observe from \autoref{fig:Generic-CT-GCL1-1-GB} a similar behavior as in \autoref{fig:Generic-CT-GCL1-1}, the guard band has only a minor effect.
However, now the second highest priority exhibits lower one-way delays in the range below 99\%.
An additional guard band combined with short slots has a negative effect, see \autoref{fig:Generic-CT-GCL1-1-GB}, because the guard band reduces the available bandwidth for the streams.

The guard band usage for \gls{gcl} configuration~1 shows even a negative impact on the priority isolation, because the guard band forces the high priority stream to wait.
On the other hand, we can now detect outliers below 99.99\% for stream~\acrshort{stream1m} which can be explained by the too short slot size (see same behavior in \autoref{fig:Generic-CT-GCL1-1}).
With an added guard band, the average latency increases by 16\% compared to the test without guard band.

If the slot size increases to~3, the behavior resembles \autoref{fig:Generic-CT-GCL1-3}; however, with one exception: the total average latency of the highest priority stream increases by 24\%, or \SI{42.8}{\micro\second} for the decade from 99\% to 99.9\%.
This can be explained by the fact of the high frequency of the stream~\acrshort{stream1h}, which is in the same region as the \gls{gcl} cycle time.
Sometimes, the transmission is delayed and therefore misses its time slot and must wait until the next one.

For configuration~4, we observe less changes.
The latency increases in the range below 50\% for \autoref{fig:Generic-CT-GCL4-1-GB} for all three streams.
The mean latency for all streams increases on the other hand, because the time for accessing the channel again increases as well.
Although, this effect is less visible for stream~\acrshort{stream1h}, since there are the most send opportunities reserved for this stream in the \gls{gcl} configuration~4.
In \autoref{fig:Generic-CT-GCL4-3-GB}, we observe only small deviations to \autoref{fig:Generic-CT-GCL4-3}, but we can detect a similar effect as for the same configuration with a slot size of~1 (shift of latency caused by the additional guard band period).

For all configurations applies: a negative effect of the guard band is the bandwidth reduction for all other streams.
However, the proportion of the guard band in relation to the overall cycle time is configurable, i.e., by extending the slot sizes the channel utilization can be enhanced.

\subsubsection{Traffic Burstiness}
Another stream characteristic is the traffic burstiness: \acrlong{stream1} and \acrlong{stream3} are mostly cyclic, whereas the \acrlong{stream2l} stream of \acrlong{stream2} has a bursty behavior, stream \acl{stream3l} is slightly bursty.
Where \gls{tas}~1 provide an equal channel bandwidth of \SI{250}{\mega\bit\per\second} for each stream, the bandwidth slightly changes if a guard-band is applied. For small slot-size each streams has still \SI{200}{\mega\bit\per\second} and for large slotsize there are \SI{230}{\mega\bit\per\second} available.
The \acrlong{stream1} have a fixed data rate of \SI{7.6}{\mega\bit\per\second}, \SI{5.1}{\mega\bit\per\second} and \SI{3.0}{\mega\bit\per\second}. So without bursts the data streams are not affected by congestion.
If the \gls{tas} has a slot size of~1, then a burst of two frames will be most likely (and a burst of three or more frames will certainly) not be forwarded immediately.
The influence of frame burstiness can be observed by comparing the first and second row in \autoref{fig:real-dataset} (see \gls{gcl} configurations). 
We observe that the stream~\acrshort{stream2l} is strongly delayed because the \gls{gcl} configuration does not respect the stream particularities.
Also, increasing the \gls{gcl} slot size to~3 does not provide enough bandwidth for stream~\acrshort{stream2l}.
In comparison, the performance of the slightly bursty stream~\acrshort{stream3l} is improved by providing more bandwidth with a slot size of~3, see second and third row of \autoref{fig:spot-dataset}.

\subsubsection{Summary}
Comparing \autoref{sec:evaluation-soft-qos} and \autoref{sec:evaluation-hard-qos} does not reveal consistent advantages of traffic shaping and hard scheduling.
The \gls{tas} does not always perform better, especially for the average latency.
For the \gls{tas}, we observe an upper (a guaranteed) limit of latency, but the \gls{pdv} varies often widely.
However, the explanation for this delay variation is simple: our senders are \underline{not} synchronized with the gates on the switch.
The measurement runs were started randomly (compared to the cycle of the switch).
Also, we have discussed in \autoref{sec:evaluation-cycle-time-measurement} that our tools for sending data depend on the host machine's precision.
Therefore, frames arrive at the switch not necessarily when the corresponding gate is open.
The frames are then enqueued and must wait until the gate opens again.
This queuing is not efficient and can significantly increase the one-way delay.

\subsection{Time-aware Traffic Shaping with Synchronized Senders}
\label{sec:evaluation-tas-sender}
\begin{figure}[bt]
    \centering
    \includegraphics{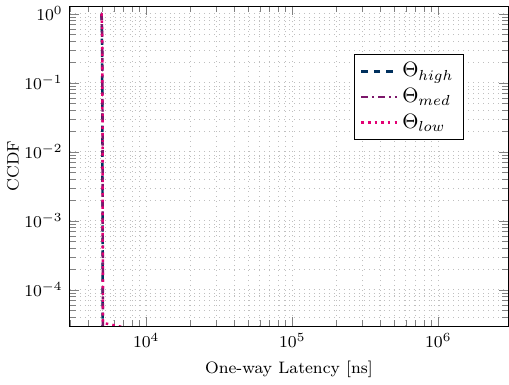}
    \caption{Measurement with \gls{tas} applied at sender and intermediate switch to synchronize sender packet transmission with switch gate states to avoid queuing delay, for \acrlong{stream1}. Ideally, Figure~\ref{fig:baseline-generic} and this figure should be identical, implying that an optimized and distributed \gls{tas} can achieve the same latencies as dedicated communication paths.}
    \label{fig:txinject}
\end{figure}

The \gls{tas} scheduling can be used for stream priority isolation preventing the streams influencing each other negatively.
But this feature comes at a price: reduced bandwidth and additional delay for all streams.
An optimally configured \gls{tas} setup should have all components configured so as to avoid collisions.
If a data source is sending traffic already periodically, the source can be synchronized to the network.

In this section, we examine for the first time a strategy to achieve near minimal latency and \gls{pdv} in a measurement testbed using \acrlong{cots} hardware.
As described in \autoref{sec:taprio-description}, the Linux kernel provides tools to use a software \gls{tas}.
A software \gls{tas} can align (time-synchronize) the sender nodes with the switch since the network testbed nodes are already time-synchronized via \gls{ptp} with the \gls{dut}.
The suggested technology uses buffering on the sender to shape the packet traffic even before it is transmitted by the sender node.
Essentially, we \textit{distribute} the \gls{gcl} configuration on the switch over the network nodes with some adjustments for path delay and other system inaccuracies~\cite[Sec.~V-D-1]{nrw18}.
This is realistic as in real systems, such as in robotics systems, the cycle of the robot nodes can be adjusted to be time-synchronized with the time-aware network architecture.
The described approach dramatically reduces the latency and \gls{pdv}, even on generic systems, such as the used COTS hardware in our TSN-FlexTest testbed.

In \autoref{fig:txinject}, we show that it is possible to synchronize a generic application running on \gls{cots} hardware with the \gls{tas} by utilizing a scheduling on the transmitter side itself.
With this configuration the packet delay caused by queuing in the network can be reduced to a minimum, compared to the baseline measurements in \autoref{fig:baseline}.
\autoref{fig:txinject} shows the CCDF of the one-way packet latency with the time synchronization distributed to the sender nodes.
From a comparison with \autoref{fig:baseline-generic} we conclude with the synchronized senders using a distributed \gls{tas}, we achieve essentially the same low one-way latencies as in our baseline measurement (when only a single stream was transmitted in \autoref{fig:baseline-generic}), but now with added cross traffic of \SI{2}{\giga\bit\per\second} transmitted at the same time in \autoref{fig:txinject}.
There is close to no \gls{pdv} for the streams from \acrlong{stream1} while the \gls{dut} is forced to drop at least half of the incoming data packets.
At the same time, the one-way latency remains at a minimum, i.e., \SI{4968}{\nano\second} and a standard deviation of \SI{40}{\nano\second} for stream~\acrshort{stream1h}.

This approach with synchronized senders using a distributed \gls{tas} allows all types of interference, such as high-bandwidth file sharing, on the same physical connection while providing exceptional performance for critical services.
However, there is a downside when using a software \gls{tas}: it is not possible to achieve very short cycles compared to the hard \gls{tas} in the \gls{dut}.
The time a gate was opened was therefore increased by a factor of 10 while still providing the same throughput.

\section{Conclusion}
\label{sec:conclusion-future-work}
We introduced TSN-FlexTest, a flexible highly-precise \gls{tsn} measurement testbed, for evaluating \gls{tsn} features.
Following a comprehensive review of the available hardware and software, TSN-FlexTest is designed with \acrfull{cots} hardware and open-source software to foster further studies, allowing lost-cost \gls{tsn} testbed measurements.
The validation of the TSN-FlexTest testbed highlighted the following features:
The underlying \gls{ptp} clock synchronization provides nanosecond precision, which enables our high-precision \gls{tsn} measurements.
The flexible cyclic traffic generator empowers researchers to reproduce various stream characteristics without the necessity of obtaining expensive devices, e.g., to simulate multi-modal feedback in the context of the \acrlong{ti}.

We conducted extensive evaluation studies with our TSN-FlexTest testbed with multiple stream sets. We measured \glspl{kpi} for widely used \acrfull{qos} configurations, including \textit{net-neutral} transmissions, \acrlong{spq}, and the usage of a \acrfull{tas}.
We found that the \gls{tas} can provide an upper bounded one-way delay, although the \acrfull{pdv} may vary significantly in certain cases, especially for sporadic data streams, e.g., video traffic.
However, adjusting the \gls{gcl} in terms of slot size and inserting guard bands can mitigate the \gls{pdv} to some extent. 
We identified the sender node behavior as the root cause of the high \gls{pdv}: randomly time-shifted data transmissions by sender nodes can introduce high delays due to possibly closed gate states at a switch.
We conducted measurements with sending nodes synchronized to the state of the \gls{tsn} switch. 
By leveraging solely \gls{cots} hardware and open-source software in the TSN-FlexTest testbed, we have thus been able to achieve a level of \gls{qos} that is comparable to a dedicated link while accommodated multiple parallel streams, with a more than 200\% over-saturation of the link.
We provide the source code of the TSN-FlexTest testbed publicly available at \href{https://github.com/5GCampus/tsn-testbed}{https://github.com/5GCampus/tsn-testbed}.

\subsection{Lessons Learned}
Our experience with the TSN-FlexTest testbed setup provided valuable lessons about tools and issues that can be helpful in future research and development:

\begin{itemize}
    \item To obtain reliable measurement results, all power-saving techniques on all \gls{cots} hardware components should be deactivated. This mostly entails adjusting the CPU clock. Since we leveraged generic CPUs and a general-purpose operating system (Ubuntu Linux), it is furthermore recommended to increase process isolation. Under Linux, it is possible to configure the scheduler to not use certain CPU cores and further to assign processes/threads to these cores. This increased process isolation improves the measurement precision by reducing interrupts from co-running services on the same machine blocking the actual measurement.
    \item Drivers may support fewer features than the official data sheet reports. For example, the documented \textit{one-step timestamping} feature of the used Intel \gls{nic} had to be manually patched into the \texttt{igb} driver. The patch is available in the \href{https://github.com/5GCampus/tsn-testbed}{https://github.com/5GCampus/tsn-testbed} repository.
    \item It is recommended to use the latest software: Particularly, the employed \texttt{tcpreplay} tool behaves differently below version \texttt{4.3.4}, where the frame transmission times become longer. At the point of time when the measurements were conducted, the Ubuntu repository did not offer the already patched version. More information on the cycle time evaluation can be found in \autoref{sec:evaluation-cycle-time-measurement}.
    \item We encountered hardware problems while stress-testing with traffic generators, such as \emph{MoonGen}. We conducted the measurements with two other \gls{tsn} switches. During high traffic loads and \gls{ptp} long-term measurement, the \gls{tsn} switches became unstable, with one device irreversibly breaking and needing to be returned to the manufacturer. Our assumption is that not all features of the switches are designed to be run under all conditions.
    \item Our results confirmed previous observations (see \autoref{sec:packgen}) that non-\gls{dpdk}-based software packet generators cannot saturate a link for all frame sizes.
    \item Handling large data sets with high precision requires several iterations with enhancements for the software used for evaluation. We want to highlight one very specific aspect of the measurement process, that can lead to deviations in the results. A Unix epoch timestamp with nanosecond resolution, as it is employed in the TSN-FlexTest testbed, should not be stored in the \textit{float} data type (see Python \texttt{PEP410} issue for more information); there is a loss of precision after 194 days of absolute time. This means the time cannot be stored anymore with a \SI{1}{\nano\second} resolution if the \enquote{time} is greater than July 14, 1970. In order to maintain correctness inside the evaluation, the more precise data type of \texttt{Decimal()} must be chosen. This is specific to the \textit{Python} language. However, this can occur in other programming languages as well and especially \textit{Python} is very well suited for evaluation because of the rich tools for statistical data analysis.
\end{itemize}

\subsection{Future Work}
The proposed \acrshort{tsn}-FlexTest testbed expands opportunities for investigating new methods and for evaluating existing methods through measurements on a real hardware testbed.
We proceed to summarize some potential future research directions. 
A first step could be to investigate the effectiveness of other \gls{tsn} standards, such as \acrfull{fp}, \acrfull{frer}, and \acrfull{srp}.
Future work could also consider a wider set of traffic profiles.
Furthermore, the TSN-FlexTest testbed topology could be extended to multiple switches and a larger number of sending and receiving nodes in future work. 

Evaluating commercial industry-grade hardware components on the market can reveal additional information about their behaviors and performance levels and may provide insights for academic researchers.
For example, using a time-coordinated CPU in the testbed could potentially improve the precision of the \gls{tsn} network, which could be investigated further.

Also, the configuration can have a significant impact on the \gls{tsn} testbed performance.
Consequently, one important future work direction is to design a framework for determining proper pre-configuration for static topologies or reconfiguring parameters according to changes in traffic profile and network resources over time.

Finally, integrating \gls{tsn} with wireless technologies can significantly improve flexibility and mobility, which can be advantageous for a wide range of new use cases~\cite{hoe2021imp,nav2020sur}.
Integrating the \gls{5gs} with a \gls{tsn} network is considered in recent 3GPP standards~\cite{3gpp17} in an architecture where the \gls{5gs} acts as a virtual \gls{tsn} bridge.
The \acrshort{tsn}-FlexTest testbed can be employed to develop a \gls{5gs} in accordance with recent 3GPP standards to enable measurement based performance evaluations for a 5G-\gls{tsn} network.

\appendices

\section{Measurements with Real-World Data Sets}
\label{appendix:evaluation-real-world-data}
\begin{figure*}
    \begin{subfigure}[t]{0.245\textwidth}
        \centering
        \includegraphics{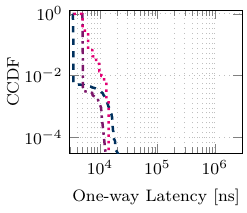}
        \caption{w/o cross traffic}
        \label{fig:real}
    \end{subfigure}
    \hfill
    \begin{subfigure}[t]{0.245\textwidth}
        \centering
        \includegraphics{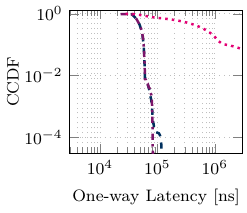}
        \caption{w/ cross traffic}
        \label{fig:real-CT}
    \end{subfigure}
    \hfill
    \begin{subfigure}[t]{0.245\textwidth}
        \centering
        \includegraphics{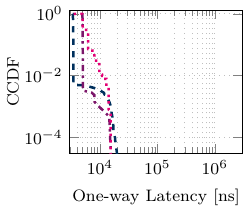}
        \caption{w/o cross traffic and SPQ}
        \label{fig:real-SPQ}
    \end{subfigure}
    \hfill
    \begin{subfigure}[t]{0.245\textwidth}
        \centering
        \includegraphics{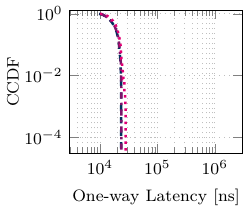}
        \caption{w/ cross traffic and SPQ}
            \label{fig:real-CT-SPQ}
    \end{subfigure}
       
    \begin{subfigure}[t]{0.245\textwidth}
        \centering
        \includegraphics{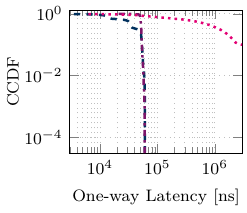}
        \caption{\gls{gcl} 1, Slot Size 1}
        \label{fig:real-CT-GCL1-1}
    \end{subfigure}
    \hfill
    \begin{subfigure}[t]{0.245\textwidth}
        \centering
        \includegraphics{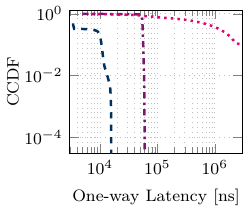}
        \caption{\gls{gcl} 2, Slot Size 1}
        \label{fig:real-CT-GCL2-1}
    \end{subfigure}
    \hfill
    \begin{subfigure}[t]{0.245\textwidth}
        \centering
        \includegraphics{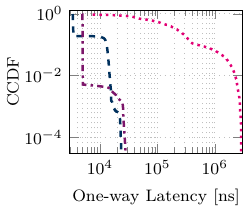}
        \caption{\gls{gcl} 3, Slot Size 1}
        \label{fig:real-CT-GCL3-1}
    \end{subfigure}
    \hfill
    \begin{subfigure}[t]{0.245\textwidth}
        \centering
        \includegraphics{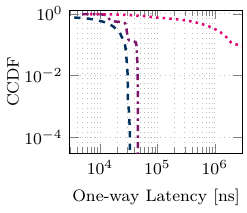}
        \caption{\gls{gcl} 4, Slot Size 1}
        \label{fig:real-CT-GCL4-1}
    \end{subfigure}
    
    \begin{subfigure}[t]{0.245\textwidth}
        \centering
        \includegraphics{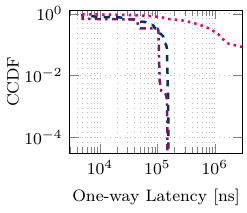}
        \caption{\gls{gcl} 1, Slot Size 3}
        \label{fig:real-CT-GCL1-3}
    \end{subfigure}
    \hfill
    \begin{subfigure}[t]{0.245\textwidth}
        \centering
        \includegraphics{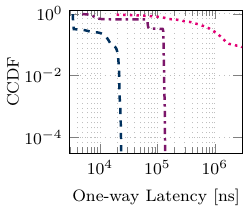}
        \caption{\gls{gcl} 2, Slot Size 3}
        \label{fig:real-CT-GCL2-3}
    \end{subfigure}
    \hfill
    \begin{subfigure}[t]{0.245\textwidth}
        \centering
        \includegraphics{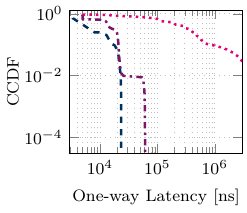}
        \caption{\gls{gcl} 3, Slot Size 3}
        \label{fig:real-CT-GCL3-3}
    \end{subfigure}
    \hfill
    \begin{subfigure}[t]{0.245\textwidth}
        \centering
        \includegraphics{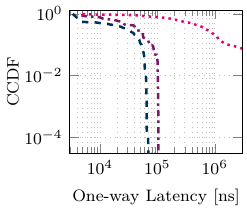}
        \caption{\gls{gcl} 4, Slot Size 3}
        \label{fig:real-CT-GCL4-3}
    \end{subfigure}
        
    \begin{subfigure}[t]{0.245\textwidth}
        \centering
        \includegraphics{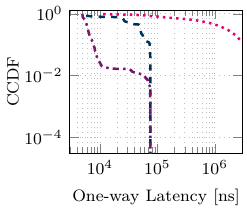}
        \caption{\gls{gcl} 1 w/ GB, Slot Size 1}
        \label{fig:real-CT-GCL1-1-GB}
    \end{subfigure}
    \hfill
    \begin{subfigure}[t]{0.245\textwidth}
        \centering
        \includegraphics{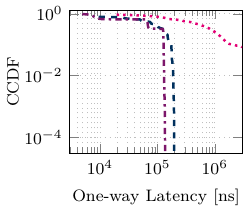}
        \caption{\gls{gcl} 1 w/ GB, Slot Size 3}
        \label{fig:real-CT-GCL1-3-GB}
    \end{subfigure}
    \hfill
    \begin{subfigure}[t]{0.245\textwidth}
        \centering
        \includegraphics{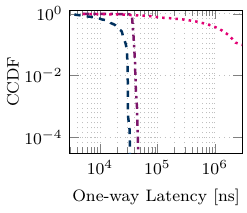}
        \caption{\gls{gcl} 4 w/ GB, Slot Size 1}
        \label{fig:real-CT-GCL4-1-GB}
    \end{subfigure}
    \hfill
    \begin{subfigure}[t]{0.245\textwidth}
        \centering
        \includegraphics{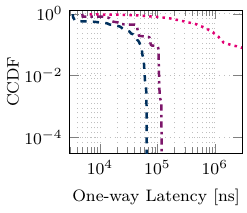}
        \caption{\gls{gcl} 4 w/ GB, Slot Size 3}
        \label{fig:real-CT-GCL4-3-GB}
    \end{subfigure}
    \caption{Measurement of \acrlong{stream2} for configurations in \autoref{sec:testbed-evaluation}:
             We measured the streams with equal priority (Figures~a and~b), with enabled \gls{spq} (Figures~c and~d), and with \gls{gcl} configurations described in \autoref{sec:evaluation-hard-qos} (see Figures~e to~p).
             We conducted measurements without (w/o) and with (w/) cross traffic and enabled Guard Band (GB) for protecting the highest priority traffic.
             The \gls{gcl} slot size was varied between~1 and~3 (see \autoref{sec:evaluation-hard-qos}).}
    \label{fig:real-dataset}
\end{figure*}

\begin{figure*}
    \begin{subfigure}[t]{0.245\textwidth}
        \centering
        \includegraphics{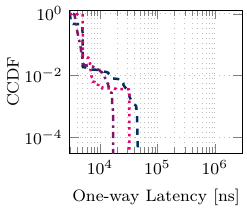}
        \caption{w/o cross traffic}
        \label{fig:spot}
    \end{subfigure}
    \hfill
    \begin{subfigure}[t]{0.245\textwidth}
        \centering
        \includegraphics{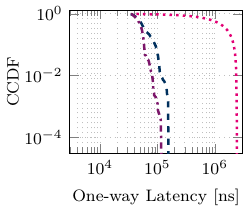}
        \caption{w/ cross traffic}
        \label{fig:spot-CT}
    \end{subfigure}
    \hfill
    \begin{subfigure}[t]{0.245\textwidth}
        \centering
        \includegraphics{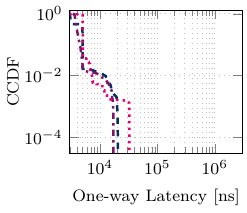}
        \caption{w/o cross traffic, SPQ}
        \label{fig:spot-SPQ}
    \end{subfigure}
    \hfill
    \begin{subfigure}[t]{0.245\textwidth}
        \centering
        \includegraphics{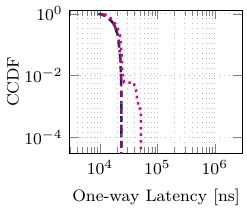}
        \caption{w/ cross traffic, SPQ}
        \label{fig:spot-CT-SPQ}
    \end{subfigure}
    
    \begin{subfigure}[t]{0.245\textwidth}
        \centering
        \includegraphics{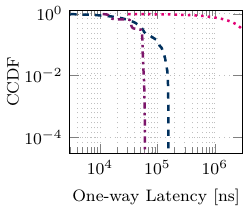}
        \caption{\gls{gcl} 1, Slot Size 1}
        \label{fig:spot-CT-GCL1-1}
    \end{subfigure}
    \hfill
    \begin{subfigure}[t]{0.245\textwidth}
        \centering
        \includegraphics{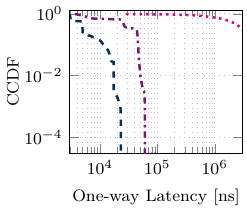}
        \caption{\gls{gcl} 2, Slot Size 1}
        \label{fig:spot-CT-GCL2-1}
    \end{subfigure}
    \hfill
    \begin{subfigure}[t]{0.245\textwidth}
        \centering
        \includegraphics{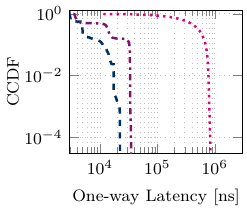}
        \caption{\gls{gcl} 3, Slot Size 1}
        \label{fig:spot-CT-GCL3-1}
    \end{subfigure}
    \hfill
    \begin{subfigure}[t]{0.245\textwidth}
        \centering
        \includegraphics{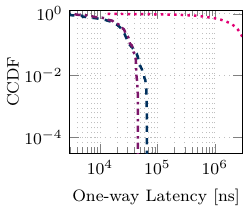}
        \caption{\gls{gcl} 4, Slot Size 1}
        \label{fig:spot-CT-GCL4-1}
    \end{subfigure}
    
    \begin{subfigure}[t]{0.245\textwidth}
        \centering
        \includegraphics{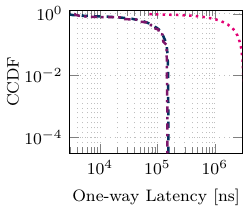}
        \caption{\gls{gcl} 1, Slot Size 3}
        \label{fig:spot-CT-GCL1-3}
    \end{subfigure}
    \hfill
    \begin{subfigure}[t]{0.245\textwidth}
        \centering
        \includegraphics{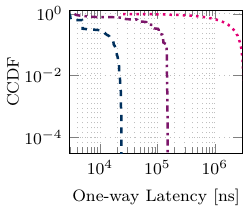}
        \caption{\gls{gcl} 2, Slot Size 3}
        \label{fig:spot-CT-GCL2-3}
    \end{subfigure}
    \hfill
    \begin{subfigure}[t]{0.245\textwidth}
        \centering
        \includegraphics{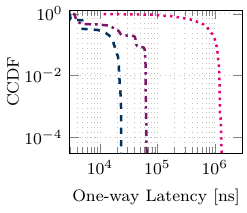}
        \caption{\gls{gcl} 3, Slot Size 3}
        \label{fig:spot-CT-GCL3-3}
    \end{subfigure}
    \hfill
    \begin{subfigure}[t]{0.245\textwidth}
        \centering
        \includegraphics{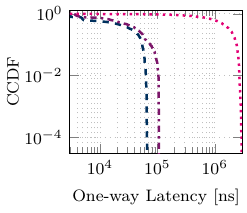}
        \caption{\gls{gcl} 4, Slot Size 3}
        \label{fig:spot-CT-GCL4-3}
    \end{subfigure}
    
    \begin{subfigure}[t]{0.245\textwidth}
        \centering
        \includegraphics{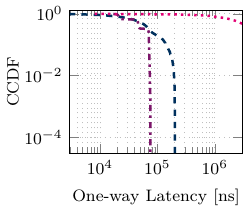}
        \caption{\gls{gcl} 1 w/ GB, Slot Size 1}
        \label{fig:spot-CT-GCL1-1-GB}
    \end{subfigure}
    \hfill
    \begin{subfigure}[t]{0.245\textwidth}
        \centering
        \includegraphics{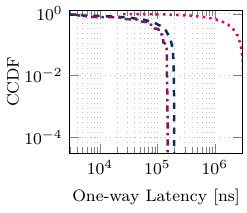}
        \caption{\gls{gcl} 1 w/ GB, Slot Size 3}
        \label{fig:spot-CT-GCL1-3-GB}
    \end{subfigure}
    \hfill
    \begin{subfigure}[t]{0.245\textwidth}
        \centering
        \includegraphics{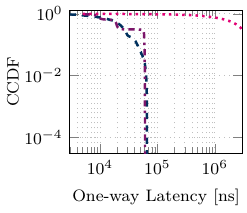}
        \caption{\gls{gcl} 4 w/ GB, Slot Size 1}
        \label{fig:spot-CT-GCL4-1-GB}
    \end{subfigure}
    \hfill
    \begin{subfigure}[t]{0.245\textwidth}
        \centering
        \includegraphics{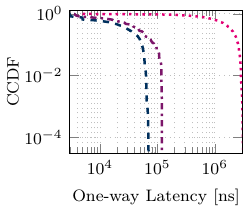}
        \caption{\gls{gcl} 4 w/ GB, Slot Size 3}
        \label{fig:spot-CT-GCL4-3-GB}
    \end{subfigure}
    \caption{Measurement of \acrlong{stream3} for configurations in \autoref{sec:testbed-evaluation}:
             We measured the streams with equal priority (Figures~a and~b), with enabled \gls{spq} (Figures~c and~d), and with \gls{gcl} configurations described in \autoref{sec:evaluation-hard-qos} (see Figures~e to~p).
             We conducted measurements without (w/o) and with (w/) cross traffic and enabled Guard Band (GB) for protecting the highest priority traffic.
             The slot size of the \gls{gcl} was varied between~1 and~3 (see \autoref{sec:evaluation-hard-qos}).}
    \label{fig:spot-dataset}
\end{figure*}

\begin{figure*}
    \begin{subfigure}[b]{0.245\textwidth}
        \centering
        \includegraphics{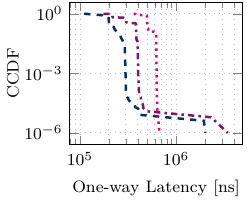}
        \caption{\gls{ccdf} of stream set~\acrshort{stream1}}
    \end{subfigure}
    \hfill
    \begin{subfigure}[b]{0.245\textwidth}
        \centering
        \includegraphics{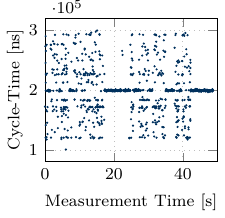}
        \caption{stream~\acrshort{stream1h}}
    \end{subfigure}
    \hfill
    \begin{subfigure}[b]{0.245\textwidth}
        \centering
        \includegraphics{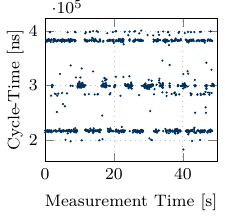}
        \caption{stream~\acrshort{stream1m}}
    \end{subfigure}
    \hfill
    \begin{subfigure}[b]{0.245\textwidth}
        \centering
        \includegraphics{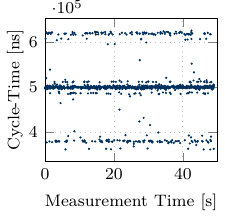}
        \caption{stream~\acrshort{stream1l}}
    \end{subfigure}
    \vspace{0.1cm}\\
    \begin{subfigure}[b]{0.245\textwidth}
        \centering
        \includegraphics{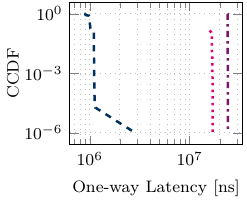}
        \caption{\gls{ccdf} of stream set~\acrshort{stream2}}
    \end{subfigure}
    \hfill
    \begin{subfigure}[b]{0.245\textwidth}
        \centering
        \includegraphics{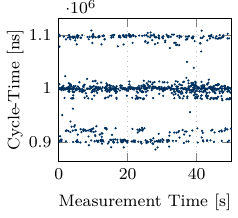}
        \caption{stream~\acrshort{stream2h}}
    \end{subfigure}
    \hfill
    \begin{subfigure}[b]{0.245\textwidth}
        \centering
        \includegraphics{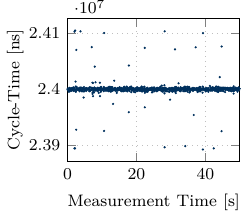}
        \caption{stream~\acrshort{stream2m}}
    \end{subfigure}
    \hfill
    \begin{subfigure}[b]{0.245\textwidth}
        \centering
        \includegraphics{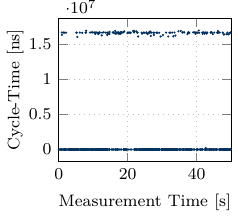}
        \caption{stream~\acrshort{stream2l}}
    \end{subfigure}
    \vspace{0.1cm}\\
    \begin{subfigure}[b]{0.245\textwidth}
        \centering
        \includegraphics{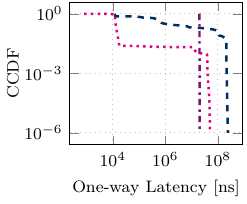}
        \caption{\gls{ccdf} of stream set~\acrshort{stream3}}
    \end{subfigure}
    \hfill
    \begin{subfigure}[b]{0.245\textwidth}
        \centering
        \includegraphics{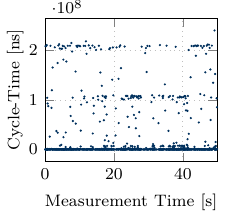}
        \caption{stream~\acrshort{stream3h}}
    \end{subfigure}
    \hfill
    \begin{subfigure}[b]{0.245\textwidth}
        \centering
        \includegraphics{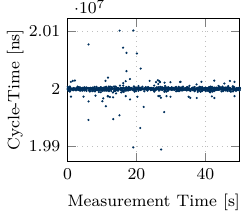}
        \caption{stream~\acrshort{stream3m}}
    \end{subfigure}
    \hfill
    \begin{subfigure}[b]{0.245\textwidth}
        \centering
        \includegraphics{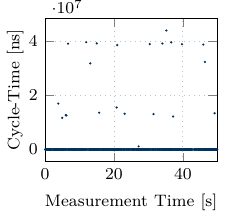}
        \caption{stream~\acrshort{stream3l}}
    \end{subfigure}
    \caption{The figure illustrates the different cycle times of the three stream sets. In each row, the first figure shows the \gls{ccdf} for a comparison of all three streams within a stream set, and then the three streams individually with a plot showing the behavior over time. Since the streams especially in stream set~\acrshort{stream2} and~\acrshort{stream3} have rather varying characteristics, the y-axis is \underline{not} fixed, otherwise, outliers may not be easily visible.}
    \label{fig:appendix-ct}
\end{figure*}

\begin{figure*}
    \begin{subfigure}[t]{0.3\textwidth}
        \centering
        \includegraphics{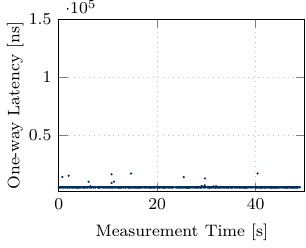}
        \caption{w/o cross traffic}
    \end{subfigure}
    \hfill
    \begin{subfigure}[t]{0.3\textwidth}
        \centering
        \includegraphics{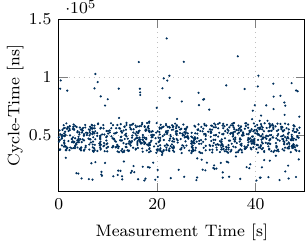}
        \caption{w/ cross traffic}
    \end{subfigure}
    \hfill
    \begin{subfigure}[t]{0.3\textwidth}
        \centering
        \includegraphics{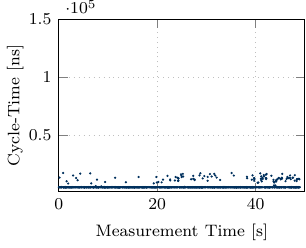}
        \caption{\gls{spq} w/o cross traffic}
    \end{subfigure}
    \vspace{0.5cm}\\
    \begin{subfigure}[t]{0.3\textwidth}
        \centering
        \includegraphics{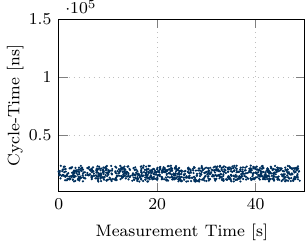}
        \caption{\gls{spq} w/ cross traffic}
    \end{subfigure}
    \hfill
    \begin{subfigure}[t]{0.3\textwidth}
        \centering
        \includegraphics{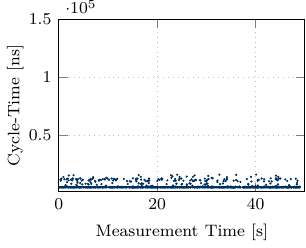}
        \caption{\gls{tas} with \gls{gcl} configuration 2 w/ cross traffic}
    \end{subfigure}
    \hfill
    \begin{subfigure}[t]{0.3\textwidth}
        \centering
        \includegraphics{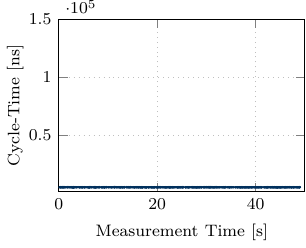}
        \caption{Distributed \gls{tas} w/ cross traffic}
    \end{subfigure}
    \caption{The figure shows the measurement for stream~\acrshort{stream1h} for different applied \gls{qos} features: no applied \gls{qos}, \gls{spq}, \gls{tas}, and distributed \gls{tas}. \autoref{fig:evaluation-no-qos}, \ref{fig:evaluation-soft-qos}, and~\ref{fig:generic-hard-qos} are using the \gls{ccdf} to visualize the measurement results. Here, the plot elucidates the behavior over time. The effects of additional cross traffic is clearly noticeable in the respective configurations. Furthermore, we can observe the advantages of all applied \gls{qos} features. However, the plots highlight the superiority of the \gls{tas}, in particular with the distributed approach.}
    \label{fig:appendix-pdv}
\end{figure*}

This appendix provides an overview of the conducted measurements with \acrlong{stream2} and \acrlong{stream3}, following the configurations described in \autoref{sec:testbed-evaluation}.
The purpose of these measurements is to gain insights into \gls{tsn} behaviors for \textit{real-world} data sets.
As described in \autoref{sec:streamdescription}, real data sources behave differently compared to our generic data set.
\autoref{fig:real-dataset} and \autoref{fig:spot-dataset} provide an overview of how stream characteristics, e.g., non-periodic and bursty traffic, can influence the one-way delay and \gls{pdv}.
\autoref{fig:appendix-ct} and \autoref{fig:appendix-pdv} allow additional insights for the cycle time and the \gls{pdv}, respectively.

\section*{Acknowledgment}
Funded in part by the German Federal Ministry for Economic Affairs and Climate Action (BMWK) projects “TICCTEC” -- grant 01MC22007A, “5G-OPERA” -- grant 01MJ22008A, and “stic5G” -- grant 01MJ22018C, and by the German Research Foundation (DFG, Deutsche Forschungsgemeinschaft) as part of Germany’s Excellence Strategy – EXC 2050/1 – Project ID 390696704 – Cluster of Excellence “Centre for Tactile Internet with Human-in-the-Loop” (CeTI) of Technische Universität Dresden.
We also would like to thank the open-source community.


\vskip -2\baselineskip plus -1fil
\begin{IEEEbiography}[{\includegraphics[width=1in,height=1.25in,clip,keepaspectratio]{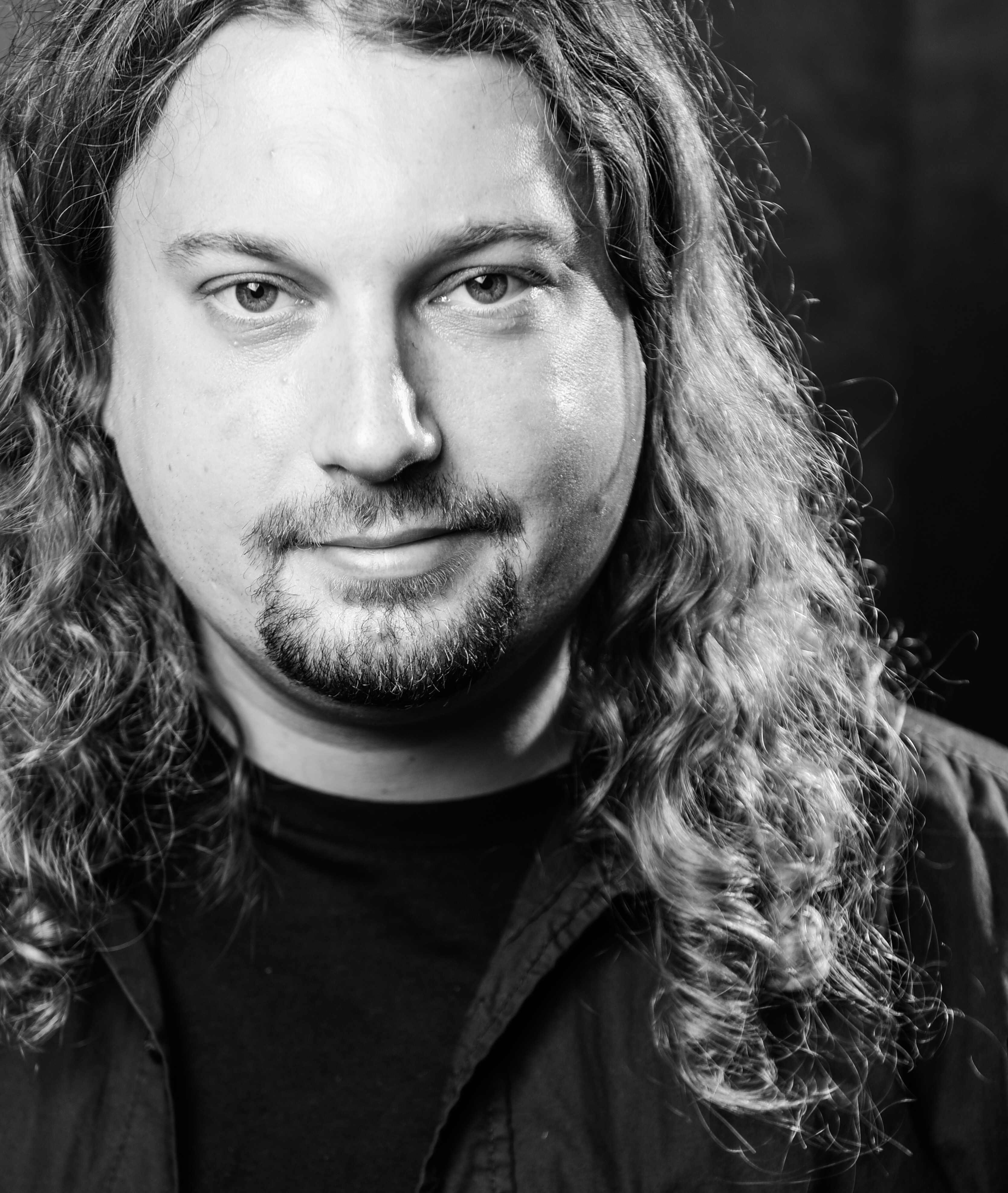}}]%
{Marian Ulbricht} is a PhD student at Technische Universität Dresden. He studied at the Hochschule für Telekommunikation Leipzig (HfTL) for Bachelor and Master in communication technology. The focus of the studies was on embedded systems and microcontroller programming. Since 2015, he is with the InnoRoute GmbH in Munich as software developer and project engineer with focus on \gls{tsn} and network node design.
\end{IEEEbiography}
\vskip -2\baselineskip plus -1fil
\begin{IEEEbiography}[{\includegraphics[width=1in,height=1.25in,clip,keepaspectratio]{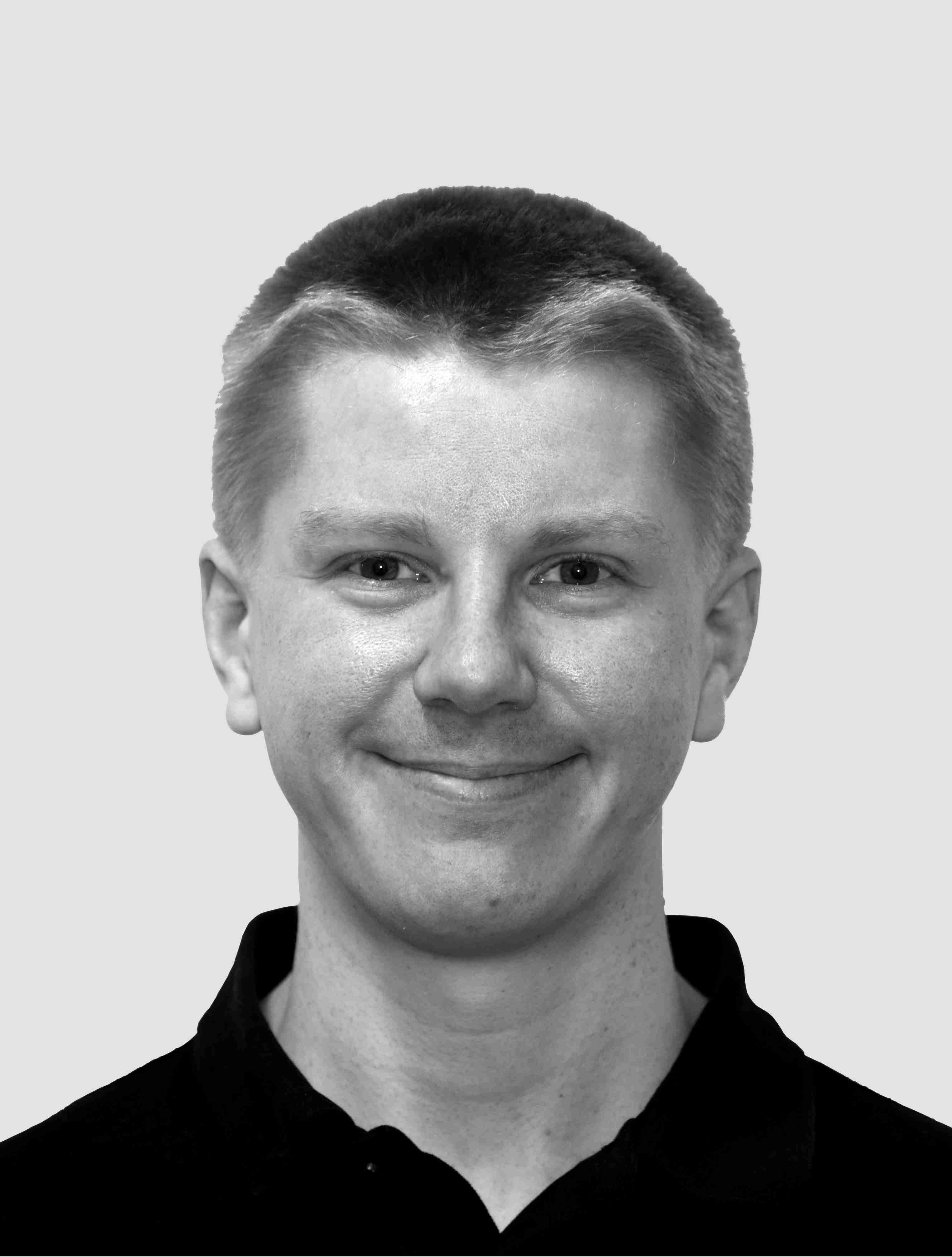}}]%
{Stefan Senk} studied at Technische Universität Dresden where he received his degree \enquote{Diplom-Ingenieur} (Dipl.-Ing.) in electrical engineering in July 2019.
Since late 2019, he is a Ph.D. student at \enquote{Deutsche Telekom Chair of Communication Networks} at TU Dresden.
He is currently working on 5G non-public networks with interest on deterministic communication.
His main research interest is Time-sensitive Networking (TSN) and the focus towards human-machine-collaboration.
\end{IEEEbiography}
\vskip -2\baselineskip plus -1fil
\begin{IEEEbiography}[{\includegraphics[width=1in,height=1.25in,clip,keepaspectratio]{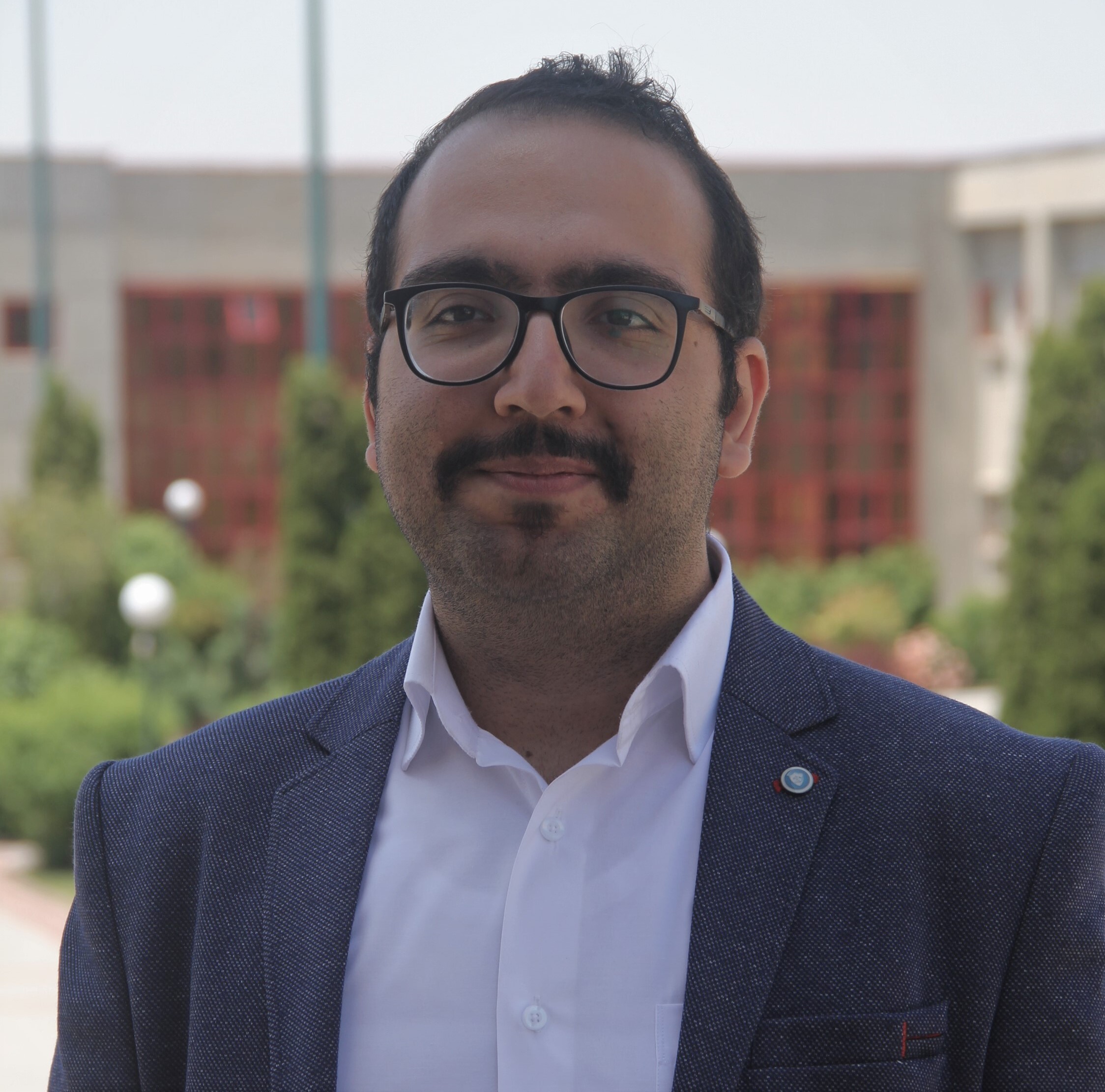}}]%
{Hosein K. Nazari} received his B.Sc. in Information Technology Engineering in 2018 from Institute for Advanced Studies in Basic Sciences (IASBS) in Zanjan, Iran. In 2021, he graduated with an M.Sc. in Computer Science from IASBS. He is currently pursuing a Ph.D. in Electrical and Computer Engineering Department of Technische Universität Dresden, Germany. His research interests include Network coding, IoT, and Time-sensitive networking.
\end{IEEEbiography}
\vskip -2\baselineskip plus -1fil
\begin{IEEEbiography}[{\includegraphics[width=1in,height=1.25in,clip,keepaspectratio]{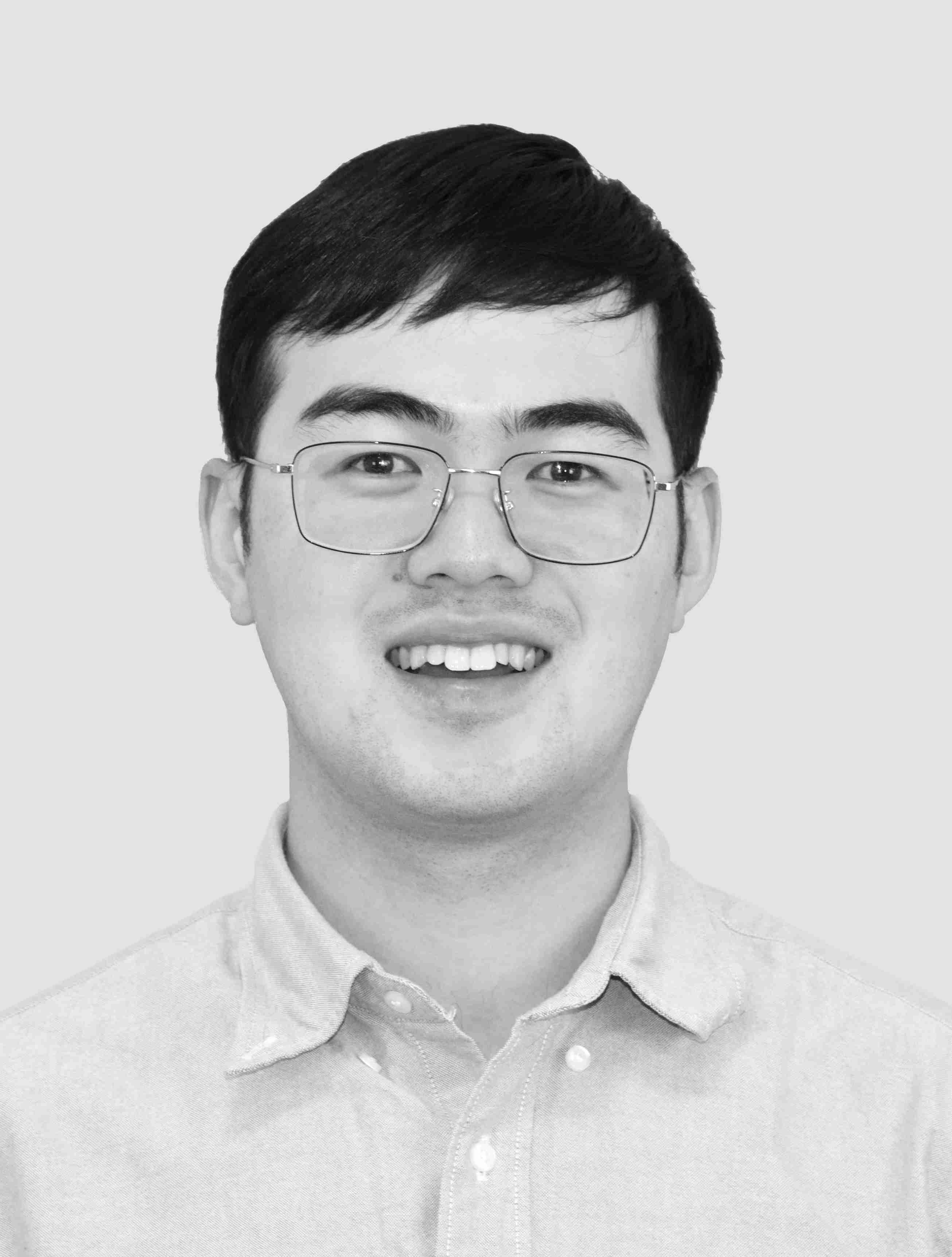}}]%
{How-Hang Liu} How-Hang Liu received his B.Sc. in Communication Engineering in 2014 from National Central University~(NCU) in Taoyuan, Taiwan. In 2016, he graduated with an M.Sc. in Communication Engineering from National Taiwan University~(NTU). He is pursuing a Ph.D. in Electrical and Computer Engineering Department of Technische Universität Dresden, Germany. His research mainly focuses on Time-sensitive networking.
\end{IEEEbiography}
\vskip -2\baselineskip plus -1fil
\begin{IEEEbiography}[{\includegraphics[width=1in,height=1.5in,clip,keepaspectratio]{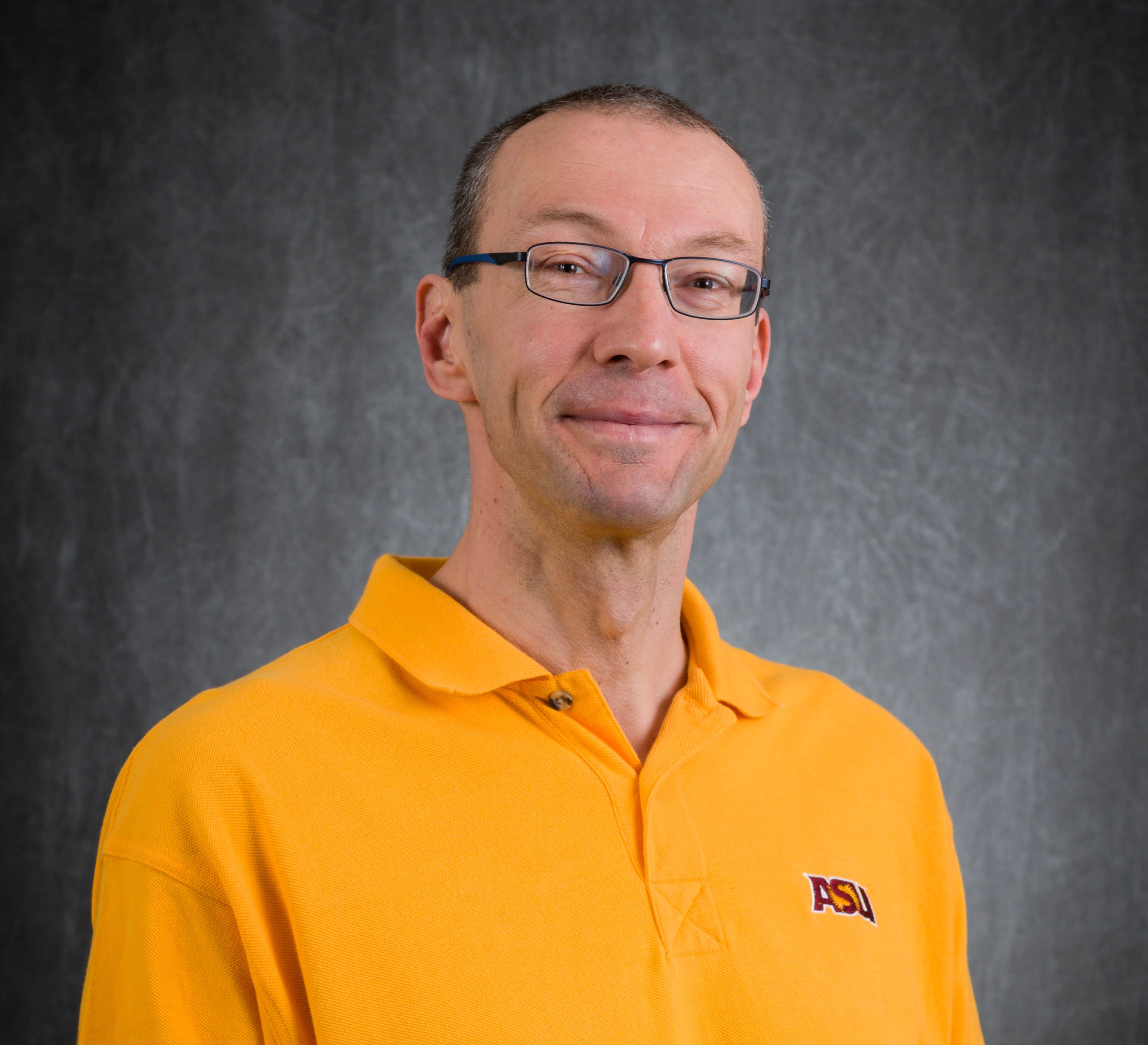}}]
{Martin Reisslein} (S'96-M'98-SM'03-F'14) received the Ph.D. in systems engineering from the University of Pennsylvania, Philadelphia, PA, USA, in 1998. He is currently a Professor with the School of Electrical, Computer, and Energy Engineering, Arizona State University (ASU), Tempe, AZ, USA. He is currently an Associate Editor for \textit{IEEE Access}, \textit{IEEE Transactions on Education}, and \textit{IEEE Transactions on Mobile Computing}.
\end{IEEEbiography}
\vskip -2\baselineskip plus -1fil
\begin{IEEEbiography}[{\includegraphics[width=1in,height=1.25in,clip,keepaspectratio]{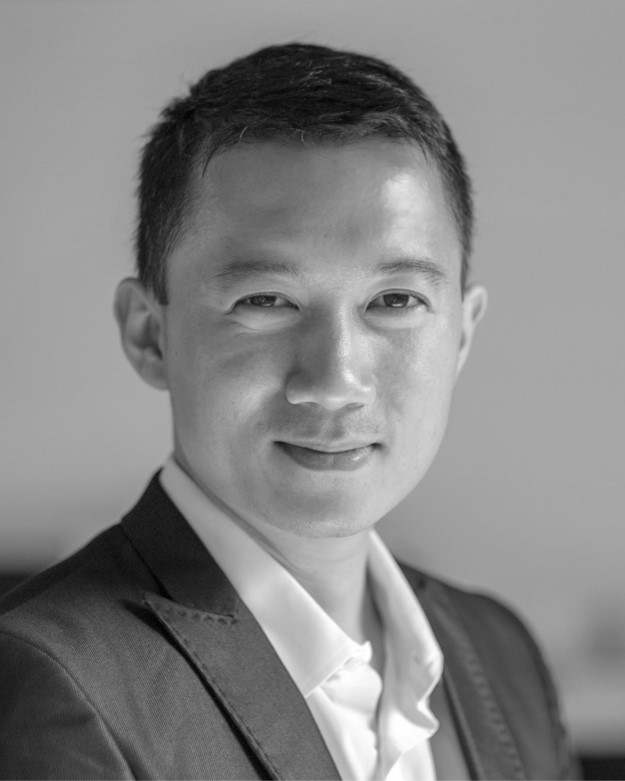}}]
{Giang T. Nguyen}~is currently an Assistant Professor, heading the Haptic Communication Systems research group at the Cluster of Excellence for Tactile Internet with Human-in-the-Loop (CeTI) and Faculty of Electrical and Computer Engineering, TU Dresden, Germany. He received a Ph.D. degree in Computer Science from TU Dresden in 2016. His research interests include network softwarization, in-network computing, and distributed systems, aiming at networked systems’ low latency, flexibility, and resilience to facilitate haptic communication.
\end{IEEEbiography}
\vskip -2\baselineskip plus -1fil
\begin{IEEEbiography}[{\includegraphics[width=1in,height=1.25in,clip,keepaspectratio]{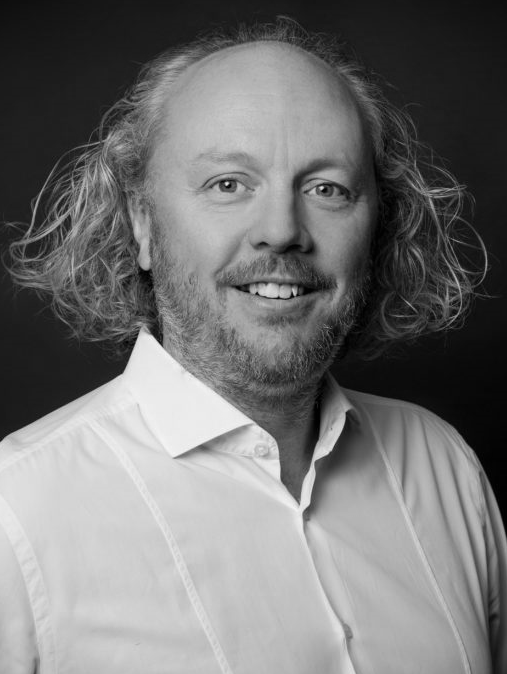}}]%
{Frank H.P. Fitzek} is a Professor and head of the Deutsche Telekom Chair of Communication Networks at the Technische Universität Dresden, coordinating the 5G Lab Germany. He is the spokesman of the DFG Cluster of Excellence CeTI. He received his diploma (Dipl.-Ing.) degree in electrical engineering from the University of Technology Rheinisch-Westfälische Technische Hochschule (RWTH) Aachen, Germany, in 1997 and his Ph.D. (Dr.-Ing.) in electrical engineering from the Technical University Berlin, Germany in 2002 and became Adjunct Professor at the University of Ferrara, Italy in the same year. In 2003 he joined Aalborg University as Associate Professor and later became Professor. In 2005 he won the YRP award for the work on MIMO MDC and received the Young Elite Researcher Award of Denmark. He was selected to receive the NOKIA Champion Award several times in a row from 2007 to 2011. In 2008 he was awarded the Nokia Achievement Award for his work on cooperative networks. In 2011 he received the SAPERE AUDE research grant from the Danish government and in 2012 he received the Vodafone Innovation prize. In 2015 he was awarded the honorary degree Doctor Honoris Causa from Budapest University of Technology and Economics (BUTE).
\end{IEEEbiography}

\end{document}